\documentclass[%
  twoside,
  reprint,
  amsmath,amssymb,
  aps,
  pra,
  nofootinbib,
  superscriptaddress,
  a4paper,
]{revtex4-2}

\usepackage{amssymb,amsmath}
\usepackage{graphicx}
\usepackage[dvipsnames]{xcolor}
\usepackage{physics}
\usepackage{bm}
\usepackage{bbm}
\usepackage[tight]{subfigure}
\usepackage[export]{adjustbox}
\usepackage{enumerate}
\usepackage{appendix}
\usepackage{fancyhdr}
\usepackage{siunitx}
\usepackage{multirow}
\usepackage{rotating,booktabs}
\usepackage[verbose]{placeins}
\usepackage{dsfont}

\usepackage[utf8]{inputenc}
\usepackage[T1]{fontenc}

\usepackage{lipsum}

\usepackage[
centering, includefoot,
text={7.1in,10.2in},
total={6.3in,8.75in},
top=0.8in, left=0.62in,
]{geometry}

\usepackage[
  bookmarks=true,
  colorlinks,
  linkcolor=blue,
  urlcolor=blue,
  citecolor=blue,
  plainpages=false,
  pdfpagelabels,
  final,
  breaklinks=true
]{hyperref}
\hypersetup{
pdftitle={Generation of non--clasical and entangled light states using intense laser--matter interactions}, 
pdfauthor={Th. Lamprou, P. Stammer, J. Rivera-Dean, N. Tsatrafyllis, M. F. Ciappina, M. Lewenstein and P. Tzallas}
}

\makeatletter \def\NAT@def@citea{\def\@citea{\NAT@separator\,}} \makeatother

\newcommand{\PT}[1]{ {\color{black} #1 }}

\begin{document}

\title{Recent developments in the generation of non--classical and entangled light states using intense laser--matter interactions}

\author{Th. Lamprou}
\affiliation{Foundation for Research and Technology-Hellas, Institute of Electronic Structure \& Laser, GR-70013 Heraklion (Crete), Greece}

\author{P. Stammer}
\affiliation{ICFO-Institut de Ciencies Fotoniques, The Barcelona Institute of
Science and Technology, Castelldefels (Barcelona) 08860, Spain}
\affiliation{Atominstitut, Technische Universität Wien, 1020 Vienna, Austria}

\author{J. Rivera-Dean}
\affiliation{ICFO -- Institut de Ciencies Fotoniques, The Barcelona Institute of
Science and Technology, Castelldefels (Barcelona) 08860, Spain}

\author{N. Tsatrafyllis}
\affiliation{Foundation for Research and Technology-Hellas, Institute of Electronic Structure \& Laser, GR-70013 Heraklion (Crete), Greece}

\author{M. F. Ciappina}
\affiliation{Department of Physics, Guangdong Technion -- Israel Institute of Technology, 241 Daxue Road, Shantou, Guangdong, China, 515063}
\affiliation{Technion -- Israel Institute of Technology, Haifa, 32000, Israel}
\affiliation{Guangdong Provincial Key Laboratory of Materials and Technologies for Energy Conversion, Guangdong Technion -- Israel Institute of Technology, 241 Daxue Road, Shantou, Guangdong, China, 515063}

\author{M. Lewenstein}
\affiliation{ICFO-Institut de Ciencies Fotoniques, The Barcelona Institute of
Science and Technology, Castelldefels (Barcelona) 08860, Spain}
\affiliation{ICREA, Pg. Lluís Companys 23, 08010 Barcelona, Spain}

\author{P. Tzallas}
\email{ptzallas@iesl.forth.gr}
\affiliation{Foundation for Research and Technology-Hellas, Institute of Electronic Structure \& Laser, GR-70013 Heraklion (Crete), Greece}
\affiliation{ELI ALPS, ELI-HU Non-Profit Ltd., Wolfgang Sandner utca 3., Szeged, H-6728, Hungary}

\date{\today}

\begin{abstract}

Non--classical and entangled light states are of fundamental interest in quantum mechanics and they are a powerful tool for the emergence of new quantum technologies. The development of methods that can lead to the generation of such light states is therefore of high importance. Recently, it has been demonstrated that intense laser--matter interactions can serve towards this direction. Specifically, it has been shown how the use of fully quantized approaches in intense laser--matter interactions and the process of high harmonic generation, can lead to the generation of high photon--number non-classical and entangled states from the far--infrared (IR) to the extreme--ultraviolet (XUV). Here, after a brief introduction on the fundamentals, we summarize the operation principles of these approaches and discuss the recent developments and future directions of non-classical light engineering using strong light fields with the potential application in ultrafast and quantum information science. These findings represent an important step in the development of novel quantum nonlinear spectroscopy methods, based on the interplay between the quantum properties of light and those of quantum matter.

\end{abstract}

\maketitle

\section{Introduction}

Strong–laser–field physics is a research direction that relies on the use of high-power lasers and has led to fascinating achievements ranging from relativistic particle acceleration (\cite{Mourou2019} and references therein) to high harmonic generation \cite{Gohle2005, Cingoz2012} and attosecond science \cite{Huillier_Nobel_2024, Krausz_Nobel_2024, Agostini_Nobel_2024}. On the other hand, quantum optics has been largely built on the use of low photon number sources and has opened the way for groundbreaking discoveries in quantum technology, advancing investigations in quantum information science including fundamental tests of quantum theory, quantum sensing, communication, computation, visual science etc. \cite{Aspect2022, Zeilinger2022, Clauser2022, Haroche2013, Wineland2013, Galindo_RMP_2002, Acin2018, Walmsay2015, Deutsch2020, Gilchrist2004, Giovannetti2004, Giovannetti2011, Jouguet2013, Lloyd1999, Ralph2003, Joo2011, Schnabel2017}. Despite the tremendous progress, until recently these two directions have remained disconnected. This is because the majority of the interactions in the strong-field regime have been successfully described by semi-classical approximations, e.g, treating the electromagnetic field classically~\cite{kulander_dynamics_1993, corkum_plasma_1993, lewenstein-theory-1994, ABC19}. Ignoring the quantum nature of the field was justified because there was no need to include its quantum properties to explain the observations. 

Because of this, advantages emerging from the connection of quantum optics with strong--laser--field physics remained unexploited for a long time. By connecting these research fields we will be able to answer the fundamental questions concerning: (a) \emph{What is the back-action of intense laser--matter interaction on the quantum state of the driving field?} (b) \emph{What is the quantum state of the radiation after the interaction with matter?} (c) \emph{Can intense laser--matter interaction be used for engineering novel non-classical light states for new applications in quantum information science?} Several groups have attempted in the past to describe different processes taking place in intense laser--matter interactions using fully quantized theoretical approaches \cite{LaGattuta2012, Wang2012, Yangaliev2020, Gauthey1995, Gombkoto2020, Varro2020, Gao2000, Hu2008, Fu2001, Kuchiev1999, Usachenko2002, Bogatskaya2017, Burenkov2010}. However, despite the important information provided in these works, none of these efforts have addressed theoretically and experimentally all of the above questions in a unified way. This has recently changed. The link between strong–laser–field physics, quantum optics, and quantum information science has been developed in the recent past \cite{LCP21, RLP22, RSM22, SRM2023, SRL22, CDF24, Gorlach2020, Stammer_PRL_2024}. For instance, experimental and theoretical studies based on fully quantized approaches and using conditioning schemes \cite{GTK2016, TKG17, TKD19, Sta22, Stammer_ArXivConditioning_2024, Lamprou_PRL_2025, Stammer_EnergyConservation2024} have shown that intense laser–matter interactions can be used for the generation of controllable high photon--number non-classical optical Schr\"{o}dinger's ``cat'' states from the far--infrared (IR) to the extreme--ultraviolet (XUV) \cite{RLP22, SRM2023, SRL22, RSP21, Sta22, RSM22}. We note that as an optical Schr\"{o}dinger ``cat'' state we refer to superpositions between different coherent states of light. \PT{The collective achievements of the community} open the way for a vast number of investigations stemming from the symbiosis of strong–laser–field physics, quantum optics, and quantum information science. Some of these, together with the generation of optical ``cat'' states in atoms, have been highlighted in our recent review article published in Ref.~\cite{Tzallas_ROPP_2024}.   

In this topical review, together with a brief summary of the past achievements, we report on the latest progress on the fully quantized description of intense light–matter interaction and the methods that have been used for the generation of high photon number non-classical light states and entangled states using strongly laser driven atoms, molecules, solids and many body quantum correlated systems. 

Such investigations go beyond the mere importance of the generation of massive (e.g.~high photon number) quantum states of light.
In contrast to traditional approaches, where the driving laser field is described classically, and the harmonics are identified with Fourier components of the matter response, here we use a full quantum electrodynamics description. This adds a new element to describe the resulting states of light and matter. It opens the possibility to talk about non-classical properties of the states of light, entanglement of different field modes, light-matter entanglement, and much more. 

The structure of the paper is as follows. For reasons of completeness of the manuscript, we start with an introduction to the fundamentals of quantum optics, introducing the necessary quantum states such as optical Schr\"{o}dinger ``cat'', squeezed and entangled light states \ref{Nonclassical} together with commonly used state characterization approaches \ref{Characterization}, as well as the fundamentals of traditional intense laser--matter interactions \ref{StrongField_Classical}. Then we \PT{briefly} discuss the quantum electrodynamics (QED) of strongly laser driven atoms, molecules and solids \ref{StrongField_QED}. Then we present the quantum operations used for the generation of optical ``cat''--like and entangled states in strongly laser driven atoms \PT{\cite{LCP21, RLP22, SRM2023, SRL22, Sta22, Tzallas_ROPP_2024}}, molecules \PT{\cite{rivera-dean_quantum_2023}} and solids \PT{\cite{rivera-dean_bloch_2023, Gonoskov_PRB_2024}} \ref{Methods6}. In sections \ref{SqueezedDepletedAtoms}, \ref{Methods8} and \ref{Methods9} we discuss the recent investigations concerning the generation of non-classical and entangled light states in depleted atoms \PT{\cite{Stammer_PRL_2024}}, excited atoms \PT{\cite{Javier_ExcitedAtoms_2024,Misha_ArXiv_2024}} and high density gas media \cite{Andrianov_2024}, as well as many body quantum correlated systems \PT{\cite{{PGR23, Tzallas_ROPP_2024,Lange_PRAe-eCorr_2024,Hamed_PRXQ_2024, Merdji_multimodeSqueezed_Arxiv2024, Gonoskov_PRB_2024}}}, respectively. In section \ref{Applications} we discuss the perspectives of this new and rapidly growing research field emphasizing on recent findings for the applicability of the high photon number non--classical light states in non--linear optics \cite{Tzur_NatPhot2023, Heimerl_NatPhys2024, Tzur_PRR2024, Wang_PRR2024, Vampa_ArXiv_2024, Maria_NatPhys_2024, Liu_PRL_2025, Lamprou_PRL_2025, Spasibko2017, Ido_NatPhoton_2023, Stammer_NatPhys_2024, stammer2024absence}. Finally in section \ref{Challenges}, we provide our view regarding the challenges of the research field and the open questions.

\section{Non--classical and entangled light states}\label{Nonclassical}

We refer to non-classical states of light as those whose properties cannot be described by classical electrodynamics. Any state that cannot be described by a statistical mixture of coherent states, or provides a quasiprobability distribution (such as Wigner function) that depicts negative values, is considered non-classical. Such states include Fock states, squeezed states, and coherent state superpositions, e.g., optical Schrödinger ``cat'' states. These states as well as the entangled light states are central to quantum technologies~\cite{Benyoucef_book_ch2}. Hereafter, our discussion will be focused on the optical Schrödinger ``cat'', squeezed and entangled light states.

\subsection{Optical Schr\"{o}dinger's ``cat'' and entangled states}

\subsubsection{Schr\"{o}dinger's ``cat'' and entangled states}
\hfill \break
To illustrate the nature of quantum mechanics, at the beginning of 1930s, Schr\"{o}dinger introduced the concept of macroscopic state superposition using a cat sealed in a box, alongside an excited atom whose decay triggers the emission of a poison that determines the cat's fate \cite{schrodinger_gegenwartige_1935}. This famous {\it Gedankenexperiment} known as Schr\"{o}dinger's cat, lies at the heart of modern quantum technologies. The scenario involves a hypothetical cat and an atom enclosed in a box. The fate of the cat (whether it is alive or dead) hinges on the final state ($\ket{1}$ or $\ket{2}$) of an randomly decayed atom. When the box is sealed, isolating the system consisting of the cat and the atom from the external environment, the cat can be considered to exist in a superposition of states being both alive ($\ket{\text{A}}$) and dead ($\ket{\text{D}}$), a consequence of its entanglement with the state of the atom. Consequently, the state of the joint system reads,
\begin{equation}\label{eq:Schodingercat}
\ket{\Psi} = \frac{1}{\sqrt{2}}(\ket{\text{A}}\otimes \ket{1} + \ket{\text{D}}\otimes \ket{2}), 
\end{equation}
which is an entangled state as cannot be expressed in a separable tensor product form, i.e., $\ket{\Psi} \neq \ket{\psi}\otimes \ket{\phi}$.

Although this \emph{Gedankenexperiment} was originally presented at the time to challenge physicist' understanding about the concept of the wavefunction, it later became foundational in discussions surrounding the interpretation of quantum mechanics. Today, states similar to the one in Eq.~\eqref{eq:Schodingercat} have been realized at the microscopic scale across a wide range of systems, including light~\cite{aspect1_experimental_1982,aspect2_experimental_1982}. Equally important is the development of methods to characterize their entanglement properties, which ultimately determine their potential for various quantum information applications, such as quantum computation~\cite{Gilchrist2004,kok_linear_2007} and quantum communication~\cite{gisin_quantum_2007}.

Although determining whether an arbitrary quantum state is entangled is known to be a difficult problem~\cite{gurvits_classical_2004}, the task can be significantly simplified in certain scenarios. For example, in the case of bipartite systems, such as in Eq.~\eqref{eq:Schodingercat}, where one system represents the cat and the other represents the atom, there are specific mathematical functions, commonly referred to as entanglement measures, that quantify the amount of entanglement present in a quantum state. One such measure is the von Neumann entropy~\cite{plenio_introduction_2007}, defined as
\begin{equation}\label{Eq:Von:Neumann:entropy}
    S(\ket{\Psi})
        = - \tr(\hat{\rho}_A\log_d(\hat{\rho}_A)),
\end{equation}
where $d$ represents the dimension of the Hilbert space of the system, and $\hat{\rho}_A \equiv \tr_B(\dyad{\Psi})$ is the reduced density matrix of $\ket{\Psi}$ after computing the partial trace over one of the subsystems. It can be shown that for any separable state of the form $\ket{\Psi} = \ket{\psi_A}\otimes \ket{\phi_B}$, $S(\ket{\Psi}) = 0$. In contrast, for all entangled states $\ket{\Psi} \neq \ket{\psi_A}\otimes \ket{\phi_B}$, the von Neumann entropy satisfies $1 \geq S(\ket{\Psi}) > 0$, with the upper bound achieved by states like those in Eq.~\eqref{eq:Schodingercat} and their locally unitarily equivalent states, $\ket{\Psi} = (\hat{U}_A\otimes \hat{U}_B)\ket{\tilde \Psi}$. However, in many cases, the local Hilbert spaces of both subsystems are exceedingly large, making it numerically challenging to compute the von Neumann entropy. In such situations, the linear entropy $S_{\text{lin}}$ (bounded in the range $0\leq S_{\text{lin}}\leq 1/d$ with $d$ the local Hilbert space dimension of the corresponding partition) serves as a suitable alternative, obtained as a first-order approximation of Eq.~\eqref{Eq:Von:Neumann:entropy}
\begin{equation}\label{Eq:Lin:Entropy}
    S_{\text{lin}}(\ket{\Psi})
        = 1 - \tr(\hat{\rho}_A^2).
\end{equation}

Although, as mentioned earlier, superpositions of the form in Eq.~\eqref{eq:Schodingercat} have been successfully achieved, preparing similar states at the macroscopic scale remains a challenging task. This difficulty arises because such states quickly decohere into a statistical mixture due to environmental interactions and associated decoherence effects. These ultimate transform pure states of the form $\hat{\rho} = \dyad{\Psi}$, with $\hat{\rho}^2 = \hat{\rho}$, into statistical mixtures or mixed states $\hat{\rho} = \sum_s p_s \dyad{\Psi_s}$, with $\hat{\rho}^2 \neq \hat{\rho}$. When dealing with mixed states, we refer to separable states as those satisfying 
\begin{equation}\label{Eq:Logarithmic:Negativity}
    \hat{\rho} 
        = \sum_{s} p_s 
            \hat{\rho}_{A,s}\otimes \hat{\rho}_{B,s},
\end{equation}
and in this case entanglement measures like the von Neumann entropy or the linear entropy cease to be valid. Alternatively, one can define the logarithmic negativity~\cite{peres_separability_1996,horodecki_separability_1996}
\begin{equation}
    E_N(\hat{\rho})
        = \log(2N_{AB}+1),
\end{equation}
where $N_{AB}$ is the negativity, that is, the sum of all negative eigenvalues of $\hat{\rho}^{T_B}$, where $\hat{\rho}^{T_B}$ represents the partial transpose of $\hat{\rho}$ with respect to one of the subsystems.

\begin{figure}
    \centering
    \includegraphics[width=1 \columnwidth]{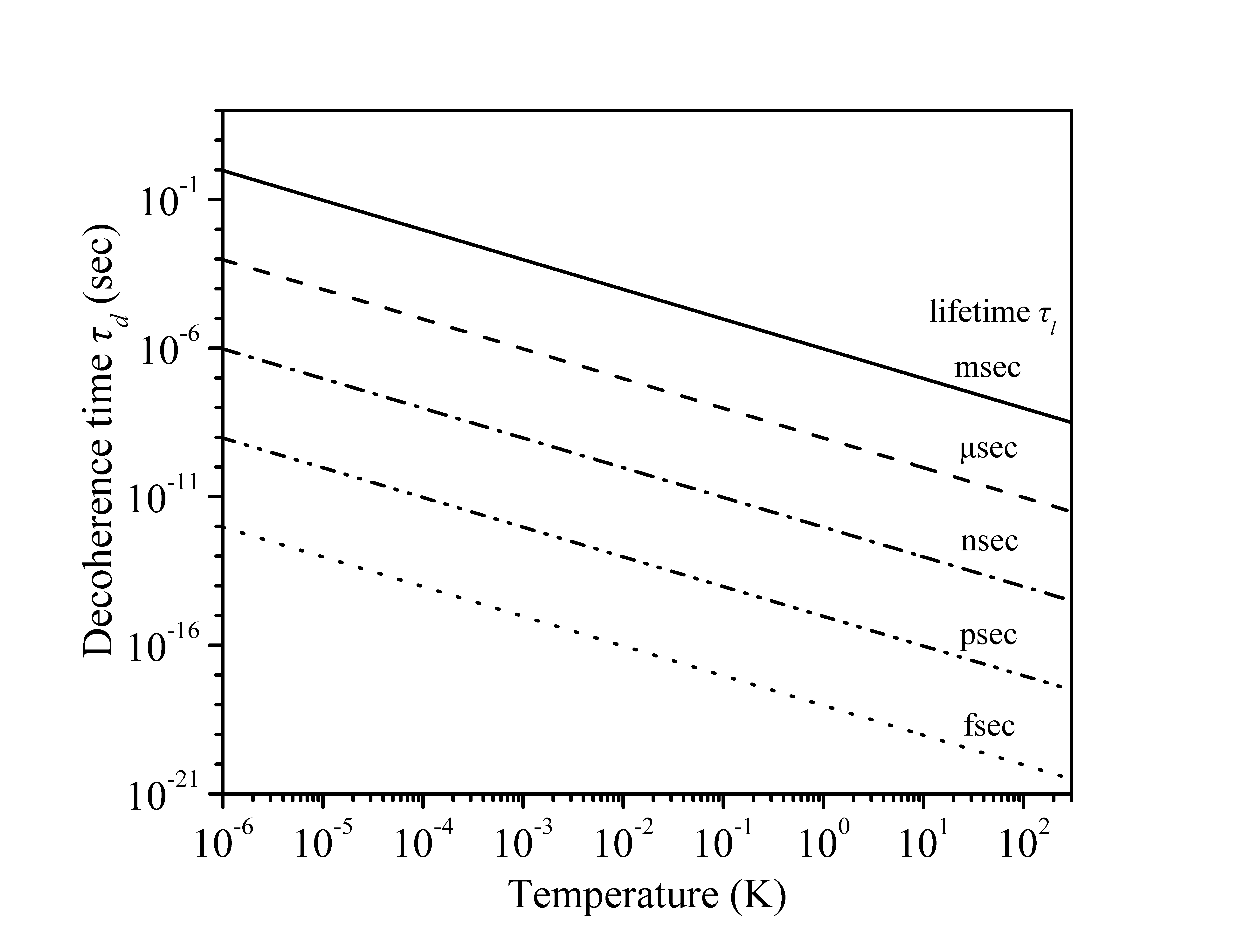}
    \caption{Decoherence times $\tau_{d}$ of a quantum superposition of atoms with mass $\sim 10^{-27}$ kg and distance $\Delta x \sim 100$ nm, as a function of temperature $T$ for lifetimes $\tau_{l}=$1 msec (solid line), 1 $\mu$sec (dash line), 1 nsec (dash--dot line), 1 psec (dash--dot--dot line) and 1 fsec (dot line). }
    \label{fig1:DecoherenceTimes}
\end{figure}

As has been nicely discussed in Ref.~\cite{Zurek_PhysicsToday_1991}, a rough estimation of the decoherence times can be provided if we consider the wave nature of a particle placed in an environment where the particles randomly collide with each other. In this case the decoherence time is approximately $\tau_{d}\approx \tau_{l} (\lambda_{dB}/\Delta x)^{2}$, where $\tau_{l}$ is the lifetime of the particle, $\Delta x$ is the distance between the particles, and $\lambda_{dB}=\hbar/ \sqrt{2mk_{B}T}$ the de Broglie wavelength. Here, $m$, $k_{B}$ and $T$ are the mass of the particle, the Boltzmann constant and the temperature, respectively. Fig. \ref{fig1:DecoherenceTimes} shows $\tau_{d}$ for quantum superposition consisted by objects in the microcosm (atoms in a 0.1$\mu$m--scale distance) as a function of temperature $T$ at different lifetimes $\tau_{l}$. Based on the rough estimations shown in Fig. \ref{fig1:DecoherenceTimes}, observing quantum superpositions in the microcosm is easier to achieve in ultra-cold and isolated systems. For systems at room temperature, their observation can be achieved by freezing the system in time, e.g., using approaches capable of resolving femtosecond or attosecond scale dynamics ~\cite{Vrakking_PRL_2021, Vrakking_PRL_2022, XFEL_SciAdv_2024}.

Evidently, for macroscopic objects (such as cats) with $m \sim $ kg, $\Delta x \sim $ cm and $\tau_{l} \sim$ years at any $T$, quantum superpositions are impossible to be measured with the current technology as they decohere into their statistical mixture in time scales $\tau_{d} < 10^{-30}$ sec. Hence, to become feasible, Schr\"{o}dinger {\it Gedankenexperiment} in macroscopic scale we should replace the cat with a physical system with its own classical states. Light, and in particular coherent states of light $|\alpha\rangle$, are considered ideal resources for engineering macroscopic light state superpositions. This statement stems from two main factors: (a) the electromagnetic environment at optical frequencies can be considered as relatively decoherence-free, and (b) coherent states provide the optimal quantum optical description of a classical field and, thus, the quantum superposition of distinguishable classical states resembles the originally envisaged Schr\"{o}dinger’s cat {\it Gedankenexperiment}. Therefore, as we describe in the next section, a superposition composed by two coherent states defines an optical ``cat'' state.

\subsubsection{Optical Schr\"{o}dinger's ``cat'' states} 
\hfill \break
The quantum description of a classically oscillating field is feasible with the formulation of coherent states of light. These states have quantum fluctuations in the two quadratures that are equal and minimize the uncertainty product given by Heisenberg’s uncertainty relation $\Delta x \Delta p = \frac {1}{2}$ with $\Delta x = \Delta p = \frac {1}{\sqrt{2}}$. They are typically denoted as $\ket{\alpha}$, they are eigenstates of the annihilation operator, i.e. $\hat{a} \ket{\alpha}=\alpha \ket{\alpha}$, and can be expanded in the Fock state basis as $\ket{\alpha} = e^{-\abs{\alpha}^2 /2} \sum_{n=0}^\infty \frac{\alpha^n}{\sqrt{n!}} \ket{n}$.  $\alpha=\lvert\alpha\rvert\exp(i\theta)$, is the complex amplitude, $\expval{n} = \bra{\alpha}\hat{N}\ket{\alpha}=|\alpha|^2$ is the mean-photon number and $\hat{N}=\hat{a}^\dagger \hat{a}$ is the photon number operator. It is easy to show that the expectation value of the electric field operator is $\bra{\alpha}\hat{E} (t) \ket{\alpha}=|\alpha| \cos(\omega t -\theta)$, which corresponds the expression of a monochromatic electromagnetic field as obtained in classical electromagnetism. For this reason coherent light states are named, mainly in the quantum optics bibliography,  as ``classical'' or ``quasi-classical'' states.

 A coherent state superposition (CSS) is given by $\ket{\Psi_{\text{CSS}}} = \sum_i \xi_i \ket{\alpha_i}$, where $\xi_i$ is the probability amplitude. Superpositions composed by two coherent states are typically referred to as optical Schr\"{o}dinger ``cat'' states. The underlying reason is that these states correspond to an optical analogue of the cat state superposition in Schrödinger's {\it Gedankenexperiment} \cite{schrodinger_gegenwartige_1935}.
In a general case, when the superposition of two coherent states has the form
\begin{equation}\label{eq:cat}
\ket{\Psi_{\text{CSS}}}\equiv\ket{\Psi_{\text{cat}}} = \ket{\alpha_1} \pm \xi \ket{\alpha_2}, 
\end{equation}
is referred as generalized (or shifted) optical ``cat'' state. 
Depending on $\xi=\langle{\alpha_{1}}|{\alpha_{2}}\rangle$, we define the states as optical ``kitten'', ``cat'' and large optical ``cat'' states for large ($\xi \xrightarrow{} 1$), medium (0 < $\xi < 1$) and small ($\xi \xrightarrow{} 0$) $\xi $, respectively, with the former being more robust against decoherence/losses~\cite{zhang_quantifying_2021, Stammer_RobustCAT_2024} compared to the large ``cat'' states .

\subsubsection{Entangled light states}
\hfill \break
So far, we have only considered a single mode of the electromagnetic field, which can be described by a state $\ket{\psi_1}$, e.g. with a Fock, coherent or squeezed state. However, if we take into account an additional mode of the field, described by the state $\ket{\psi_2}$, we can now consider the total system of the field composite of the two modes. 
Since we now have two modes, which can, for instance, be different spatial or frequency modes, we can speak about a bipartite system in which each field mode represents one subsystem. The total state of the field, in this case, would be given by the tensor product $\ket{\psi_1} \otimes \ket{\psi_2}$, i.e., a separable state. However, this is not the most general expression we can have, as one can always expand a given bipartite quantum state $\ket{\Psi}$ as, 
\begin{align}
	\ket{\Psi} = \sum_{ij} c_{ij} \ket{i}\otimes \ket{j},
\end{align} 
where $\ket{i}$ and $\ket{j}$ denote some basis sets for both subsystems, and $c_{ij} = \bra{ij} \ket{\Psi}$. In this context, we say that the state $\ket{\Psi}$ is entangled whenever it cannot be written in a separable form, i.e., $\ket{\Psi} \neq \ket{\psi_1}\otimes \ket{\psi_2}$.

\subsubsection{Squeezed light states}
\hfill \break
For squeezed states the quantum fluctuations of light are not equally distributed between the field quadratures. They are minimum uncertainty states, which have reduced fluctuations (compared to the coherent or vacuum state) in one quadrature and increased in the other. They are produced by non--linear interactions, with the parametric down conversion process in a crystal being one of the most commonly used methods (see Ref.~\cite{Andersen2016} and references therein). This interaction can be described in terms of the Hamiltonian $\hat{H}=\hbar\chi(\hat{a}^2 - {{}\hat{a}^\dagger}^2)$, and its unitary evolution introduces the so-called \emph{squeezing operator}, given by $\hat{S}(k)=\exp[-\frac{k}{2}(\hat{a}^2 - {{}\hat{a}^\dagger}^2)]$. In this expression, $k=2i\chi \tau$ is the squeezing parameter ($k \in \mathbb{R}$), $\chi$ the non-linear coupling parameter and $\tau$ the interaction time. By applying $\hat{S}(k)$ to a vacuum state $\ket{0}$, we obtain the \emph{squeezed-vacuum} state

\begin{equation} \label{Eq:SV}
\begin{aligned}
\ket{\text{SV}}=\hat{S}(k)\ket{0},
 \end{aligned}
\end{equation}
while the displaced squeezed vacuum state $\ket{\text{DSV}}$ can be obtained by applying the displacement operator $\hat{D}(\alpha)$ on $\ket{\text{SV}}$. In this case Eq.~\eqref{Eq:SV} reads, 
\begin{equation} \label{Eq:DSV}
\begin{aligned}
\ket{\text{DSV}}=\hat{D}(\alpha)\hat{S}(k)\ket{0}.
 \end{aligned}
\end{equation}

\section{Quantum state characterization of light} \label{Characterization}
\begin{figure}
    \centering
   \includegraphics[width=0.9 \columnwidth]{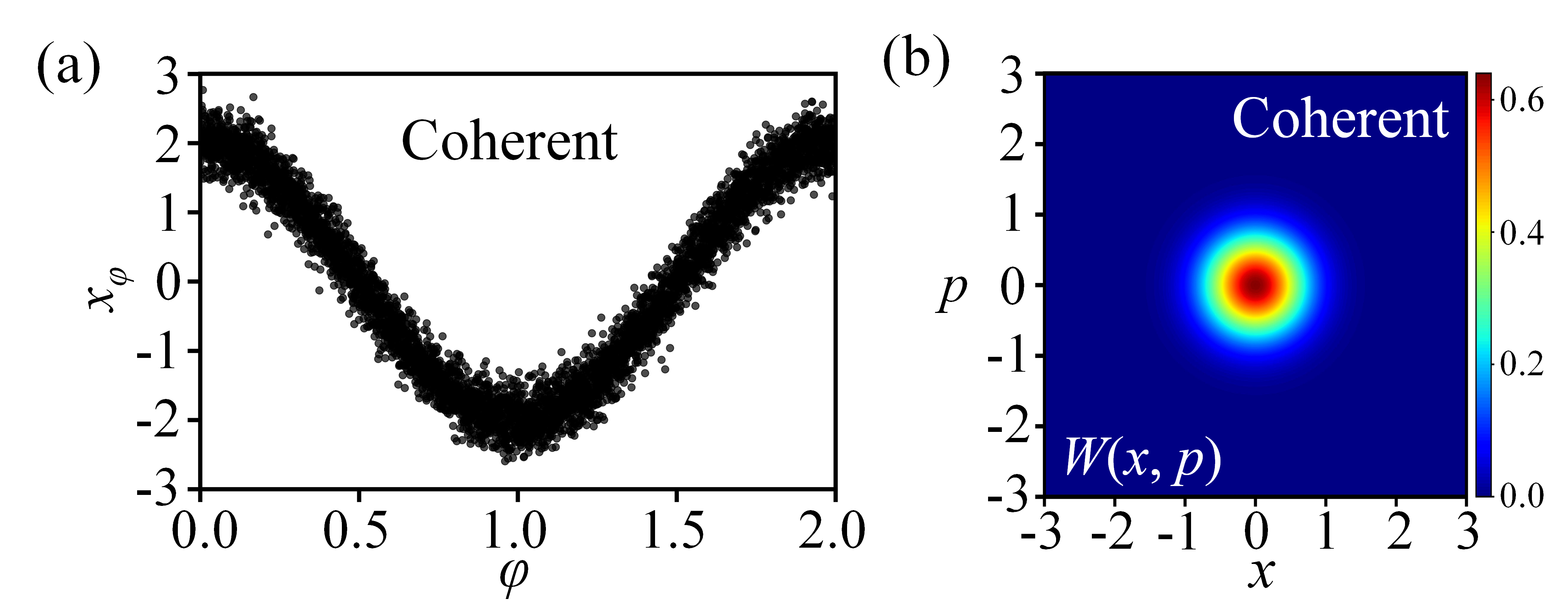}
    \caption{(a) Calculated $x_{\varphi}$ of a coherent light state. (b) $W(x,p)$ (centered at $|\alpha|$) corresponding to (a).}
    \label{fig2:QT}
\end{figure}

\PT{One of the most commonly used ways to witness quantum signatures of a light state is to obtain the Wigner function $W(x, p)$ in phase space $(x, p)$. This can be achieved by means of quantum tomography (see for example~\cite{Schleich_book_2001, BSM97, LR09, Bachor_book_2019, Lvovsky_MaxLik_alg2004, Herman80, Leonhardt_book_1997, Banaszek_PRA_1999}), which relies on the use of a homodyne detection techniques.
The method provides access to the measurement of the values of the field quadrature $\hat{x}_{\varphi}=\cos(\varphi) \hat{x}+ \sin(\varphi) \hat{p}$ where $\varphi$ is the field phase. The homodyne trace contains the information needed to characterize the light state. Fig.~\ref{fig2:QT}(a) shows a homodyne trace calculated for a coherent light state. The reason is, that repeated measurements of $\hat{x}_{\varphi}$ at each $\varphi$ provide the probability distribution $P_{\varphi}(x_{\varphi})=\langle{x_{\varphi}}|{\hat{\rho}}|{x_{\varphi}}\rangle$ of its eigenvalues $x_{\varphi}$, with $\hat\rho$ being the density operator of the light state to be characterized and $|{x_{\varphi}}\rangle$ is the eigenstate of $\hat{x}_{\varphi}$ with eigenvalue $x_{\varphi}$. The density matrix $\hat \rho$ provides complete information about the light state. The estimation of the density matrix is feasible by introducing the ensemble of measured quadrature values ($x_{\varphi},\varphi$) to the \textit{Maximum--likelihood} scheme presented in Ref.~\cite{Lvovsky_MaxLik_alg2004}. With these values, the mean photon number of the light state can be obtained by the diagonal elements $\hat \rho_{nn}$ of the density matrix $\hat \rho$ (expressed in the Fock basis), and the relation $\langle{n}\rangle=\sum n\hat \rho_{nn}$.}

\subsection{$W(x,p)$ of coherent states}\label{Sec:Wigner coherent}
The $W(x,p)$ of a coherent state, is that of the vacuum shifted by $(x_{0}, p_{0})$. It depicts a Gaussian distribution of the form $W(x,p) = \frac{1}{\pi} \exp[-(x-x_0)^2-(p-p_0)^2]$ which in a more compact form reads,
\begin{equation}\label{Eq:Wigner:Coher:Compact}
\begin{aligned}
W(\beta) = \frac{2}{\pi} e^{-2|\beta - \alpha|^{2}}.
 \end{aligned}
\end{equation} 
In the above equation we have used the transformation $\it x \equiv \Re[\beta-\alpha]$ and $\it p \equiv \Im[\beta-\alpha]$ (where $\beta=(x+ip)/\sqrt{2}$ is a variable), which centers the Wigner function at the origin for $\beta=\alpha$, $x=0$ and $p=0$. Figure~\ref{fig2:QT}(b) shows an example of calculated $W(x,p)$ of a coherent light state corresponding to the homodyne trace of Fig.~\ref{fig2:QT}(a).

\subsection{$W(x,p)$ of optical ``cat'' states}\label{Sec:Wigner cat}
\begin{figure}
    \centering
    \includegraphics[width=0.7 \columnwidth]{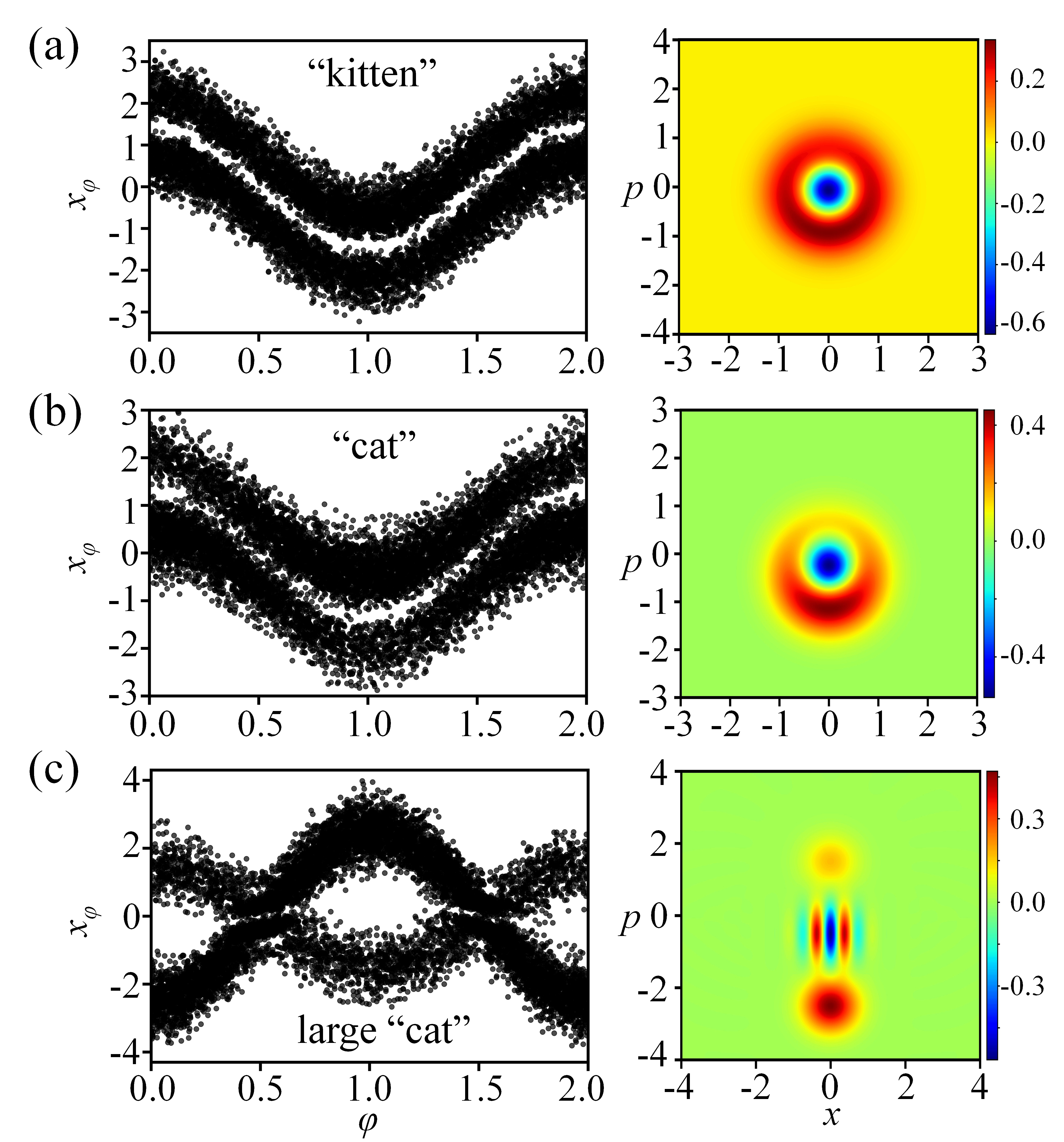}
    \caption{Calculated homodyne traces (left panels) and the corresponding $W(x,p)$ (right panels) of an optical ``kitten'' (a), ``cat'' (b) and large optical ``cat'' states (c) of the form $\ket{\Psi_{\text{cat}}} = \ket{\alpha_1} - \xi \ket{\alpha_2}$. In (a), $\xi=0.98$, $\alpha_{1}=1.3$ and $\alpha_{2}=1.5$. In (b), $\xi=0.78$, $\alpha_{1}=0.8$ and $\alpha_{2}=1.5$. In (c), $\xi=0.6$, $\alpha_{1}=-2.5$ and $\alpha_{2}=1.5$.}
    \label{fig3:CAT}
\end{figure}

Applying the transformation used in Eq.~\eqref{Eq:Wigner:Coher:Compact}, the $W(x,p)$ of a generalized optical ``cat'' state reads,
\begin{equation} \label{eq:WignerPlot}
\begin{aligned}
W(\beta)
			=\ &\dfrac{2}{\pi N}
			 \Big[ e^{-2\lvert\beta - \alpha - \chi\rvert^2}
			 + e^{-\lvert\chi\rvert^2}e^{-2\lvert\beta - \alpha\rvert^2}\\
			 &- \big(
			 		e^{2(\beta - \alpha)\chi^*}
			 		+ e^{2(\beta - \alpha)^*\chi}
			 	\big)
			 	e^{-\lvert\chi\rvert^2}e^{-2\lvert\beta - \alpha\rvert^2}
			 	\Big],
 \end{aligned}
\end{equation}
where $N = 1 - e^{-\lvert\chi\rvert^2}$ is the normalization factor for $\ket{\Psi_{\text{cat}}} = \ket{\alpha_1} - \xi \ket{\alpha_2}$ with $\ket{\alpha_1}=\ket{\alpha_2+\chi}$. The $W(x,p)$ (Fig.~\ref{fig3:CAT}) depicts a strong interference pattern with negative values in the region where the two coherent states, $\ket{\alpha_1}$ and $\ket{\alpha_2}$, overlap. Homodyne traces and the corresponding $W(x,p)$ of an optical ``kitten'', ``cat'' and large ``cat'' states are shown in Figs.~\ref{fig3:CAT}(a), (b) and (c), respectively. The $W(x,p)$ of an optical ``kitten'' and ``cat'' state depicts a ring-shaped structure with negative values around the center of the distribution. The $W(x,p)$ of large optical ``cat'' states, depicts two pronounced Gaussian-like maxima $(x_{i}, \pm p_{i})$, where $\pm x_i =\mel{\alpha_i}{\hat{x}}{\alpha_i}$ and $\pm p_i =\mel{\alpha_i}{\hat{p}}{\alpha_i}$, with a fringes pattern around the center of the distribution. 

\subsection{$W(x,p)$ of squeezed light states}\label{Sec:Wigner Squeezed}

\begin{figure}
    \centering
  \includegraphics[width=0.7 \columnwidth]{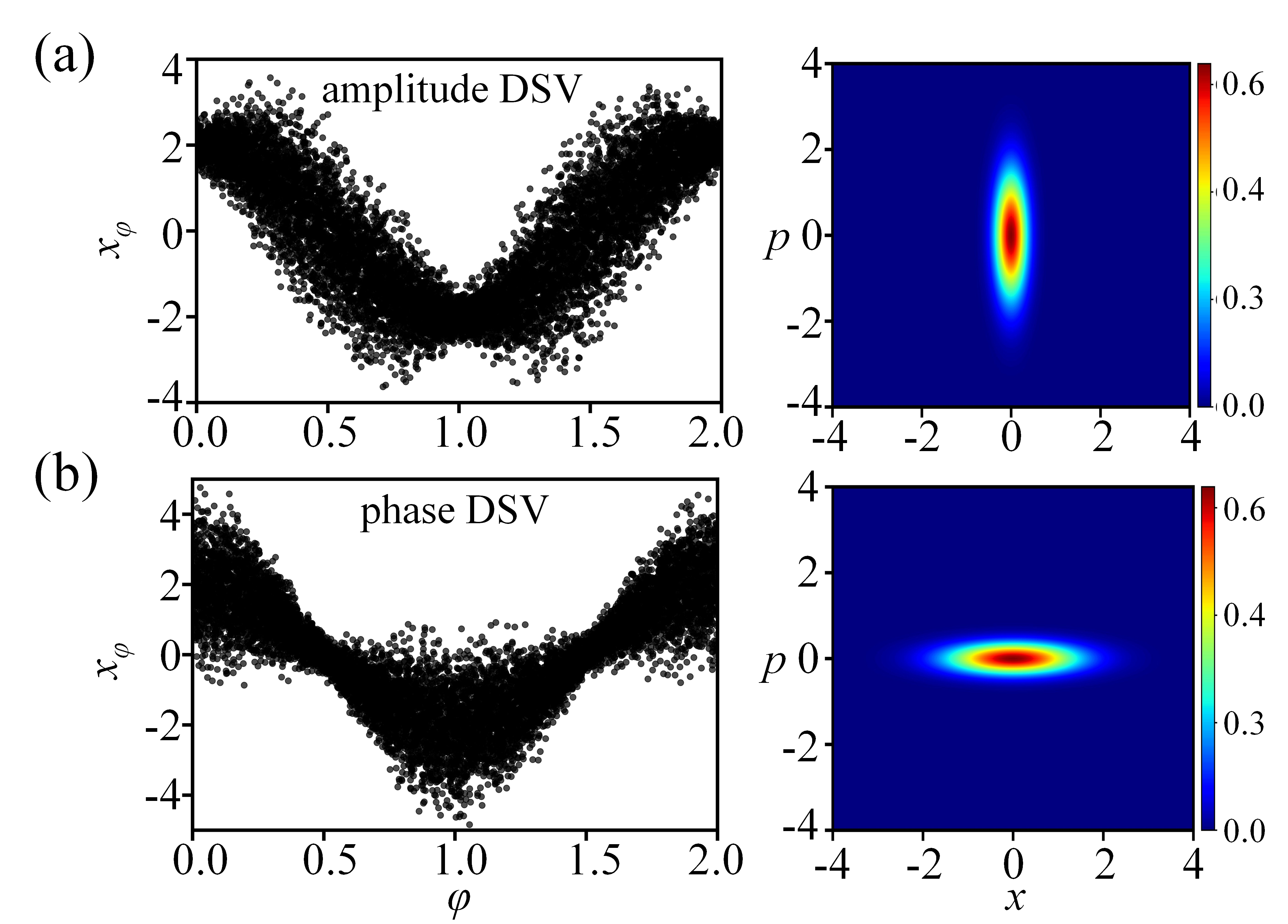}
    \caption{Calculated homodyne traces (left panels) and the corresponding $W(x,p)$ (right panels) of an amplitude (a) and phase (b) squeezed states $\ket{\text{DSV}}$ with $|\alpha|=2$ and squeezing factor $k=0.7$ and $k=- 0.7$, respectively.}
    \label{fig4:Squeezed}
\end{figure}

The $W(x,p)$ of the squeezed light states corresponds to a Gaussian which has been ``squeezed'' along one of its quadratures and stretched in the other. Applying the transformation used in Eq.~\eqref{Eq:Wigner:Coher:Compact}, the $W(x,p)$ reads, 
\begin{equation} \label{Eq:WignerSqueezed}
\begin{aligned}
W_{DSV}(\beta) = \frac{2}{\pi} e^{-2\abs{\frac{(\beta-\alpha)}{e^{-2k}}+\frac{(\beta^{*}-\alpha^{*})}{e^{2k}}}^{2}}.
 \end{aligned}
\end{equation}
Figure \ref{fig4:Squeezed} shows a calculated homodyne trace (left panel) and the corresponding Wigner function (right panel) of an amplitude and phase squeezed state $\ket{\text{DSV}}$.

\subsection{Decoherence effects in optical cat states}\label{Sec:Wigner analysis}

The non-classicality of the measured light state can be obtained by quantitatively evaluating its quantum features (such as its negativity) using the reconstructed Wigner function. The negativity is associated with the contrast of the fringe pattern created by quantum interference effects between the coherent states participating in the coherent state superposition. One of the reasons why the negativity can be reduced compared to the ideal case is the presence of decoherence effects. Although the optical coherent state superpositions can be considered as (almost) decoherence-free compared to the particle state superposition, the presence of optical losses during the propagation in optical elements affects the non-classicality of these light states. This leads to a Wigner function with reduced negativity or reduced fringe contrast. The decoherence can be quantitatively studied using a noise model that introduces photon losses due to the interaction with a Gaussian reservoir \cite{RLP22, Stammer_RobustCAT_2024, leonhardt_quantum_1993}, via the von Neumann entropy $S$, or using the fidelity $F$ (see e.g. Ref.~\cite{zhang_quantifying_2021}).

\subsection{Noise model for photon losses}
In the following, we focus on optical losses occurring within an empty environment, that is, in the absence of any external radiation source. In such conditions, any light propagating through this medium will inevitably experience damping, resulting in a reduced final intensity compared to its initial value. If the medium is particularly  sensitive to the frequency of the source, the damping rate increases and, over sufficiently long enough times, the environment may completely absorb the incoming radiation. In this extreme case, the final state of our source would correspond to a vacuum state $\ket{0}$, regardless of the initial quantum optical state. Furthermore, when the environment exhibits no nonlinearity, the interaction must preserve the Gaussianity of the input quantum optical state. This implies that if the initial state is Gaussian, it will remain Gaussian after the light-environment interaction. 

\begin{figure}
    \centering
    \includegraphics[width=1 \columnwidth]{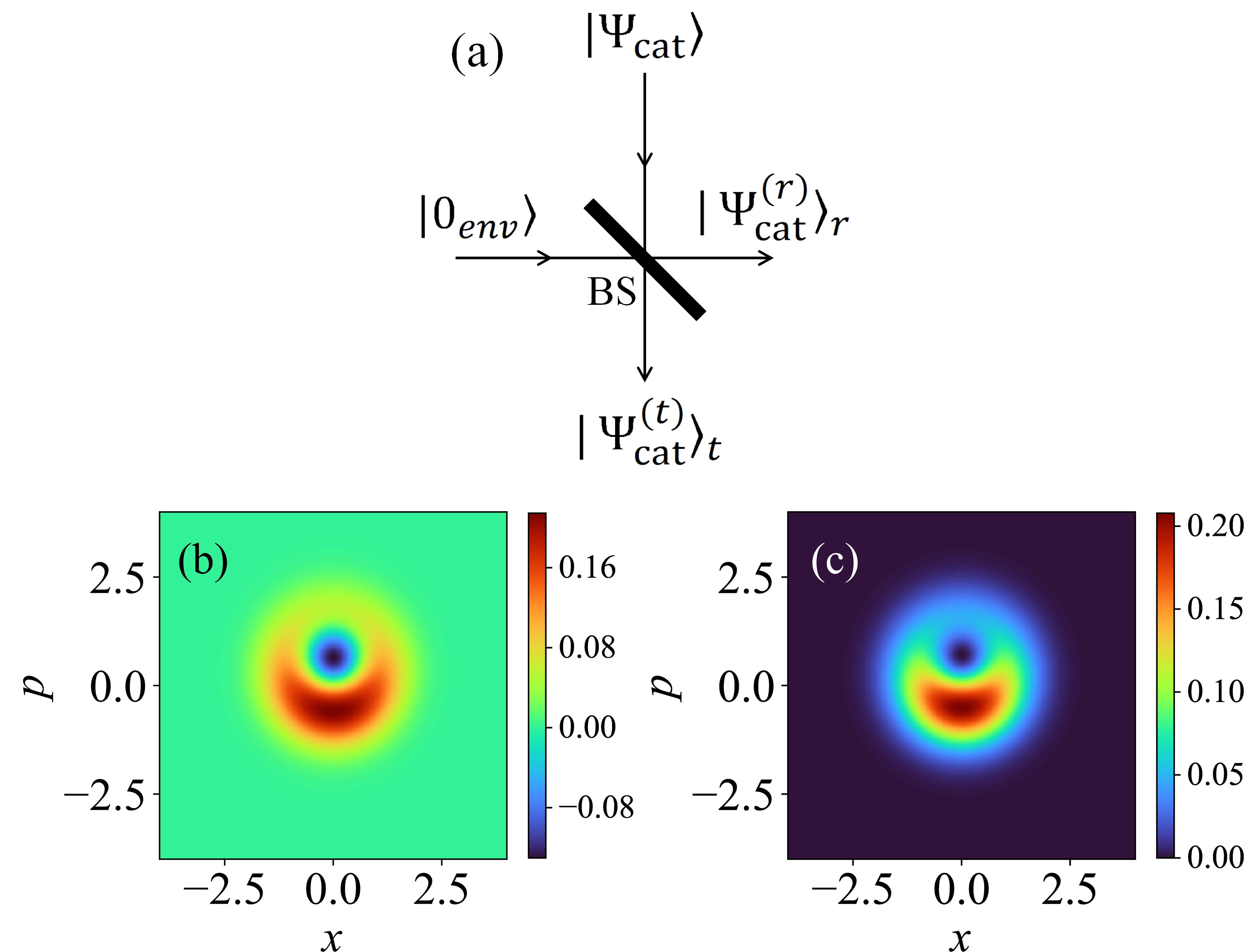}
    \caption{(a) Model used for calculating the interaction of an optical ``cat state $\ket{\Psi_{\text{cat}}}$ with the environment. BS is a beam splitter through which we introduce the losses from the environment. $\lvert{\Psi_{\text{cat}}^{(r)}}\rangle_{r}$ and $\lvert{\Psi_{\text{cat}}^{(t)}}\rangle_{t}$ are the reflected and transmitted light states. (b) and (c) Calculated $W(x,p)$ of an optical ``cat'' state with $\ket{\alpha+\delta\alpha}-\xi\ket{\alpha}$ with $|\alpha|=0$, $|\delta\alpha|=0.7$, $\xi=0.78$, and $\eta = 0.75$ (25\% losses) in panel (b) while $\eta = 0.5$ (50\% losses) in panel (c).}
    \label{fig5:NoiseModel}
\end{figure}

In the literature, environments exhibiting this property are commonly referred to as Gaussian reservoirs~\cite{leonhardt_quantum_1993}. These reservoirs map coherent states $\ket{\alpha}$ to coherent states with reduced amplitude $\ket{\alpha\sqrt{\eta}}$, with $0 \leq \eta \leq 1$. An environment with $\eta = 1$ corresponds to a perfectly transmissive medium with no losses, while $\eta = 0$ represents a completely absorbing medium. As demonstrated in Ref.~\cite{leonhardt_quantum_1993}, the interaction with such environments can be modelled using a beam splitter, where one input mode corresponds to the input quantum optical state, and the other input mode represents the quantum state of the environment (see Fig.~\ref{fig5:NoiseModel}~(a)) which, in this case, is the vacuum state $\ket{0_{\text{env}}}$. Consequently, the unitary operator that effectively describes the interaction between the source and the environment can be expressed as
\begin{equation}
    \hat{U}(\theta)
        = \exp[\theta
            \big(
                \hat{a}^\dagger \hat{a}_{\text{env}}
                - \hat{a}\hat{a}^\dagger_{\text{env}}
            \big)
            ],
\end{equation}
with $\theta = \arccos(\sqrt{\eta})$. Therefore, in the absence of knowledge about the state of the environment after the interaction, the quantum optical state of the source is given by
\begin{equation}\label{Eq:final:noise}
    \hat{\rho}_s
        = \tr_{\text{env}}
            \Big[
                \hat{U}(\theta) 
                \dyad{\Psi_0}
                \hat{U}^\dagger(\theta)
            \Big],
\end{equation}
where $\ket{\Psi_0} = \ket{\psi_0}\otimes\lvert0_{\text{env}}\rangle$ denotes the initial state.

If the initial state of the input source is $\ket{\psi_0} = \ket{\alpha}$, then Eq.~\eqref{Eq:final:noise} becomes $\hat{\rho}_s = \dyad{\alpha \sqrt{\eta}}$, corresponding to a pure Gaussian state with reduced amplitude compared to $\ket{\psi_0}$. This follows from the relation $\hat{U}(\theta) \ket{\alpha}\otimes \lvert 0_{\text{env}}\rangle = \ket{\alpha \sqrt{\eta}}\otimes \ket{-\alpha\sqrt{1-\eta}}$, which shows that the final state of the environment depends on both amplitude $\abs{\alpha}$ and phase $\varphi$, where $\alpha = \abs{\alpha}e^{i\varphi}$, of the initial coherent state. Consequently, different initial coherent states produce different quantum optical states for the environment. This means that, for a general coherent state superposition $\ket{\psi_0} = a \ket{\alpha} + b \ket{\beta}$, the interaction with the environment creates quantum correlations between the system and the environment, leading to the entangled state $\ket{\Psi} = a \ket{\alpha\sqrt{\eta}}\otimes\ket{-\alpha\sqrt{1-\eta}} + b \ket{\beta\sqrt{\eta}}\otimes\ket{-\beta\sqrt{1-\eta}}$. After tracing out the environmental degrees of freedom, the resulting state is a mixed state, given by
\begin{equation}\label{Eq:noisy:cat}
    \begin{aligned}
    \hat{\rho}_s
        &= \abs{a}^2 \dyad{\alpha\sqrt{\eta}}
        + \abs{b}^2 \dyad{\beta \sqrt{\eta}}
        \\&\quad
        + a b^* \tilde{\xi}
            \dyad{\alpha\sqrt{\eta}}{\beta\sqrt{\eta}}
        + a^* b \tilde{\xi}^*
            \dyad{\beta\sqrt{\eta}}{\alpha\sqrt{\eta}},
    \end{aligned}
\end{equation}
where $\tilde{\xi} = \braket{-\alpha\sqrt{1-\eta}}{-\beta\sqrt{1-\eta}}$.

As previously mentioned, in the case $\eta = 0$ the mixed state in Eq.~\eqref{Eq:noisy:cat} reduces to the vacuum state, which corresponds to a Gaussian state with a fully positive Wigner function. Therefore, one can expect that for intermediate regimes $0<\eta <1$, the Wigner functions will exhibit reduced negativity as $\eta \to 0$. This behavior is illustrated in Fig.~\ref{fig5:NoiseModel}~(b) and (c), where a cat state of the form $\ket{\alpha + \delta\alpha} - \xi \ket{\alpha}$ propagates through a medium that is inducing 25\% ($\eta = 0.75$) and 50\% ($\eta = 0.5$) of losses, respectively. As observed, the lower the transmissivity of the medium becomes, the less pronounced the Wigner negativities become, disappearing entirely when 50\% of the radiation is absorbed (panel (c)), while still retaining the overall volcano shape of the Wigner functions.~It was further shown in Ref.~\cite{Stammer_RobustCAT_2024} that the coherent state superposition of the form from Eq.~\eqref{eq:cat} are more robust in the metrological task of phase estimation. These states show an enhanced robustness against photon loss compared to high photon number even or odd cat states, which makes them an inherently more robust quantum information carrier.

\subsection{Fidelity and relative entropy of an optical cat state}

Together with the negativity of the Wigner function, a useful way to provide a quantitative information about the quality of an optical ``cat'' state is via the fidelity $F$ or relative entropy $S$.

In quantum mechanics, the $F$ quantifies how close are the density matrices of two quantum states $\rho_{1}$ and $\rho_{2}$. In other words, it provides the probability of one state to be identical with the other. It is defined as $F(\rho_{1}, \rho_{2})=(\text{tr}(\sqrt{\rho_{1}\rho_{2}}))^2$, which for pure states $\rho_{1}= \dyad{\psi_{1}}$ and $\rho_{2}=\dyad{\psi_{2}}$, takes the form $F(\rho_{1},\rho_{2})=\text{tr}(\rho_{1} \rho_{2})=\bra{\psi_{1}}\rho_{2}\ket{\psi_{1}}$. The fidelity is bounded in the range $0\leq F \leq 1$ with the value of 1 corresponding to the ideal matching between the states. Hence, the fidelity of an optical ``cat'' state $\rho_{\text{exp}}$ measured by an experiment (as has been described in Sec \ref{Characterization}) can be obtained by using the relation $F(\rho_{\text{exp}},\rho_{\text{cat}})=\text{tr}(\rho_{\text{exp}} \rho_{\text{cat}})$ where $\rho_{\text{cat}} = \dyad{\Psi_{\text{cat}}}$ is the ideal (theoretical) optical ``cat'' state used to compare the experimental outcome. 

Similarly to the fidelity, a quantitative comparison between two states can be obtained using the quantum relative entropy $S_{\text{rel}}$ of $\rho_{\text{exp}}$ with respect to $\rho_{\text{cat}}$, with $S_{\text{rel}}(\rho_{\text{exp}}||\rho_{\text{cat}}) = \tr (\rho_{\text{exp}} \log \rho_{\text{exp}} - \rho_{\text{exp}} \log \rho_{\text{cat}})$ \cite{Nielsen_BookEntropy_2010}.

\section{Intense laser--matter interaction} \label{StrongField_Classical}

\begin{figure}
    \centering
  \includegraphics[width=1 \columnwidth]{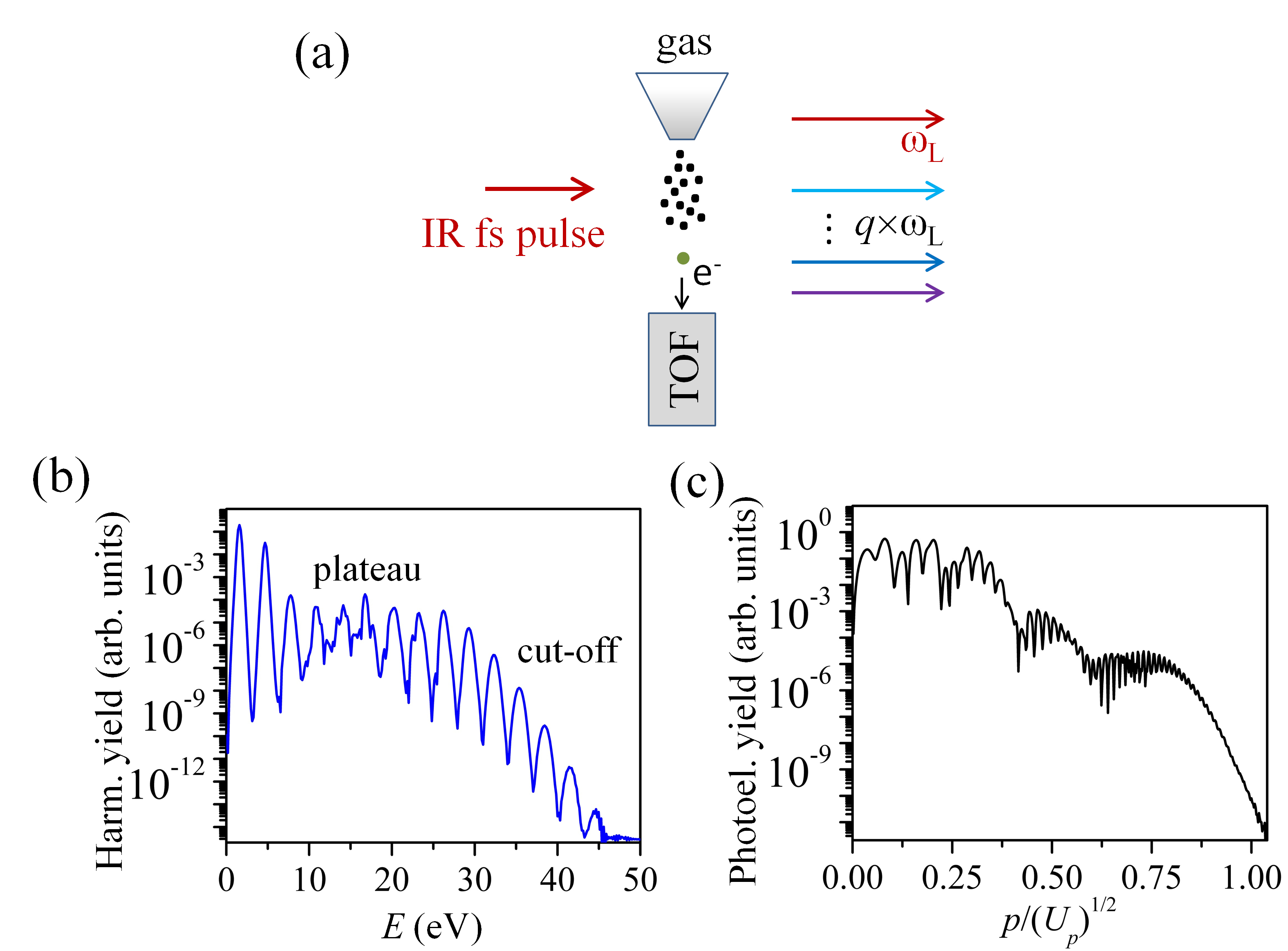}
    \caption{(a) A scheme typically used for investigating intense laser--matter interactions. A linearly polarized intense IR fs laser pulse of frequency $\omega_{L}$ is focused into a target medium (here for simplicity we have used atoms) where the high order harmonics of frequency $q\omega_{L}$ and photo electrons are generated. The high harmonics can be measured by photodetectors and spectrometers (not shown in the figure) while the ionization products (ions, photoelectrons) can be measured by means of photoelectron spectrometers such as time-of-flight (TOF)- and Velocity--Map--Imaging (VMI)--type of arrangements. The high harmonics co-propagate with driving IR field, and the photoelectrons are typically emitted along the direction of the driving IR field polarization. (b) and (c) Characteristic HHG and photoelectron spectrum generated in atoms, reproduced from refs ~\cite{SRM2023, Tzallas_ROPP_2024}. The HHG spectrum exhibits a plateau structure that lasts until the cutoff region. The photoelectron spectrum includes the direct electron emission which lasts until $\approx 0.6 p/\sqrt{U_{p}}$ and the rescattered electron emission up to $\approx 0.9 p/\sqrt{U_{p}}$. The spectra have been calculated when Xenon atoms interact with a $30$ fs IR laser pulse of intensity $8 \times 10^{13}$ W/cm$^2$.}
    \label{fig6:IntenseLaserMatter}
\end{figure}

The development of high-power femtosecond (fs) lasers has allowed scientists to explore interactions in the strong field limit, where the laser’s electric field is comparable to, or even stronger than, the field keeping the electrons bound in atoms. This can be quantitatively defined by the Keldysh parameter $\gamma$, when $\gamma\lesssim 1$. In case of intense laser--atom interactions $\gamma=\sqrt{I_p/(2U_{p})}$ \cite{keldysh_ionization_1965, Perelomov1966, Reiss1980, Ammosov1986, Faisal_book_1987}. In solids, the Keldysh parameter differs from atoms gases by a factor $1/\sqrt{2}$, i.e. $\gamma=\sqrt{E_g/(4U_{p})}$ \cite{Krausz_RMPSolids_2018}. Here, $I_p$ is the ionization potential of the atoms, $E_{g}$  is the energy band gap of the material, $U_{p}=E_L^2/(4\omega_{L}^2) \approx 9.33 \cdot 10^{-14} \ [I_{L}(\text{W}/\text{cm}^2)\lambda_{L}^{2}(\mu \text{m})$] is the ponderomotive energy, i.e., the average oscillation energy of the electron in a laser field with amplitude $E_L$ and frequency $\omega_{L}$. For gas phase and solid targets, the interaction is in the strong field limit when the pulse intensity is typically in the range $I_{L}>10^{14}$ W/cm$^2$ and $I_L>10^{11}$ W/cm$^2$, respectively. Interactions in this intensity region led to the observation of a number of new phenomena in all states of matter. Among these, are the processes of tunneling ionization ~\cite{keldysh_ionization_1965, Perelomov1966, Reiss1980, Ammosov1986, Faisal_book_1987, chin_tunnel_1985}, above threshold ionization (ATI) ~\cite{agostini_freefree_1979}, high-energy ATI (HATI)~\cite{paulus_plateau_1994}, non--sequential double ionization (NSDI) \cite{lhuillier_multiply_1983, walker_precision_1994, moshammer_momentum_2000, weber_correlated_2000}, and high harmonic generation (HHG) ~\cite{ferray-multiple-1988, mcpherson-studies-1987} where the low-frequency photons of a driving laser field are converted into photons of higher frequencies. Figure~\ref{fig6:IntenseLaserMatter} shows an example of an HHG and photoelectron spectrum generated by the interaction of Xenon atoms with intense IR fs laser pulses.  

These processes are important in strong laser physics and have been used in numerous fascinating achievements in atomic, molecular and optical physics and attosecond science \cite{Huillier_Nobel_2024, Krausz_Nobel_2024, Agostini_Nobel_2024}. Until recently, the laser-matter processes have been successfully described by classical \cite{corkum_plasma_1993} or semi-classical strong-field approximations \cite{lewenstein-theory-1994, ABC19}, which treat the electromagnetic field classically. As has been highlighted in the introduction of this  Review, this has recently changed. Theoretical and experimental investigations, conducted using fully quantized approaches, have shown that the intense laser–-matter interactions and the HHG/photoionization process can be used for the generation of optical Schrödinger ``cat'' and entangled light states. 

In the next sections (Sec. \ref{StrongField_QED}--\ref{Methods9}), we review some of the main tools and ideas underlying these investigations. We discuss, the recently developed fully quantized approaches that have been used for the description of intense laser--matter interactions and the generation of non-classical and massively entangled states for applications in quantum information science. The discussion will be focused on the interaction of intense IR laser pulses with ground state depleted and excited atoms, molecules, solids and many body quantum correlated systems.

\section{Quantum electrodynamics of intense laser--matter interactions} \label{StrongField_QED}
  
\begin{figure}
    \centering
    \includegraphics[width=0.5 \columnwidth]{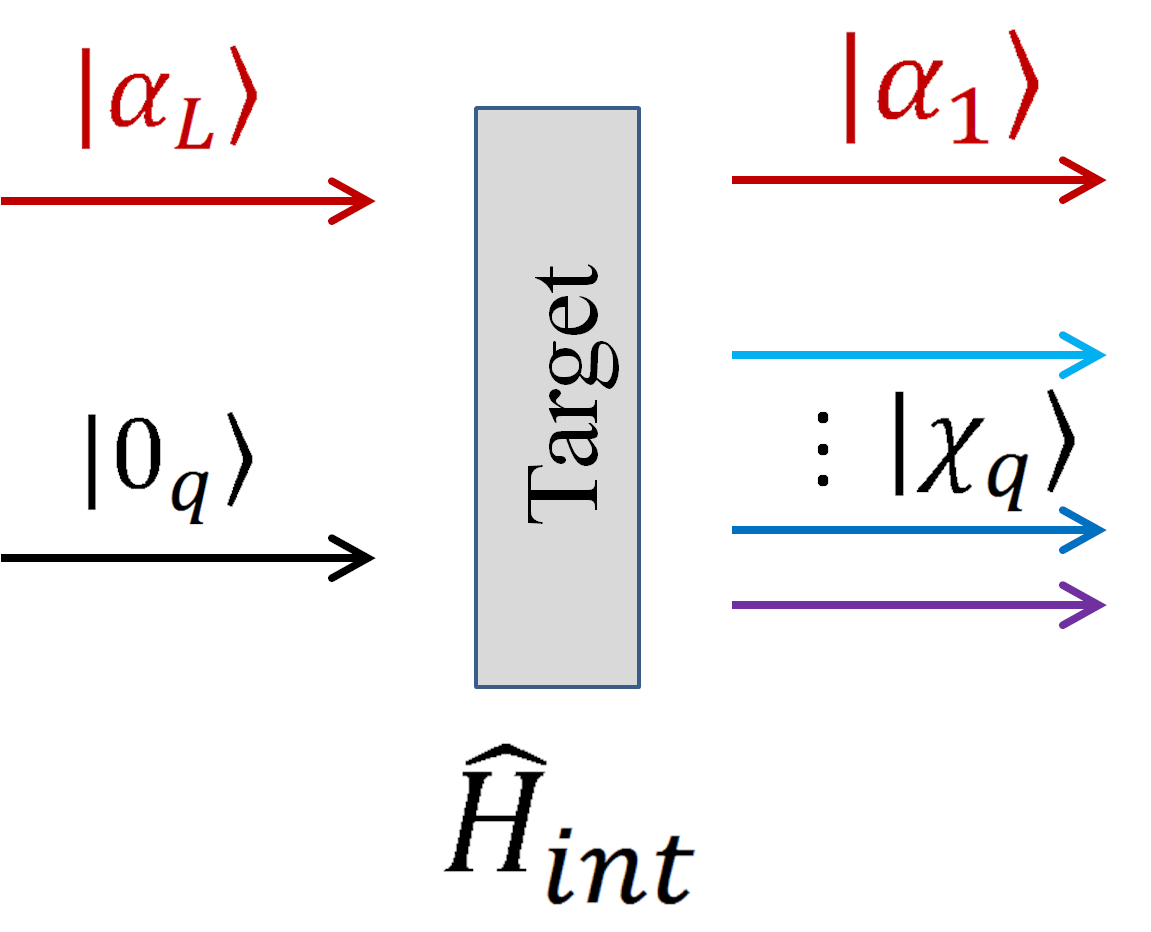}
    \caption{A schematic of the fully quantized description of intense laser--matter interaction where the electromagnetic radiation and the interaction are treated quantum mechanically. In this plot, the IR field before and after the interaction, are in the coherent states $\ket{\alpha_{L}}$ and $\ket{\alpha_{1}}$, respectively, with $|\alpha_{1}|<|\alpha_{L}|$ due to energy conservation. Before the interaction the harmonic are in vacuum state $\ket{0_{q}}$ and are also shifted to coherent states due to the HHG process $\ket{\chi_{q}}$ with amplitudes $|\chi_{q}|$.}
    \label{fig7:QEDScheme}
\end{figure}

In this section we discuss the case where the depletion of the ground state of the system from the driving IR laser field is negligible. The fully quantized description of intense laser---matter interaction (shown schematically in Fig.~\ref{fig7:QEDScheme}) has been extensively discussed in Refs.~\cite{LCP21, RLP22, SRM2023, RLP22, SRM2023, SRL22, RSP21, RSM22, Tzallas_ROPP_2024, CDF24}. Here, we summarize the main findings. The driving IR field mode before and after interaction are in the coherent states $\ket{\alpha_{L}}$ and $\ket{\alpha_{1}}\equiv\ket{\alpha_{L}+\delta\alpha_{L}}$, respectively. Due to energy conservation in the HHG area $|\alpha_{1}|<|\alpha_{L}|$ with the negative $\delta\alpha_{L}$ representing the IR energy losses. 
The field state before the interaction ($t=0$) is $\ket{\Phi(0)} = \ket{\alpha}\bigotimes^{\text{N}_{\text{c}}}_{q=2} \ket{0_q}$ where $N_{c}$ is the cut-off harmonic order and the harmonics are initially in the vacuum. Considering that the target medium before the interaction is in the ground state $\ket{g}$, the state of the joint (light and target medium) system before the interaction is $\ket{\Psi(0)} = \ket{g}\otimes\ket{\Phi(0)}= \ket{g}\otimes \ket{\alpha}\bigotimes^{\text{N}_{\text{c}}}_{q=2} \ket{0_q}$.

The Hamiltonian describing the interaction is given by $\hat{H} = \hat{H}_{s} + \hat{H}_f + \hat{H}_{\text{int}}$. $\hat{H}_{s}$ is the Hamiltonian of the matter system, $\hat{H}_f$ is the Hamiltonian of the electromagnetic field, and $\hat{H}_{\text{int}}$ describes the interaction between light and matter. $\hat{H}_f$ for linearly polarized fields is $\hat{H}_f=\sum_{q} \hbar\omega_{q}\hat{a}_q^\dagger \hat{a}_q$ with $\hat{a}_q^\dagger$($\hat{a}_q$) the creation (annihilation) operator acting on the field mode with frequency $\omega_{q}$ ($q=1$ corresponds to the fundamental IR driving field). Under the length-gauge and dipole approximation, $\hat{H}_{\text{int}} = \hat{\vb{d}}\cdot \hat{\vb{E}}$, with $\hat{\vb{d}}$ the dipole moment operator and $\hat{\vb{E}} =  i \vb{g}(\omega_L) \sum_{q=1}^{N_c} (\hat{a}_q - \hat{a}_q^\dagger)$ the electric field operator. $|\vb{g}(\omega_L)| \propto \sqrt{\omega_{L}/V_{eff}}$ is the coefficient that enters into the expansion of the laser electric modes and depends on the effective quantization volume $V_{eff}$ \cite{Grynberg_Book_2010}. The dynamics of the system is described by the Schrödinger equation, 

\begin{equation}\label{Eq:Sch:transf}
    i\hbar \dv{\ket{\Psi(t)}}{t} 
        = \hat{H}(t) \ket{\Psi(t)}.
\end{equation}

In the next subsections (\ref{QED_Atoms}--\ref{QED_Solids}) we describe the solution of the above equation in case of strongly laser driven atoms, molecules and solids, and we show how it provides access to the generation of non-classical and entangled light states ranging from the far--IR to the XUV (Sec. \ref{Methods6}--\ref{Methods8}).

\subsection{QED of strongly laser driven atoms} \label{QED_Atoms}

\begin{figure}
    \centering
    \includegraphics[width=0.7 \columnwidth]{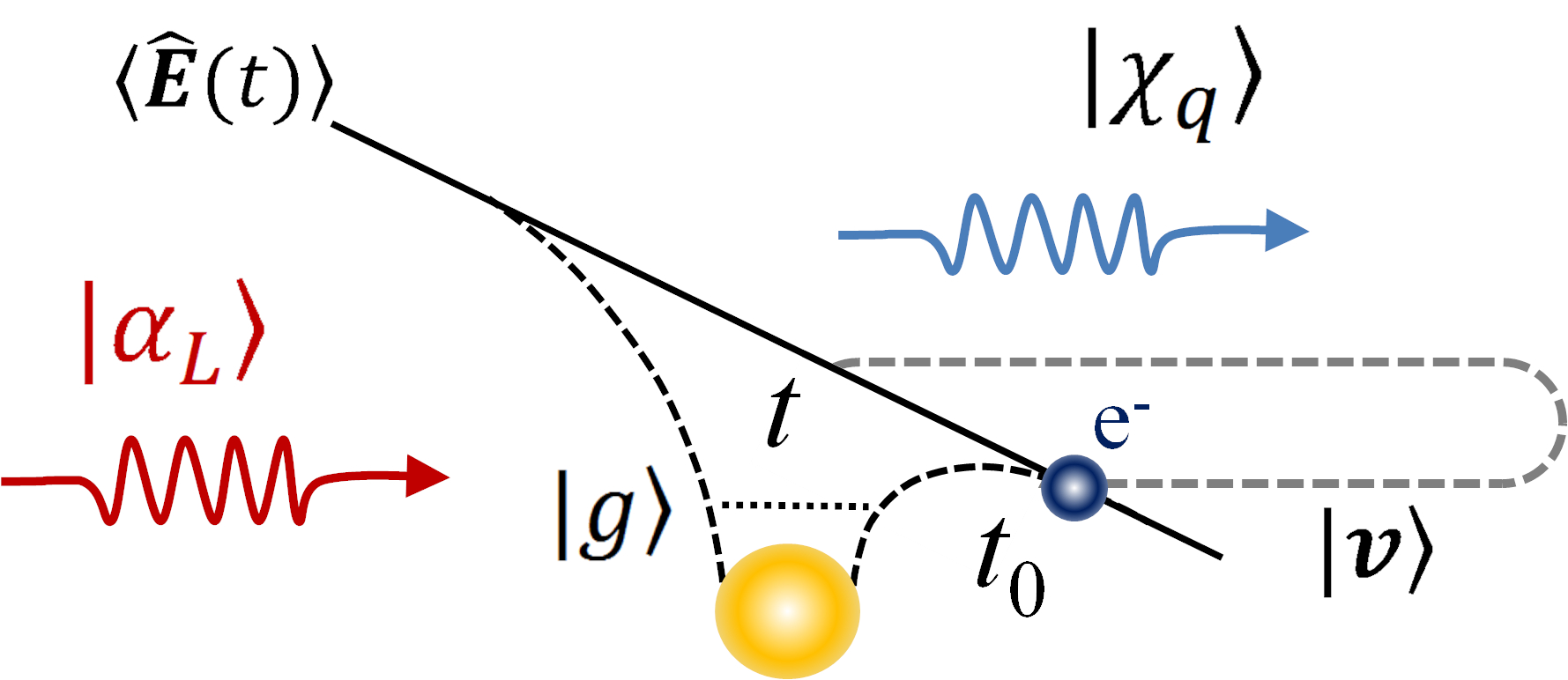}
    \caption{A schematic of the fully quantized description of intense laser--atom interaction and HHG process. For simplicity, the photoelectron emission processes are not shown. The driving IR field distorts the atomic potential. The HHG process in a single atom is initiated by an electron’s tunnelling to the continuum, its subsequent acceleration in the intense laser field and finally its recombination to the ground state of the atom. $\ket{\alpha_{L}}$ and $|\chi_{q}|$ are the coherent states of the driving IR laser field and the emitted harmonic modes. $\ket{g}$ and $\ket{\mathbf{v}}$ is the ground state of the atom and the state of the electron in the atomic continuum with velocity $\mathbf{v}$. $t_{0}$ and $t$ is the ionization and recombination times, respectively. $\langle{\hat{\mathbf{E}}(t)}\rangle$ is the mean value of the electric field operator.}
    \label{fig8:QEDAtoms}
\end{figure}

The QED of strongly laser driven atoms is pictorially given in Fig.~\ref{fig8:QEDAtoms}. We consider the situation of a single active electron (SAE) participating in the dynamics initially found in the atomic ground state $\ket{g}$. To simplify the description of the dynamics we perform a set of unitary transformations \cite{LCP21, SRM2023, Tzallas_ROPP_2024} bringing us to a more convenient frame of reference. In this frame, the quantum optical description of the dynamics is given by the interaction Hamiltonian $H_{I}(t) = - \vb{d}(t) \vb{E}_{Q}(t)$ where $\vb{E}_{Q}(t) = - i\vb{g} f(t) \sum_{q=1}^{N} \sqrt{q} (a_{q}^{\dagger} e^{i \omega_{q} t}-a_{q} e^{-i \omega_{q} t})$ and $\vb{d}(t)$ denotes the time-dependent dipole moment. Here, $f(t)$ is a dimensionless function introduced for using a discrete set of modes and accounts for the finite duration of the driving laser pulse. The final form of the Time Dependent Schr\"{o}dinger Equation (TDSE) then reads

\begin{equation}\label{Eq:trans:TDSE}
    i\hbar \pdv{\ket{\Psi(t)}}{t}
        = \vb{d}(t)\cdot \vb{E}_{Q}(t)\ket{\Psi(t)}.
\end{equation}

\par 
The HHG process from a single atom begins with an electron tunneling into the continuum, followed by its acceleration in the intense laser field and eventual recombination with the atom's ground state. To characterize the quantum optical state after the HHG process, where the electron returns to the ground state, we project Eq. \eqref{Eq:trans:TDSE} onto $\ket{g}$ and denote the state of light as  $\ket{\Phi (t)} = \braket{g}{\Psi (t)}$. We then introduce the SFA version of the identity, $\mathbbm{1} = \dyad{g} + \int \dd{\text{v}} \dyad{\text{v}}$, neglecting the contribution of the excited bound states as they barely affect the dynamics under the SFA \cite{lewenstein-theory-1994}. Furthermore, by neglecting the contribution of the electronic continuum states $\ket{ \text{v} }$, since the continuum amplitude is much smaller than the ground state amplitude, and that the depletion of the ground state is small \cite{lewenstein-theory-1994, Stammer_PRL_2024}, we find that the dynamics of $\ket{\Phi (t)}$ can be described by

\begin{equation}\label{QED_ATOMSeq:TDSE_Av_dipole}
    i\hbar \pdv{\ket{\Phi(t)}}{t}
        = \expval{\vb{d}(t)}\cdot \vb{E}_{Q}(t)\ket{\Phi(t)}.
\end{equation}

The interaction is described by a multimode displacement operator

\begin{equation}\label{QED_ATOMSeq:Displ_OP_no_dipole_correlations}
    K_{HHG}\simeq \mathcal{T} \exp[i\int_{0}^{t} \dd{t'} \expval{\vb{d}(t')}\vb{E}_{Q}(t')] = D[\chi_{q}]
\end{equation}

\noindent where the coherent displacements appearing in the last equation are proportional to the Fourier transform of the dipole moment expectation value, $\chi_{q} \propto \int_{0}^{t}\dd{t'} f(t') \expval{\vb{d}(t')}e^{i \omega_{q} t'}$. The final field state $\ket{\Phi}$ is, therefore, given by a product of coherent states

\begin{equation}\label{QED_ATOMSeq:Final_Field_State}
   \ket{\Phi (t)} = K_{HHG}\ket{\Phi(0)} = e^{i\phi(t)} \bigotimes_{q=1}^{N} D [\chi_{q}(t)] \ket{0_{q}}
\end{equation}

\par The approximation leading to Eq.~\eqref{QED_ATOMSeq:Final_Field_State} has the underlying assumption of vanishing correlations in the dipole moment operator \cite{sundaram_high-order_1990, SRL22, Sta22}, where in typical experiments with intensities $\leq 10^{14} \text{W/cm}^{2}$ holds true. We note that, so far, we have been working under single-atom dynamics, and therefore the generated shift $\chi_{q}(t)$ is very small. Nevertheless, this solution can be extended to a more realistic scenario, where we have $N_{at}$ atoms participating in the dynamics in a phase-matched way. This scenario leads to the enhancement of the coherent displacement, i.e. $\chi_{q} \to N_{at}\chi_{q}(t)$.

\subsection{QED of strongly laser driven molecules} \label{QED_Molecules}

\begin{figure}
    \centering
    \includegraphics[width=0.7 \columnwidth]{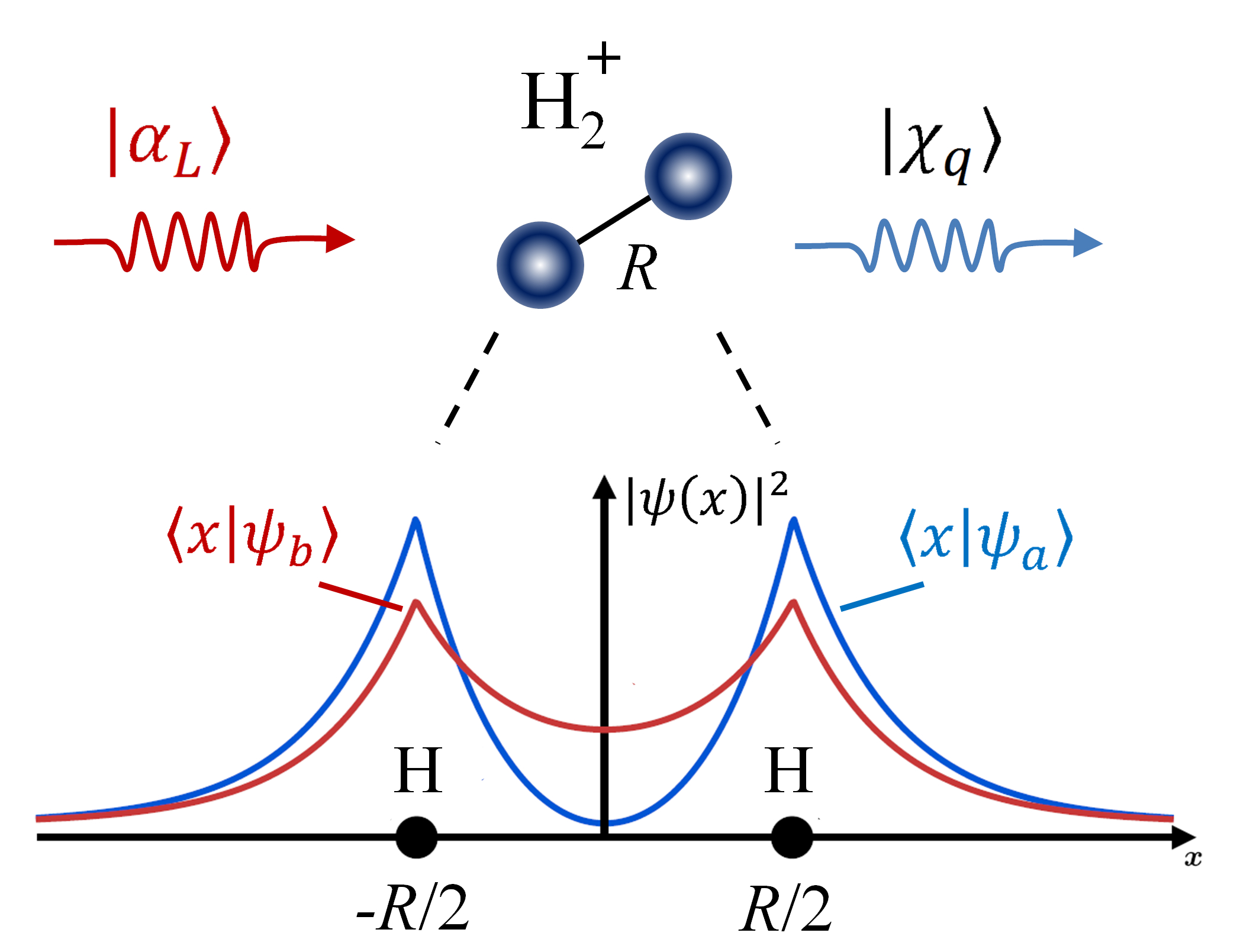}
    \caption{A schematic of the fully quantized description of strongly laser driven $\text{H}_{2}^{+}$ molecule and HHG process. For simplicity, the photoelectron emission processes are not shown. The HHG process in a single molecule is similar to atoms i.e. it is initiated by an electron’s tunnelling to the continuum, its subsequent acceleration in the intense laser field and finally its recombination to the ground state of the molecule. $\ket{\alpha_{L}}$ and $|\chi_{q}|$ are the coherent states of the driving IR laser field and the emitted harmonic modes. Here, $\ket{\psi_{b}}$ and $\ket{\psi_{a}}$ are the bonding and antibonding states of the $\text{H}_{2}^{+}$, and $R$ denotes the internuclear distance, also referred to as the bond length. The figure has been reproduced from ref. ~\cite{rivera-dean_quantum_2023}.}
    \label{fig9:QEDMolecules}
\end{figure}

The QED of strongly laser--driven molecules has been theoretically demonstrated in Ref.~\cite{rivera-dean_quantum_2023} using H$^+_2$ molecules. HHG in molecular systems is intrinsically different from its atomic counterpart, primarily due to the additional degrees of freedom that molecules offer.~Factors such as molecular orientation with respect to the polarization axis~\cite{lein_role_2002,lein_orientation_2003}, and the number of atomic cores in the system~\cite{suarez_high-order-harmonic_2017,suarez_rojas_strong-field_2018}, strongly influence the inherent HHG dynamics and, ultimately, the resulting spectra. In this section, we focus on the simplest molecule one can consider: the H$^+_2$ molecule (see Fig.~\ref{fig9:QEDMolecules}), which consists of two hydrogen nuclei separated by a distance $R$--the bond length--and a single active electron. Despite its simplicity, this molecule is complex enough to exhibit HHG dynamics distinct from those in atomic systems. For example, the electron can ionize from one center and recombine with either the same center or the opposite one, leading to interference effects observable in the HHG spectrum~\cite{lein_interference_2002,figueira_de_morisson_faria_high-order_2002,suarez_high-order-harmonic_2017}. Here, we briefly discuss how these mechanisms affect the quantum optical state of the system~\cite{rivera-dean_quantum_2023}.

In addition to the dipole and single-active electron approximations, it is common to also employ the Born-Oppenheimer approximation, when dealing with molecules~\cite{born_zur_1927}, which considers the atomic positions to remain fixed during the electron dynamics. This approximation assumes that the atomic positions remain fixed during the electron dynamics, which is justified by the large mass ratio between nuclei and electrons, typically on the order of $10^3$. As a result, nuclear dynamics occur on much longer timescales than electronic dynamics. For instance, in the case of H$_2^+$ molecules, vibrations due to the repulsion between the two atomic centers occur on timescales of approximately 10 fs~\cite{poll_vibrational_1966,silvera_solid_1980}, whereas HHG electron dynamics typically unfold on timescales around 1 fs. Consequently, under the dipole, single-active electron, and Born-Oppenheimer approximations, the Hamiltonian governing the HHG interaction can be expressed as $\hat{H} = \hat{H}_{\text{mol}} + \hat{H}_{\text{int}} + \hat{H}_f$, where $\hat{H}_{\text{mol}}$ represents the Hamiltonian for an electron interacting with the H$_2^+$ molecular potential.

Within these approximations, the interaction between light and molecules closely resembles that with atomic systems, with the main distinction being the molecular Hamiltonian. Therefore, it is crucial for this analysis to have an accurate characterization of the molecular system. While this is in general a complex task, especially as the molecule becomes larger, for H$_2^+$ molecules the Linear Combination of Atomic Orbitals (LCAO) method provides a simple yet reasonably good characterization of the molecular eigenstates~\cite{finkelstein_uber_1928,atkins_book}. This method expresses the molecular eigenstates as a linear combination of a finite set $\{\Lambda_{i,j}\}$ of atomic orbitals $j$ at different centers $i$, i.e. $\ket{\psi}=\sum_i \sum^N_j c_{i,j}\ket{\Lambda_{i,j}}$. Increasing the number of atomic orbitals $N$ yields a more accurate description of the molecular eigenstates. Once the set of atomic orbitals is fixed, the task reduces to finding the probability amplitudes $c_{i,j}$ that minimize the mean energy of the molecule, a problem that can be approached through a variational principle~\cite{atkins_book_ch6}. 

For simple systems, such as H$_2^+$, the required number of atomic orbitals needed to faithfully represent the molecular eigenstates is relatively small. To describe the lowest energy molecular orbitals of H$_2^+$, it suffices to consider the set $\{\ket{g_1},\ket{g_2}\}$, representing the hydrogen atom ground state for each of the two centers. The ground state, known as the bonding state, is then $\ket{\psi_{\text{b}}} = 1/\sqrt{2}[\ket{g_1}+\ket{g_2}]$, while the first excited state, known as the anti-bonding state, is $\ket{\psi_{\text{a}}} = 1/\sqrt{2}[\ket{g_1}-\ket{g_2}]$.~This bonding and anti-bonding notation arises because, in the former, the electron is delocalized between the two atomic centers, while in the latter, it is localized at the two centers (see Fig.~\ref{fig9:QEDMolecules}). These two states can also be used to define a set of localized states at each center, specifically $\ket{R} = 1/\sqrt{2}[\ket{\psi_b} + \ket{\psi_{a}}]$ and $\ket{L} = 1/\sqrt{2}[\ket{\psi_b} - \ket{\psi_{a}}]$, corresponding to states localized at the right and left molecular centers, respectively.

Following a similar description to that in Sec.~\ref{QED_Atoms}, the Schrödinger equation describing the light-matter interaction dynamics can be written as in Eq.~\eqref{Eq:trans:TDSE}. In this case, the initial state of the joint light-matter system is $\ket{\Psi(t_0)} = \ket{\psi_b}\otimes \ket{\alpha}\otimes \ket{\{0\}_{q>1}}$. To solve it, we go slightly beyond the standard SFA and include the first excited state, with the approximate resolution of the identity being $\mathbbm{1} = \dyad{\psi_{b}} + \dyad{\psi_{a}} + \int \dd v  \dyad{v}$. This extension is necessary because, as mentioned earlier, HHG in molecules can involve recombination localized at different atomic centers~\cite{lein_interference_2002,figueira_de_morisson_faria_high-order_2002,suarez_high-order-harmonic_2017}, which requires both the ground and first excited states to be considered within the LCAO. This leads to a set of coupled differential equations, which can be approximately solved by
\begin{equation}
    \ket{\Psi(t)}
        \approx \ket{\psi_b} \otimes \ket{\Phi_b(t)}
            + \ket{\psi_a} \otimes \ket{\Phi_a(t)},
\end{equation}
where we have disregarded the contribution of the continuum terms similarly as in Sec.~\ref{QED_Atoms}. However, the molecular continuum states are indeed considered when solving the HHG electron dynamics. In this expression, $\ket{\Phi_b(t)}$ and $\ket{\Phi_a(t)}$ represent the quantum optical states associated to the bonding and anti-bonding components, respectively, and are approximately given by
\begin{align}
    & \ket{\Phi_b(t)}
        = \hat{\vb{D}}\big(\boldsymbol{\chi}_b(t,t_0)\big)
            \ket{\bar{0}},\label{Eq:QO:bonding}
    \\&
    \ket{\Phi_a(t)}
        = -i \int^t_{t_0} \dd \tau
            \hat{\vb{D}}\big(\boldsymbol{\chi}_a(t,\tau)\big)
            \hat{M}_{ab}(\tau)
            \hat{\vb{D}}\big(\boldsymbol{\chi}_b(\tau,t_0)\big)
            \ket{\bar{0}},\label{Eq:QO:antibonding}
\end{align}
where $\chi^{(q)}_i(t,t_0) \propto \int^t_{t_0} \mel{\psi_i}{\boldsymbol{\epsilon}_L\cdot\hat{\vb{r}}}{\psi_i}e^{iq\omega_L t}$ and $\hat{M}_{ab}(t) \equiv \mel{\psi_a}{\hat{\vb{r}}}{\psi_b}\cdot \hat{E}(t)$. 

Equations \eqref{Eq:QO:bonding} and \eqref{Eq:QO:antibonding} represent two very distinct processes. The first resembles those found in atomic systems, where the electron undergoes a parametric process, resulting in a displacement on the quantum optical degrees of freedom. In contrast, the second presents additional events, where at time $\tau$ the electron transitions to the anti-bonding state---mediated by the operator $\hat{M}_{ab}(t)$---and evolves within this state until the final time $t$, with the quantum optical state experiencing a shift of $\boldsymbol{\chi}_a(t,\tau)$. These different dynamics affect the final quantum optical state in distinct ways, ultimately leading to non-classical features in the field modes, as will be described later. However, for these effects to become significant, the contribution of a large number of molecules, $N_{\text{mol}}$, must be considered. Since transitions from bonding to antibonding states are less likely than bonding-bonding transitions, the $N_{\text{mol}}$ state can be expressed as~\cite{rivera-dean_quantum_2023}
\begin{equation}\label{Eq:QO:Final:Molecules:Nmol}
    \ket{\Psi(t)} 
        = \dfrac{1}{\sqrt{\mathcal{N}}}
            \big[
                \ket{\bar{\psi}_b}
                    \otimes \ket{\bar{\Phi}_b}
                + \sqrt{N_{\text{mol}}}
                    \ket{\bar{\psi}_a}
                    \otimes \ket{\bar{\Phi}_a}
            \big],
\end{equation}
where $\mathcal{N}$ is the normalization factor, $\ket{\bar{\psi}_b}$ represents the joint state of the $N_{\text{mol}}$ molecules in the bonding state, while $\ket{\bar{\psi}_a} = (1/\sqrt{N_{\text{mol}}})\sum_{k=1}^{N_{\text{mol}}} \ket{\bar{\psi}_{a_k}}$ with $\ket{\bar{\psi}_{a_k}}$ denoting the state of all molecules in the bonding state except the $k$th one, which is in the antibonding state. In this case, we find for the quantum optical components
\begin{align}
    & \ket{\bar{\Phi}_b(t)}
        \approx \hat{\vb{D}}\big(\boldsymbol{\bar{\chi}}_b(t,t_0)\big)
            \ket{\bar{0}},\label{Eq:QO:bonding:Nmol}
    \\&
    \ket{\bar{\Phi}_a(t)}
        \approx -i 
            \hat{\vb{D}}\big(\boldsymbol{\bar{\chi}}_b(t,t_0)\big)
            \int^t_{t_0}
            \dd \tau
            \hat{\bar{M}}_{ab}(t)
            \ket{\bar{0}},\label{Eq:QO:antibonding:Nmol}
\end{align}
where $\boldsymbol{\bar{\chi}}_b(t,t_0) = N_{\text{mol}}\boldsymbol{\chi}_b(t,t_0)$ and $\hat{\bar{M}}_{ab}(t) = \hat{M}_{ab}(t) + \mel{\psi_a}{\hat{\vb{r}}}{\psi_b}\cdot \mel{\boldsymbol{\bar{\chi}}_b(t,t_0)}{\hat{\vb{E}}(t)}{\bar{\boldsymbol{\chi}}_b(t,t_0)}$.

\subsection{QED of strongly laser driven semiconductor solids} \label{QED_Solids}
\begin{figure}
    \centering
    \includegraphics[width=0.9 \columnwidth]{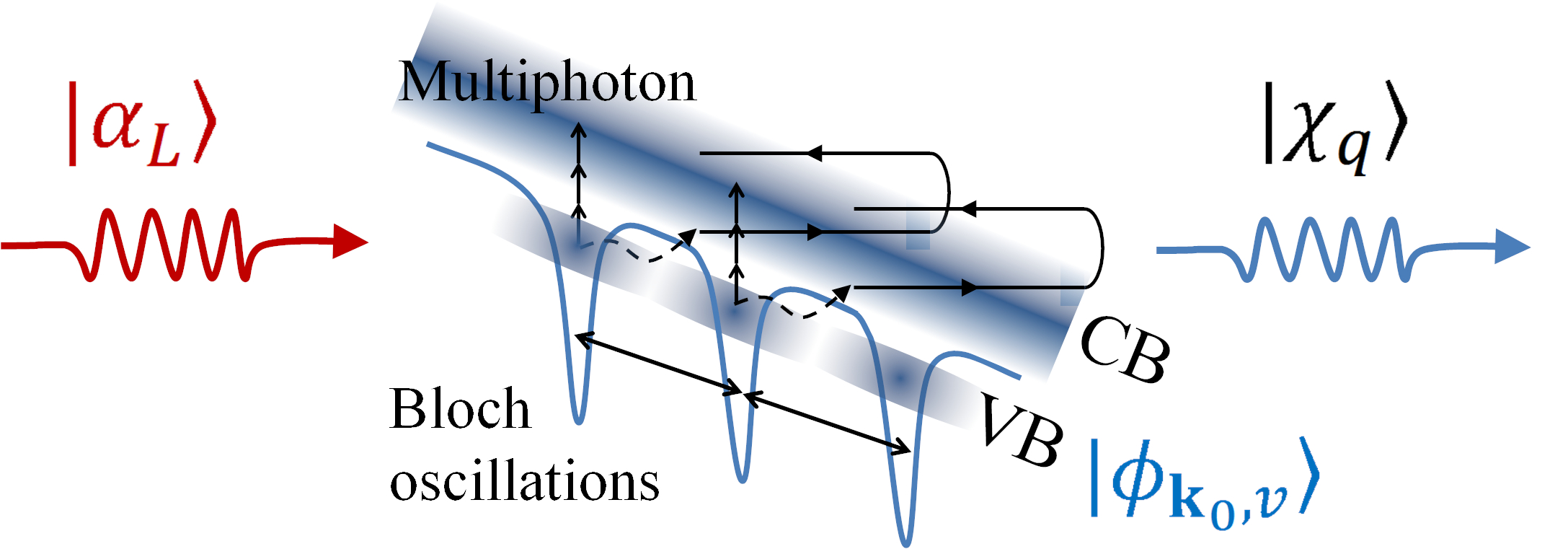}
    \caption{A schematic of the fully quantized description of strongly laser driven solids and HHG process. An intuitive physical picture of the process can be given using the concept of the electron trajectories. When an intense linearly polarized laser field (with $\hbar\omega_{L}\ll E_{g}$) interacts with a crystal, the electron escapes the valence band (VB) to the conduction band (CB), it subsequently accelerates in the CB (likewise the hole in the VB) gaining energy from the driving field, and recombines with the hole (in a multicenter recombination process along the crystal lattice) within the same cycle of the driving field. This recombination process generates high harmonics. $\ket{\alpha_{L}}$ and $|\chi_{q}|$ are the coherent states of the driving IR laser field and the emitted harmonic modes. Here, $\ket{\phi_{\textbf{k}_0, v}}$ is the ground state of the system where the electron is initially located in the VB with crystal momentum $\textbf{k}_0$.}
    \label{fig10:QEDSolids}
\end{figure}

The QED of strongly laser--driven semiconductor solids has been theoretically demonstrated in Refs.~\cite{rivera-dean_bloch_2023, Gonoskov_PRB_2024}. Here we will describe the approach followed by J. Rivera--Dean {\it et al.}, in Ref.~\cite{rivera-dean_bloch_2023}. The HHG process in semiconductor materials bears similarities to that in the gas phase, with the key distinction being that in semiconductors, the electron is constrained by the material's band structure~\cite{vampa_theoretical_2014,vampa_semiclassical_2015}, which ultimately dictates the properties of the emitted HHG radiation. Consequently, the specific crystal orientation along which the HHG process is driven~\cite{ghimire_observation_2011}, as well as the various dephasing mechanisms present in the system~\cite{vampa_theoretical_2014,kilen_propagation_2020}, play a crucial role in determining the electron dynamics. These dynamics can either occur within a given band (intraband dynamics or Bloch oscillations~\cite{bloch_uber_1929}) or involve transitions between different bands (interband dynamics~\cite{vampa_theoretical_2014,vampa_semiclassical_2015}). The latter mechanism resembles the three-step model found in atomic systems (see Fig.~\ref{fig10:QEDSolids}). This variety of dynamics, along with the dependence of HHG radiation on the driving conditions, has been extremely useful for gaining insights into semiconductor materials~\cite{goulielmakis_high_2022}, nanomaterials~\cite{Ciappina2017} and quantum materials~\cite{zong_emerging_2023}.

From an operational perspective, HHG in semiconductor materials require different excitation conditions compared to the gas phase. For typical semiconductors such as ZnO or GaSe~\cite{ghimire_observation_2011,schubert_sub-cycle_2014}, the laser sources used are typically in the mid-IR regime, with photon energies much smaller than the material's bandgap, and laser intensities on the order of $10^{11}-10^{12}$ W/cm$^2$, which are below the material's damage threshold. Despite these distinct conditions, HHG radiation has been observed from strong-laser-driven semiconductors extending up to the XUV regime~\cite{vampa_observation_2018}, opening the possibility of developing attosecond light sources~\cite{Nayak_NatCom_2025} on a much smaller scale, especially given that the typical thicknesses of solid samples in HHG experiments are around the $\sim$500 $\mu$m range. This makes semiconductor materials excellent candidates developing sources of non-classical light, which can be integrated into more complex setups designed for specific quantum information science applications~\cite{Ralph2003,gisin_quantum_2007}.

Here, we briefly describe how the quantum optical state is affected by the various strong-field-driven electron dynamics occurring in semiconductor materials~\cite{rivera-dean_bloch_2023}. We work under both the single active electron approximation and dipole approximation, which we expect to be valid given that typical lattice sites where electron dynamics occur are in the order of 1 \AA, much smaller than the $\sim\!1-10$ $\mu$m wavelength of the driving field. With these considerations, the Hamiltonian characterizing the laser-solid interaction can be written as $\hat{H} = \hat{H}_{\text{cr}} + \hat{H}_{\text{int}} + \hat{H}_f$, where $\hat{H}_{\text{cr}}$ denotes the crystal Hamiltonian.

In the following, we consider a two-band model for the solid, with the eigenstates of the crystal Hamiltonian given by the Bloch states $\{\ket{\phi_m(\vb{k})}\}$ with eigenvalues $\{E_m(\vb{k})\}$, where $\vb{k}$ denotes the crystal momentum and $m$ represents the band ($v$ for valence band and $c$ for conduction band). The initial state is taken as $\ket{\Psi(t_0)} = \ket{\phi_v(\vb{k}_0)}\otimes \ket{\alpha_L}\otimes \ket{\{0\}_{q>2}}$, with the electron initially in the valence band occupying the Bloch state with crystal momentum $\vb{k}_0$. Furthermore, we express the interaction Hamiltonian as $\hat{H}_{\text{int}} = [\hat{\vb{r}}_{\text{tra}} + \hat{\vb{r}}_{\text{ter}}]\cdot \hat{\vb{E}}(t)$, where $\hat{\vb{r}}_{\text{tra}}$ and $\hat{\vb{r}}_{\text{ter}}$ represent the contributions to the dipole operator arising from the intraband and interband dynamics, respectively. To describe the dynamics, we perform similar unitary transformations to those in Sec.~\ref{QED_Atoms}, along with an additional unitary transformation $U_{\text{tra}}(t) = e^{i\vb{A}_{\text{cl}}(t)\cdot\hat{\vb{r}}_{\text{tra}}}$, where $\vb{A}_{\text{cl}}(t)$ denotes the classical vector potential, allowing us to work in terms of the canonical crystal momentum $\vb{K} = \vb{k} - \vb{A}_{\text{cl}}(t)$, spanned by the basis set $\{\ket{\vb{K},m}\}$. In the interaction picture with respect to the semiclassical Hamiltonian $H_{\text{sc}}(t) = \hat{H}_{\text{cr}} + \hat{\vb{r}}_{\text{ter}} \cdot \vb{E}_{\text{cl}}(t) $, the TDSE describing the light-matter interaction dynamics, up to first order in $g(\omega_q)$, reads
\begin{equation}\label{Eq:QO:TDSE:Solids}
    i\pdv{\ket{\Psi'(t)}}{t}
        = \big[
                \hat{\vb{r}}_{\text{ter}}
                    \cdot \hat{\vb{E}}(t)
                -i \hat{\vb{v}}_{\text{tra}}
                    \cdot \hat{\vb{A}}(t)
            \big]\ket{\Psi'(t)},
\end{equation}
where $\hat{\vb{v}}_{\text{tra}} = [\hat{\vb{r}}_{\text{tra}}, \hat{H}_{\text{cr}}]$, and the time-dependence in the dipole operators arises upon transformation with respect to $\hat{U}_{\text{sc}}(t)$.

A solution to Eq.~\eqref{Eq:QO:TDSE:Solids} can be always expressed as
\begin{equation}
    \ket{\Psi'(t)}
        = \sum_{m} \int \dd \vb{K}
            \ket{\vb{K},m}\otimes \ket{\Phi_m(\vb{K},t)},
\end{equation}
where $\ket{\Phi_m(\vb{K},t)}$ denotes the quantum optical component associated to the state $\ket{\vb{K},m}$. Upon introducing this expansion in Eq.~\eqref{Eq:QO:TDSE:Solids}, and assuming a linearly polarized driving field that vanishes as $t\to \pm \infty$, such that $\vb{k}$ and $\vb{K}$ asymptotically converge, the solution can be expressed up to first order in $g(\omega_q)$ as
\begin{equation}\label{Eq:QO:state:HHG:solids}
    \begin{aligned}
    \ket{\Psi'(t)}
        &= \hat{\vb{D}}
            \big(
                \boldsymbol{\chi}_v(\vb{k}_0,t,t_0)
            \big)
                \ket{\vb{k}_0,v}
                \otimes 
                \ket{\bar{0}}
            \\&\quad
            -i\int \dd\tau
                \hat{\vb{D}}
                \big(
                    \boldsymbol{\chi}_c(\vb{k}_0,t,\tau)
                \big)
                \hat{M}_{c,v}(\vb{k}_0,\tau)
            \\&\hspace{2cm}
                \times
                \hat{\vb{D}}
                \big(
                    \boldsymbol{\chi}_v(\vb{k}_0,\tau,t_0)
                \big)
                \ket{\vb{k}_0,c}\otimes \ket{\bar{0}},
    \end{aligned}
\end{equation}
where $\hat{M}_{i,j}(\vb{K},t) \equiv m_{i,j}^{(\text{ter})}(\vb{K},t)\hat{E}(t) + m^{(\text{tra})}_{i,j}(\vb{K},t) \hat{A}(t)$ is an operator accounting for the transitions between valence and conduction band, and where the displacements are given by $\chi^{(q)}_{i}(\vb{K},t,t_0)\propto \int^t_{t_0} \dd \tau [-m_{i,i}^{(\text{ter})}(\vb{K},\tau)f_q(\tau) + m_{i,i}^{(\text{tra})}(\vb{K},\tau)F_q(\tau)]$, with $f_q(t) \equiv f(t)e^{i\omega_q t}$ and $F_q(t) = \int \dd t f_q(t)$ where $f(t)$ represents the envelope of the applied laser field. The terms $m^{(\text{tra})}_{i,j}(\vb{K},t)$ and $m^{(\text{ter})}_{i,j}(\vb{K},t)$ represent the matrix elements of the operators $\hat{\vb{r}}_{m}^{(\text{ter})}(t)$ and $\hat{\vb{r}}_{m}^{(\text{tra})}(t)$. These matrix elements are the ones obtained in semiclassical analyses using the Semiconductor Bloch Equations (SBEs)~\cite{vampa_theoretical_2014,yue_introduction_2022}. These elements introduce the so-called interband and intraband currents, which ultimately define the HHG spectral features. Specifically, it has been observed that the intraband current dominates the low-harmonic spectral region (below the material's bandgap), while interband transitions primarily influences the high-harmonic spectral region (above the bandgap of the material)~\cite{vampa_theoretical_2014,vampa_semiclassical_2015}. When computing these components, we can account for the finite dephasing times $T_2$, typically around 1 fs, which are commonly employed in solid-state-HHG to reconcile numerical spectra (in the absence of dephasing) with experimental results~\cite{vampa_theoretical_2014,vampa_semiclassical_2015}. Although these small dephasing times are standard practice, they remain a point of debate within the community~\cite{goulielmakis_high_2022}.

Given that in HHG events the electron eventually returns to the valence band, we restrict the analysis to this subspace by introducing the operator $\hat{P}_v = \int \dd \vb{K} \dyad{\vb{K},v}$, and tracing out the other degrees of freedom. By applying this operation to Eq.~\eqref{Eq:QO:state:HHG:solids}, and after transforming back to the laboratory frame, the final quantum optical state can be approximately written as
\begin{equation}\label{Eq:QO:Solids:Final}
    \ket{\Phi_v(\vb{K},t)}
        = \hat{\vb{D}}
            \big(
             \boldsymbol{\chi}_v(\vb{k}_0,t,t_0)
            \big)
                \ket{\bar{0}},
\end{equation}
which corresponds to a product of coherent states, similar to those obtained in atomic systems, but with coherent state amplitudes influenced by both interband and intraband dynamics occurring within the solid. In writing this expression, we have implicitly neglected contributions that could introduce entanglement between the electron and the field degrees of freedom. While such features cannot be disregarded a priori, it was shown in Ref.~\cite{rivera-dean_Book_2024} that, for semiconductor materials such as ZnO, these effects are negligible.

Finally, although this analysis has been done within the single-active electron approximation, the HHG signal only becomes important when multiple electrons contribute collectively. However, unlike atomic systems, electrons in solid-state systems are densely packed, leading to inevitable many-body interactions. In semiconductor materials, these electron-electron interactions are weak and can be effectively modeled through couplings such as dephasing. This allows for treating each electron as an independent entity, thus extending Eq.~\eqref{Eq:QO:Solids:Final} as
\begin{equation}\label{Eq:QO:Solids:Final:MB}
    \ket{\Phi_v(\vb{K},t)}
        = \hat{\vb{D}}
            \bigg(
             N_z \int \dd \vb{K} \
             \boldsymbol{\chi}_v(\vb{K},t,t_0)
            \bigg)
                \ket{\bar{0}},
\end{equation}
where the integral in $\vb{K}$ arises from the fact that the valence band of the semiconductor is fully occupied. Here, $N_z$ denotes the number of Brillouin zones excited by the laser field, ultimately determined by the excitation conditions such as the laser beam width, the crystal alignment and, in case of transmission geometries, on the sample's width.

\section{Generation of optical ``cat'' and entangled light states using intense laser--matter interaction} \label{Methods6}

The quantum operations that have been used for the generation of non-classical have been discussed in Refs.~\cite{LCP21, RLP22, SRM2023, RSM22, TKG17, SRL22, Sta22, Stammer_ArXivConditioning_2024, Stammer_EnergyConservation2024, Tzallas_ROPP_2024}. These operations are associated to conditioning measurements on the interaction products such as photoelectrons and high-order harmonics. These can lead to the generation of optical ``cat'' and entangled states between different frequency modes. Here, we summarize the operation principles and the main findings using as target media atoms, molecules and solids in interactions where the intensity of the driving field does not deplete the ground state of the medium. 

\begin{figure}
    \centering
    \includegraphics[width=0.6 \columnwidth]{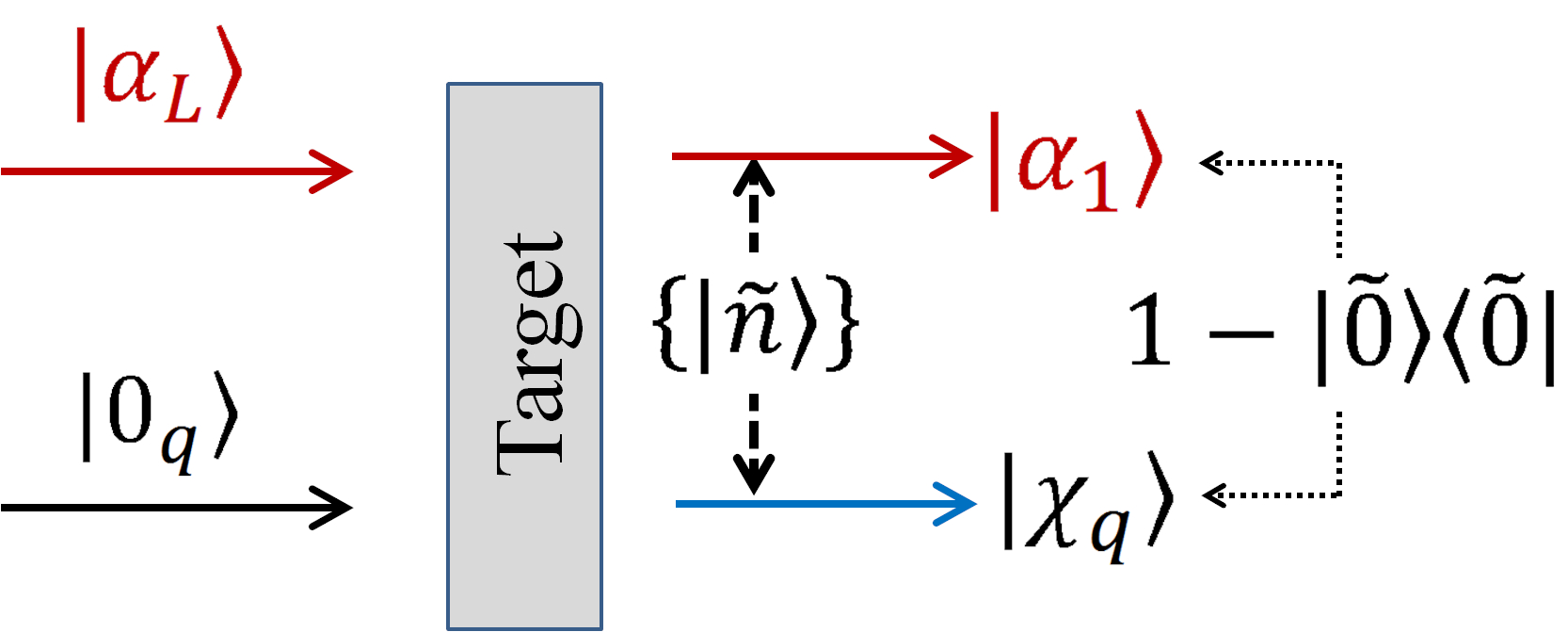}
    \caption{Conditioning scheme for the generation of optical ``cat'' and entangled light states. $\ket{\alpha_{L}}$ and $\ket{\alpha_{1}}$ are the IR states before and after the interaction, respectively, with $|\alpha_{1}|<|\alpha_{L}|$ due to energy conservation. $\ket{0_{q}}$ and $\ket{\chi_{q}}$ is vacuum $\ket{0_{q}}$ and coherent $\ket{\chi_{q}}$ of the harmonics before and after the interaction, respectively. $\ket{{\Tilde{n}}}$ is the excited multimode wavepacket which accounts the correlations between $\ket{\alpha_{1}}$ and $\ket{\chi_{q}}$. $\mathds{1}-\dyad{\Tilde{0}}$ is the conditioning operator applied for the creation of the optical ``cat'' states.}
    \label{fig11:Conditioning_scheme}
\end{figure}

As has been discussed in the previous section and shown schematically in Figs.~\ref{fig7:QEDScheme} and~\ref{fig11:Conditioning_scheme}, the light modes after the interaction are in coherent states. The generation of the optical ``cat'' states conceptually relies on the projection of the outgoing from the medium states on their part that has not been affected by the interaction. For the HHG process this can be formally captured by the projection operator $\mathds{1}-\dyad{\Tilde{0}}$.
Such conditioning operations are possible as a consequence of the light-matter interaction. Since the interaction with the matter system is energy conserving the photon exchange between the different frequency modes induce correlations between the field modes. These correlations can be captured by introducing a set of wave-packet modes $\{\ket{\tilde{n}}\}$ (with $\Tilde{n}=\tilde{0}$ representing the absence of HHG excitations), which inherently correlates the different modes as the excitations are a consequence of the same electron dynamics. It has been found theoretically and experimentally \cite{TKG17, LCP21, RSM22, SRL22, Sta22, Stammer_ArXivConditioning_2024, Stammer_EnergyConservation2024}, that the conditioning procedure is then sufficiently described by the set of positive operator-valued measures (POVM) $\{\mathds{1}-\dyad{\Tilde{0}},\dyad{\Tilde{0}}\}$, describing the case when the HHG process and the corresponding IR depletion have occurred or not, respectively~\cite{Sta22}. We refer to $\mathds{1}-\dyad{\Tilde{0}}$ as the conditioning on HHG. Applying this operator to the state $\ket{\alpha_{1}}\otimes_{q}\ket{\chi_{q}}$ we create a massively entangled state between all field modes,
\begin{equation}\label{Eq:HHG:cond1}
    \ket{\Phi_{\text{IR,HHG}}}
        = \ket{\alpha_1 }
            \bigotimes^{N_{\text{c}}}_{q=2}
                \ket{\chi_q}
            -\xi_1 \ket{\alpha_L} \bigotimes^{N_{\text{c}}}_{q=2}
                \xi_q\ket{0_q},
\end{equation}
where $\xi_{1}=\langle\alpha_{L}|\alpha_{1}\rangle$ and $\xi_{q}=\prod_{q=2}^{N_c} \langle\chi_q|0_q\rangle$. 

\subsection{Optical ``cat'' and entangled states using strongly laser driven atoms}
In this section we summarize our experimental and theoretical findings \cite{LCP21, RLP22, RSM22, Sta22, SRM2023} on how intense laser--atoms interactions and conditioning operations on the HHG and ATI processes can lead to the generation of high mean photon number optical ``cat'' states with controllable quantum features.

\subsubsection{Optical ``cat'' states in the IR spectral region}

\begin{figure}
    \centering
    \includegraphics[width=0.9 \columnwidth]{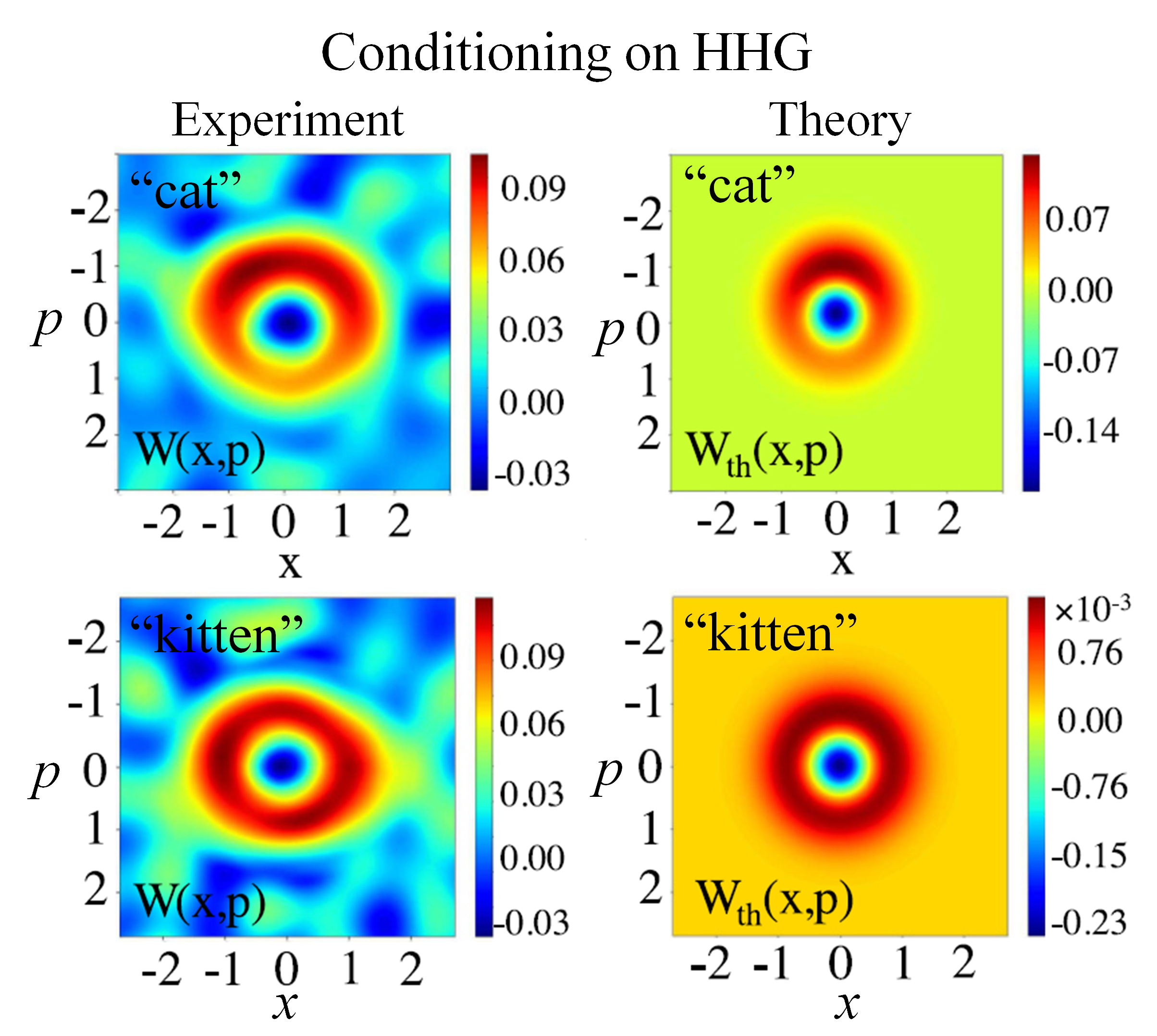}
    \caption{The left panel shows the $W(x,p)$ of the experimentally measured IR ``cat'' (large $|\delta\alpha|$) and ``kitten'' (small $|\delta\alpha|$) states generated by conditioning on the HHG process induced by intense laser--atom interactions. The right panel shows theoretically calculated Wigner functions of the corresponding states. The Figure has been reproduced from ref. \cite{RLP22}}
    \label{fig12:WignerHHG_CAT_Kitten}
\end{figure}

The generation of the optical ``cat'' state in the IR spectral range can be obtained by projecting $\ket{\Phi_{\text{IR,HHG}}}$ shown in Eq.~\eqref{Eq:HHG:cond1} onto the state in which the harmonics are found. In this case Eq.~\eqref{Eq:HHG:cond1} reads, 
\begin{equation}\label{Eq:HHG:cond}
    \ket{\Phi_{\text{IR}}}
        = \ket{\alpha_1 }
            -\xi_1 \ket{\alpha_L},
\end{equation}
and corresponds to a generalized optical ``cat'' state. The results of these measurements (Wigner function) together with the theoretical calculations are shown in Fig.~\ref{fig12:WignerHHG_CAT_Kitten}. We show the ability to control the quantum features of the state by controlling the density of atoms in the interaction region which consequently affects the shift $\delta\alpha$ of the coherent state. The upper panel of Fig.~\ref{fig12:WignerHHG_CAT_Kitten} corresponds to the generation of an optical ``kitten'' state (large $|\delta\alpha|$), while the lower panel shows an optical ``kitten' state resulted by reducing the gas density in the HHG region i.e. (small $|\delta\alpha|$). 
The implementation of the conditioning process leading to Eq.~(\ref{Eq:HHG:cond}), can be achieved by means of IR/XUV photon correlation approaches \cite{TKG17, Sta22, Stammer_ArXivConditioning_2024, Stammer_EnergyConservation2024}.
We note that an IR optical ``cat'' state with controllable quantum features can also be obtained by conditioning on the ATI/HATI processes \cite{RSM22}.   

\subsubsection{Optical ``cat'' states from the far--IR to XUV spectral region.}

\begin{figure}
    \centering
    \includegraphics[width=0.9 \columnwidth]{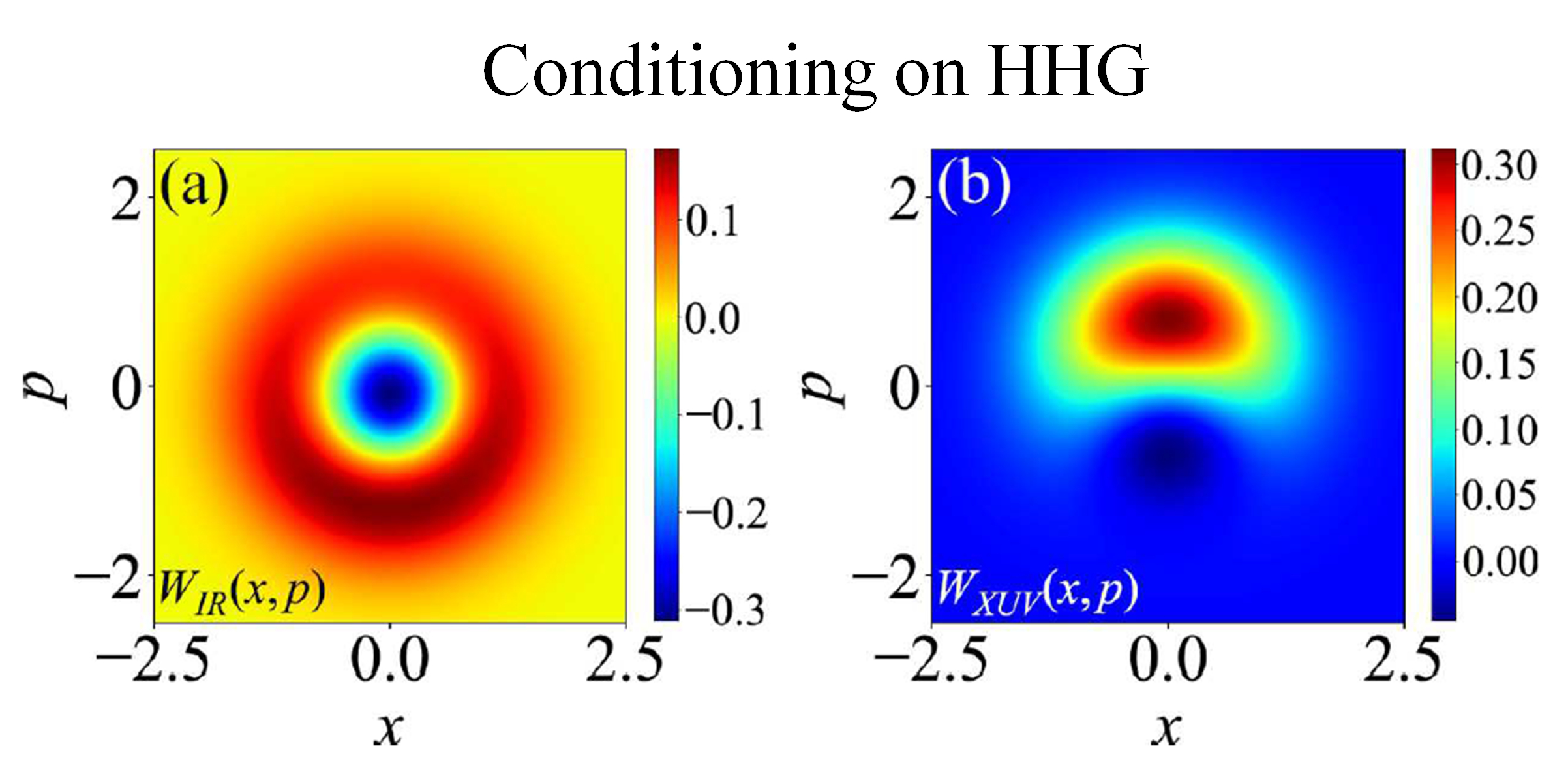}
    \caption{Calculated $W(x,p)$ of an IR (a) and XUV (b) ``cat'' states generated by conditioning on the HHG process induced by intense laser--atom interactions. The Figure has been reproduced from ref. \cite{SRL22}}
    \label{fig14:WignerHHG_CAT_XUV}
\end{figure}

Building upon the multimode formulation of the HHG process, we are able to further extend this approach to generate nonclassical optical states in extreme wavelength regimes \cite{SRL22}. In particular, the HHG process enables the generation of entanglement across various frequency modes of the optical field, spanning from the far-infrared to the XUV region. Specifically, interchanging the role of the fundamental with the harmonics in the field state \eqref{Eq:HHG:cond1} and subsequently measuring the harmonic modes $q'\neq q$, we can obtain a superposition of a coherent state with the vacuum in the XUV spectral regime
\begin{equation}\label{CAT_IRXUVeq:XUV_CAT}
    \ket{\Psi_{q}} = \ket{\chi_{q}} - e^{-\gamma} \ket{0_{q}}
\end{equation}
\noindent where $\gamma = \abs{\delta \alpha_{L}}^{2} + \Omega' + \frac{1}{2}\abs{\chi_{q}}^{2}$ with $\Omega' = \sum_{q' \neq q} \abs{\chi_{q'}}^{2}$. In Fig.~\ref{fig14:WignerHHG_CAT_XUV} we show the Wigner function of the fundamental (a) and the $q$-th mode (b) field where we can observe the deviation from the Gaussian distribution and small negative regions \cite{SRL22}. Furthermore, the opposite shift in imaginary part reflects the correlation between the field modes.

\subsubsection{Entanglement between the field modes}\label{Sec:Entanglement:Atoms}

The generation of high photon number entangled coherent states can be achieved via a two-color driving field \cite{SRL22}. In this scheme, the initial state of the field is $\ket{ \alpha_{ \omega_{1} } } \ket{ \alpha_{ \omega_{2} } }$ and after the HHG process $\ket{ \alpha_{ \omega_{1} } + \delta\alpha_{ \omega_{1} } }\ket{ \alpha_{ \omega_{2} } + \delta\alpha_{ \omega_{2} } }\otimes_{q}^{N}\ket{ \chi_{q} }$. Conditioning on the HHG process leads to the following state,
\begin{equation}\label{QedAtoms_EntanglEq:2color_state_cond}
    \begin{aligned}
        \ket{ \Phi_{HHG}^{(\omega_{1}, \omega_{2})} (t)} = &  \ket{ \alpha_{ \omega_{1} } + \delta\alpha_{ \omega_{1} } }\ket{ \alpha_{ \omega_{2} } + \delta\alpha_{ \omega_{2} } }\\
        & - \xi^{(\omega_{1}, \omega_{2})} \ket{ \alpha_{ \omega_{1} } }\ket{ \alpha_{ \omega_{2} } }
    \end{aligned}
\end{equation}

\noindent where $\delta\alpha_{ \omega_{1,2} }$ is the depletion of the corresponding driving mode. The factor $\xi^{(\omega_{1}, \omega_{2})}$ is a complex number which depends on the driving field amplitudes and the amount of their depletion. 

\par
\begin{figure}
    \centering
    \includegraphics[width=0.8 \columnwidth]{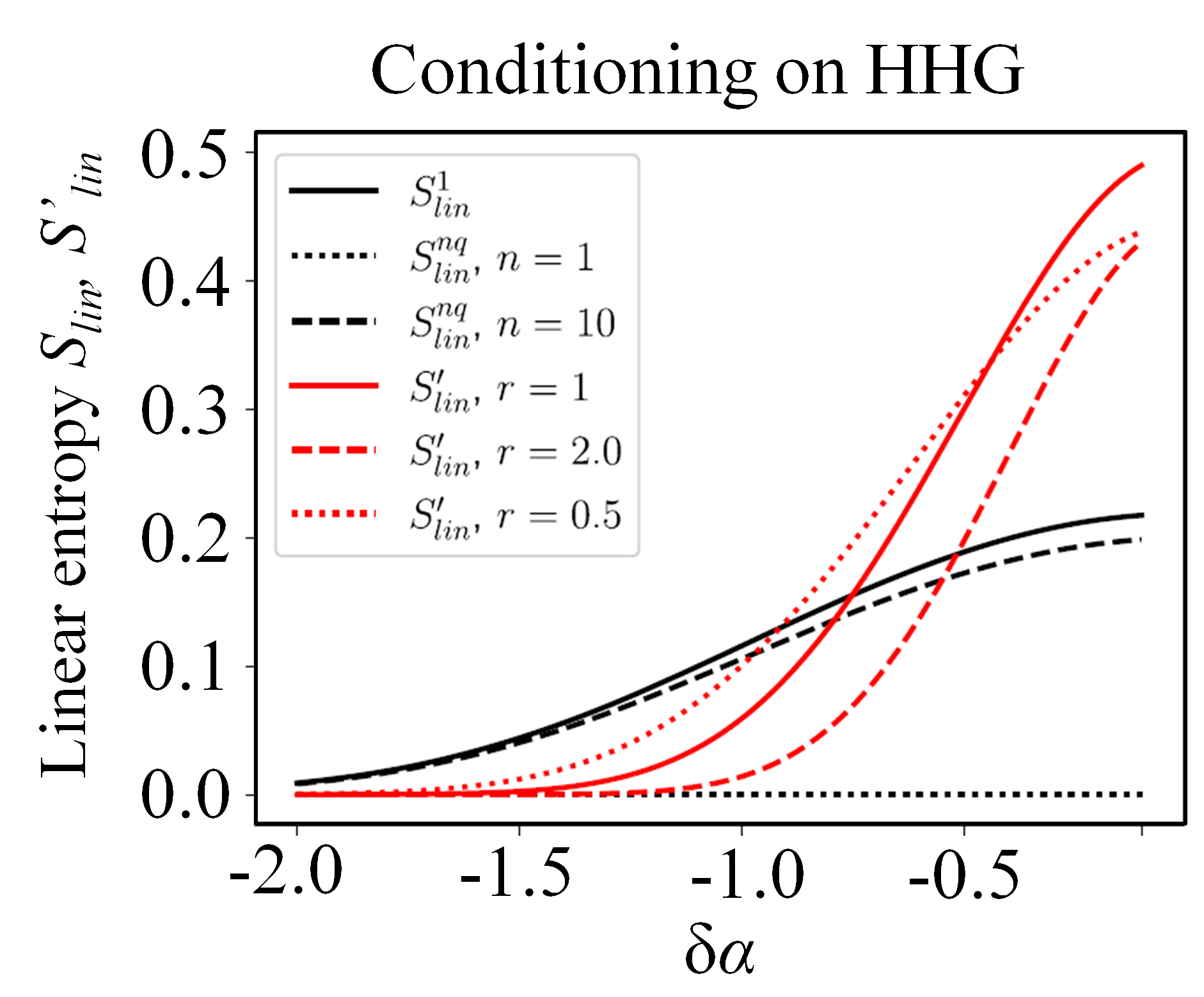}
    \caption{Entanglement classification between the field modes after conditioning on HHG induced by intense laser--atom interaction using a single color IR driving field of frequency $\omega_{L}$, and two--color driving fields of frequency $\omega_{L}+\omega_{2}$ with $\omega_{2}=2\omega_{L}$. (black lines) Linear entropy $S_{lin}$ between the modes for single--color driving field. $\delta\alpha$ corresponds to the shift of the fundamental $\omega_{L}$ driving field. The $S^1_{lin}$ denoted with black solid line shows the entanglement between the fundamental and all harmonic modes. The $S^{nq}_{lin}$ denoted with black dot and dashed lines shows the entanglement between $n$ harmonic modes and all other modes (including the fundamental).  (red lines) Linear entropy $S'_{lin}$ between the two modes of two--color driving field for different values of $r$. $r=|\delta\alpha_{2}|/|\delta\alpha|$ is the ratio of the shifts between the two frequency modes of the two-color driving field. The Figure has been reproduced from ref. \cite{SRL22}}
    \label{fig15:Entanglement}
\end{figure}

 
In Fig.~\ref{fig15:Entanglement} we show the degree of entanglement between the field modes in the single- and two-color HHG schemes \cite{SRL22} using the liner entropy $S_{\text{lin}} = 1 - \Tr(\rho^{2})$. In the single-color HHG scheme (see Eq.~\eqref{Eq:HHG:cond1}), we examine the entanglement between the fundamental driving field with all harmonic modes ($S^{1}_{\text{lin}}$) and the entanglement of $n$ harmonic modes with all remaining modes ($S^{nq}_{\text{lin}}$). For high harmonics generated by a two-color driving field, Fig.~\ref{fig15:Entanglement} shows the entanglement between the modes of the driving field.

\subsection{Optical ``cat''--like and entangled states using strongly laser driven molecules} \label{CatMolecules}
In this section we discuss our recent theoretical work~\cite{rivera-dean_quantum_2023} on the generation of non--classical and entangled light states using strongly laser driven H$_2^+$ molecules. Using the HHG process, we discuss the possibility of creating non--classical states in specific harmonics. We also discuss how the quantum features these states depend on the electronic states of the molecule, an issue which connects the electronic state dynamics with the quantum state of the light field after the interaction.

\subsubsection{Non-classical states in the IR spectral region}

The more complex dynamics in molecular systems allows for the generation of non-classical states of light without the need for post-selection operators, as is the case of atomic systems. In fact, it is the final electronic state that enables the heralding of non-classical states of light in the different field modes. While events where electrons end up in the bonding state $\ket{\psi_b}$ lead to a product of coherent states, as described in Eq.~\eqref{Eq:QO:bonding:Nmol}, events where one of the electrons ends up in the anti-bonding state result in Eq.~\eqref{Eq:QO:antibonding:Nmol}. Under appropriate conditions, this state can exhibit non-classical features if the operator $\hat{\bar{M}}_{ab}(t)$ is linear with respect to the creation and annihilation operators acting on the field modes. This results in a frequency-entangled state which, in the displaced frame with respect to $\hat{\vb{D}}(\bar{\boldsymbol{\chi}}(t,t_0))$, is given as a quantum superposition where a single quantum optical excitation takes place along one of all possible field modes generated during the HHG process. Specifically, when focusing on the quantum optical state of a specific harmonic mode $q$, and tracing out the others, we obtain 
\begin{equation}\label{Eq:QO:q:antibonding:Molecules}
    \hat{\rho}^{(q)}_{a}
        = \tr_{q'\neq q}
            \big( \dyad{\bar{\Phi}_{a}(t)} \big).
\end{equation}

\begin{figure}
    \centering
    \includegraphics[width=1 \columnwidth]{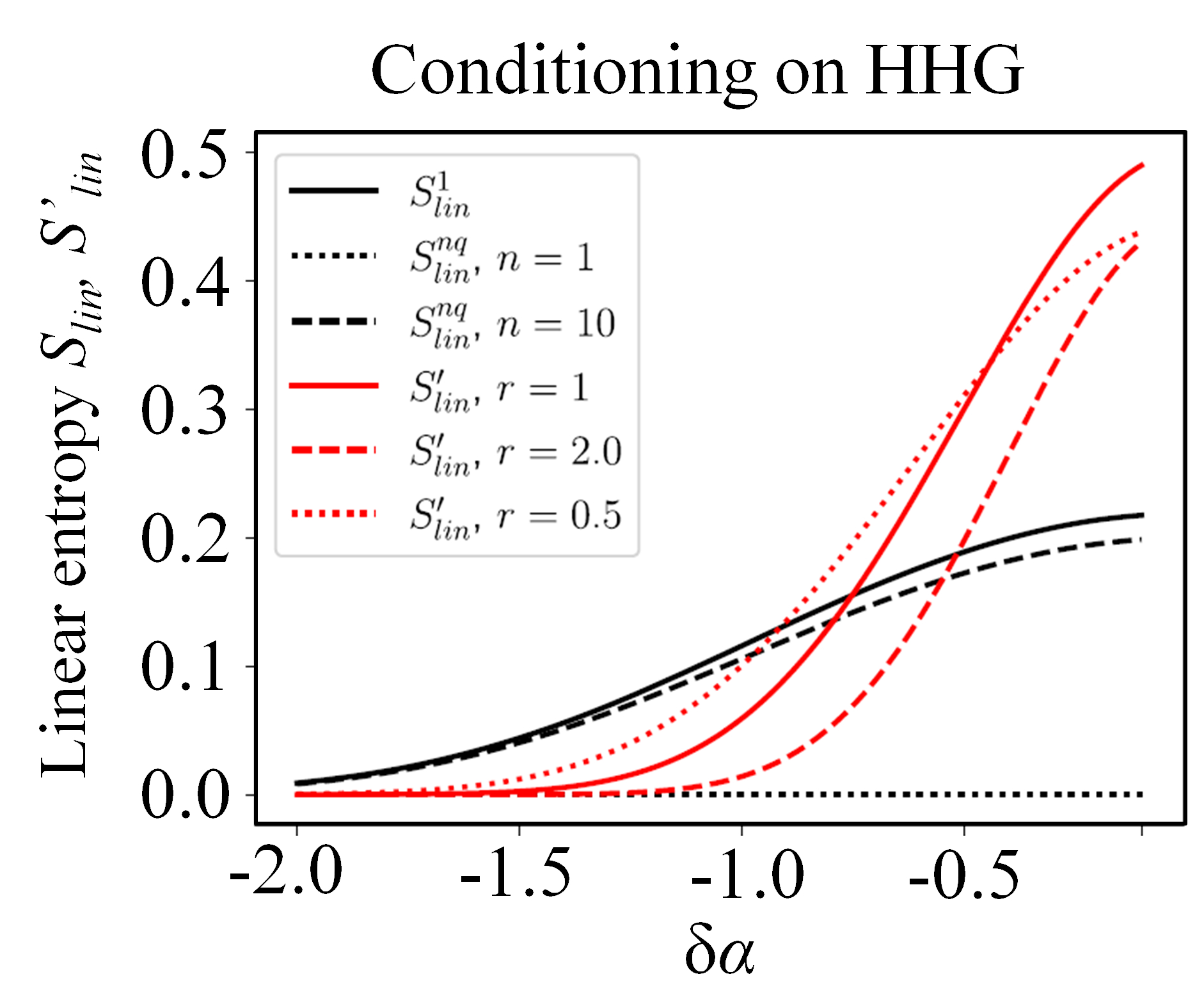}
    \caption{Wigner function $W(x,p)$ for different harmonic modes when considering events where the electron has ended up in the antibonding state. This Figure has been adapted from Ref.~\cite{rivera-dean_quantum_2023}.}
    \label{fig16:WignerCAT_H2}
\end{figure}

In Fig.~\ref{fig16:WignerCAT_H2}, we present the Wigner function of Eq.~\eqref{Eq:QO:q:antibonding:Molecules} for two different harmonic modes (corresponding to the different rows) and for various bond lengths (corresponding to the different columns). These conditions produce Wigner functions with varying shapes and values, which are not necessarily restricted to positive ones. Specifically, for $R=2.0$ a.u. and $R=3.5$ a.u. bond lengths, we observe negativities in the Wigner function for harmonics $q=4$ and $q=6$, respectively, while Gaussian Wigner functions appear for the opposite modes. For the $R = 2.5$ a.u., no negativities are found, though the Wigner functions retain a volcano-like shape, which suggests the presence of significant entanglement between the different field modes (see Sec.~\ref{Sec:Wigner analysis}). Notably, negative Wigner functions are only found for even harmonic orders, which are those generated in HHG events where the electron begins in the bonding state and ends up in the anti-bonding one, i.e., states of opposite parity.

\subsubsection{Entanglement between the field modes.}

The structure of the state in Eq.~\eqref{Eq:QO:Final:Molecules:Nmol} is quite rich in terms of entanglement. On the one hand, the observation of different quantum optical states depending on the final electronic state suggests the presence of light-matter entangled states. On the other hand, the delocalized harmonic excitations in Eq.~\eqref{Eq:QO:antibonding:Nmol} indicate potential entanglement between different light modes. Since this text focuses on entanglement features between different frequency modes, we concentrate on the latter and refer to Ref.~\cite{rivera-dean_quantum_2023} for a more comprehensive analysis.

When assuming that we, as observers, have no knowledge of which state has the electron recombined with, the final quantum optical state is given by
\begin{equation}
    \begin{aligned}
    \hat{\rho}_f(t)
        &= \tr_{e}\big(\dyad{\Psi(t)})
        \\& 
        = \dfrac{1}{\mathcal{N}}
            \bigg[
                \dyad{\bar{\Phi}_b(t)}
                + N_{\text{mol}} \dyad{\bar{\Phi}_a(t)}
            \bigg],
    \end{aligned}
\end{equation}
that is, as a mixed state with contributions arising from both the bonding and anti-bonding quantum optical components.~As such, to characterize the entanglement features we utilize the logarithmic negativity shown in Eq.~\eqref{Eq:Logarithmic:Negativity}. More specifically, we focus on the entanglement between one field mode $q$ and the rest. However, due to the multidimensional nature of our state, which leads to computationally demanding calculations, we rely on the following lower bound
\begin{equation}\label{Eq:Log:Neg:Molecules}
    E_N^{(q)}\big(\hat{\rho}_f(t)\big)
        = \log_2\Big(2\abs{\min_i \lambda_i}+1\Big),
\end{equation}
where $\lambda_i$ corresponds to the lowest eigenstate of the partial transpose of $\dyad{\bar{\Phi}_a(t)}$ with respect to one of the subsystems.

\begin{figure}
    \centering
    \includegraphics[width=1 \columnwidth]{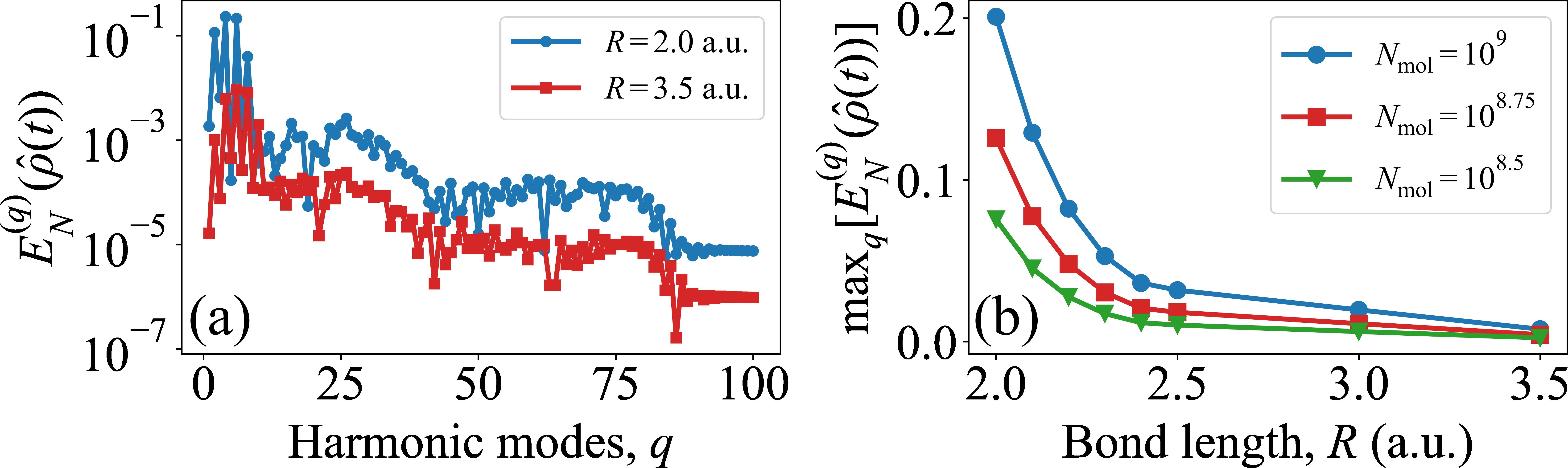}
    \caption{Entanglement features between the harmonic mode $q$ and the rest after driving H$_2^+$ with a strong laser field. In (a), a lower bound to the logarithmic negativity is shown as a function of the harmonic order. The different curves correspond to various bond lengths. In (b), the maximum amount of entanglement is shown as a function of bond length, with the different curves representing different numbers of molecules.~This figure has been adapted from Ref.~\cite{rivera-dean_quantum_2023}.}
    \label{fig17:Entanglement_H2}
\end{figure}

Figure~\ref{fig17:Entanglement_H2}~(a) displays the results from the evaluation of Eq.~\eqref{Eq:Log:Neg:Molecules} for different field modes and various bond lengths. As observed, the entanglement structure exhibits a pattern similar to that of molecular HHG spectra, where we identify the presence of different plateau structures, followed by a well-defined cutoff. For $q \leq 15$, peaks are observed for even harmonic orders which, as mentioned earlier, arise from the bonding-anti-bonding transitions. This underscores the necessity of the anti-bonding component to observe non-vanishing spectral features. As the bond length increases, such events become less likely, resulting in a reduction in the overall amount of entanglement, as can seen by comparing the blue and red curves. This trend is further highlighted in Fig.~\ref{fig17:Entanglement_H2}~(b), which shows the maximum amount of entanglement as a function of bond length for different values of $N_{\text{mol}}$.

\subsection{Optical ``cat'' and entangled states using strongly laser driven semiconductor solids} \label{CatSemiconductors}

The generation of optical Schrödinger ``cat'' states and entangled states using strongly laser-driven semiconductor solids marks a significant breakthrough in quantum optics and quantum information science. This has been theoretically demonstrated in QED of strongly laser--driven semiconductor solids~\cite{rivera-dean_bloch_2023}. The backaction of the high-order harmonic generation (HHG) process on the driving laser field leads to the production of nonclassical states of light. Additionally, semiconductor-driven HHG can produce entangled states by correlating harmonic modes with the input field. This entanglement can be engineered across multiple frequency modes, leading to highly entangled quantum states. These phenomena leverage the unique electronic band structure and coherence properties of semiconductors, making them ideal for the generation and manipulation of quantum states.

\subsubsection{Optical ``cat'' states in the mid-IR spectral region}

Overall, and within the considered approximations, the quantum optical state after HHG events in semiconductor materials (see Eq.~\eqref{Eq:QO:state:HHG:solids}) shares the form as that obtained in atomic systems. The key difference lies in the HHG amplitudes: for semiconductor materials, these amplitudes are influenced by both interband and intraband dynamics that the electron undergoes in these systems. As a result, the same conditioning procedure used in atomic systems (see Sec.~\ref{Methods6}) can also be applied in the semiconductor scenario~\cite{TKD19}, producing the quantum superposition described in Eq.~\eqref{Eq:HHG:cond}.

\begin{figure}
    \centering
    \includegraphics[width=1 \columnwidth]{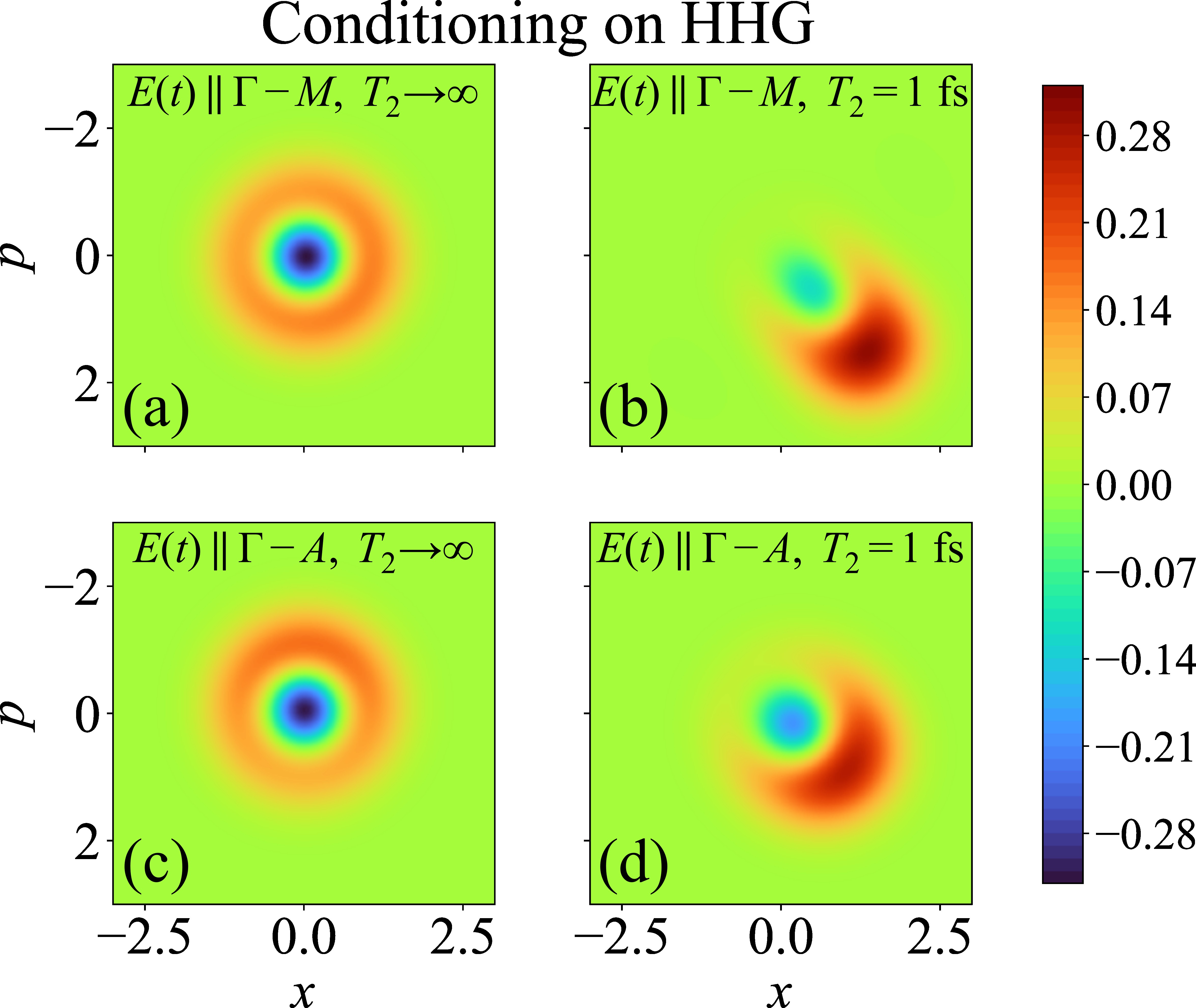}
    \caption{Calculated $W(x,p)$ of the mid--IR ($3.25$ $\mu$m) ``cat'' states generated by conditioning on HHG produced by, both inter-- and intraband transitions, in strongly laser driven semiconductors.~The laser pulse duration is 96 fs and the field strength in the interaction region is $\approx 0.5 $ V/\AA.  In (a) and (b) the Wigner function has been calculated at different dephasing times $T_2$ when the laser polarization is parallel to $\Gamma-M$ orientation of the crystal. Panels (c) and (d) shows the $W(x,p)$ at the same $T_2$ as in (a) and (b), when the laser polarization is parallel to $\Gamma-A$ crystal orientation. The Figure has been adapted from Ref.~\cite{rivera-dean_bloch_2023}.}
    \label{fig18:CATSolids}
\end{figure}

Similar to the atomic case, Fig.~\ref{fig18:CATSolids} displays the Wigner function of the solid-state-HHG-conditioned state under various conditions.~As noted earlier, the HHG outcomes are heavily influenced by the excitation conditions and the solid-state system characteristics, such as the dephasing time. Here, we illustrate how these factors shape the properties of the resulting Wigner function. Specifically, we examine cases where the laser field is polarized along different crystal directions: (a) and (b) along the $\Gamma-M$ direction, and (c) and (d) along the $\Gamma-A$ direction. Two dephasing times are considered, $T_2=\infty$ and $T_2=1$ fs. In all cases, non-classical features manifest as Wigner function negativities, significantly influenced by the considered parameters. From the comparison between (a) and (c) with (b) and (d), two key conclusions arise. First, shorter values of $T_2$ result in more distinct and imbalanced coherent state superpositions, with reduced Wigner function negativities due to the increased magnitude of $|\bar{\chi}^{(q=1)}_{v}(t,t_0)|$. Second, the displacement varies with crystal direction, with excitations along the $\Gamma-M$ direction producing larger displacements for the same $N_z$.

In this analysis, distinct values for the numbers of Brillouin zones are considered for the different crystal orientations. In particular, for the $\Gamma-M$ direction ((a) and (b)), we used $N_z=6.6\times10^6$, while for the $\Gamma-A$ direction ((c) and (d)), we used $N_z=1.6\times10^{7}$. Larger values of $N_z$ lead to more spread and imbalanced coherent state superpositions, resulting in Gaussian-like Wigner functions. In contrast, smaller $N_z$ values yield distributions similar to those of displaced single-photon excitations. These parameters also highlight a key distinction between the two crystal directions: for the same number of Brillouin zones, excitations along the $\Gamma-M$ direction produce more spread and imbalanced superpositions compared to those along the $\Gamma-A$ direction. This behavior arises from the differences in band structures, where electron mobility in the valence band of the $\Gamma-A$ direction does not involve the same energy exchanges as in the $\Gamma-M$ direction. As a result, fewer electrons (and thus a smaller $N_z$) are needed to achieve the same displacement along the $\Gamma-M$ direction as along the $\Gamma-A$ direction.

\subsubsection{Entanglement between the field modes.}

As discussed in Sec.~\ref{Methods6}, the introduction of the conditioning operation leads to a massively entangled state between the field modes. Here, we discuss how the entanglement properties of the state get affected by the electron dynamics of the solid sample. For that purpose, similar to Sec.~\ref{Sec:Entanglement:Atoms}, we use the linear entropy as entanglement measure (see Eq.~\eqref{Eq:Lin:Entropy}), which is particularly useful in this case provided that we are dealing with coherent states.

\begin{figure}
    \centering
    \includegraphics[width=1.0 \columnwidth]{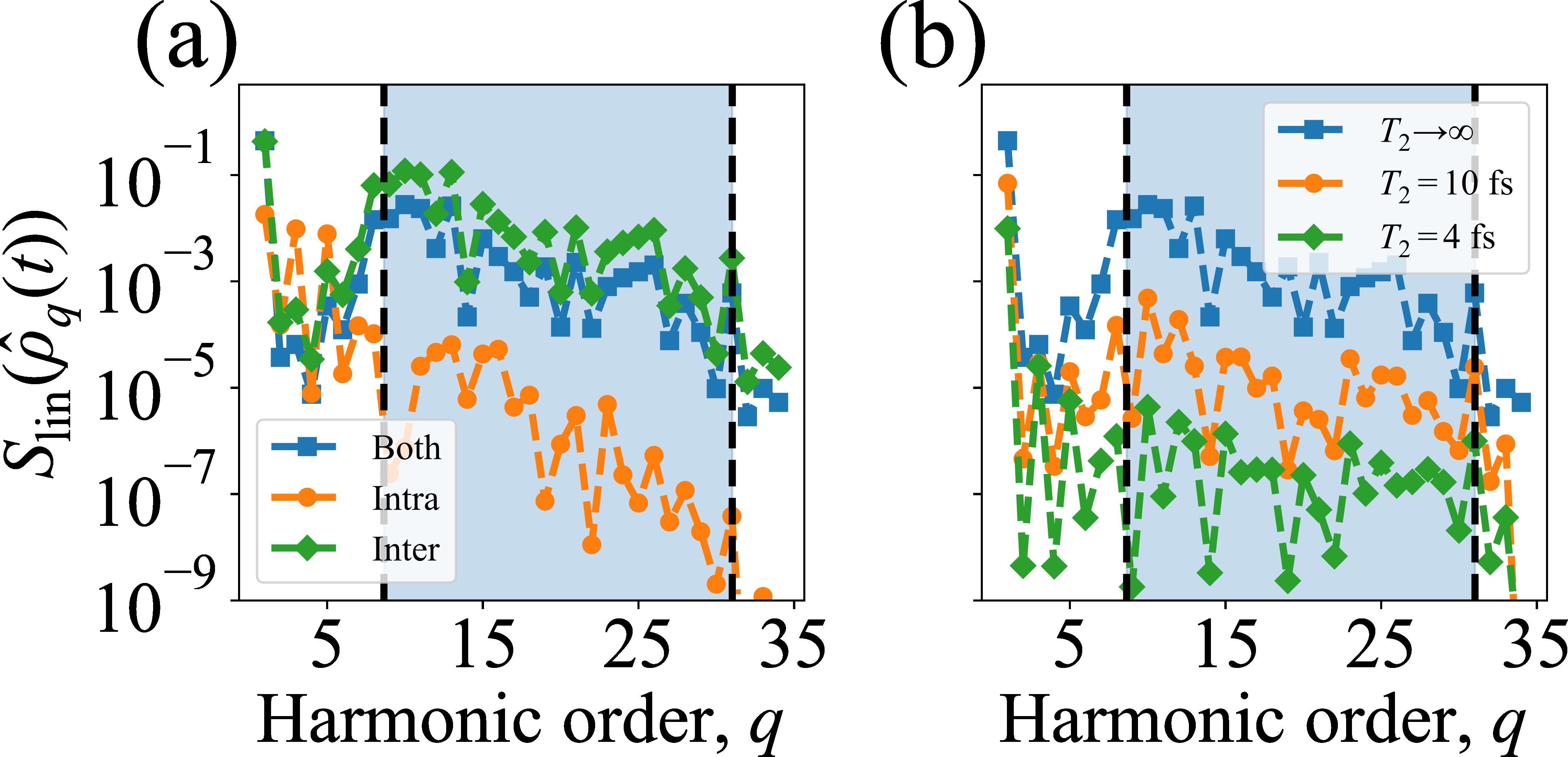}
    \caption{Entanglement features of the quantum optical state obtained when introducing conditioning operations on strong-field-driven ZnO along the $\Gamma-M$ crystal orientation.~Panel (a) depicts how the dependence of the entanglement on the distinct strong-field driven dynamics.~Panel (b) shows the dependence with the dephasing time.~The dashed blue region highlights the non-perturbative part of the HHG spectrum, comprised in this case between the lowerst and the maximum energy bandgap of ZnO. The Figure has been adapted from Ref.~\cite{rivera-dean_bloch_2023}.}
    \label{fig19:Entanglement_Solids}
\end{figure}

In Fig.~\ref{fig19:Entanglement_Solids}, we present the amount of entanglement as a function of the harmonic order. In panel~(a), we display how the different dynamics contribute to the amount of entanglement, while in panel~(b) we present the influence of the dephasing time. Overall, the entanglement presents a structure alike to typical HHG spectra, with a plateau region followed by a sharp cutoff, in this case located at the maximum energy bandgap (black dashed arrow located at the right of both panels). In fact, when looking at panel~(a), we see that the dominant contribution to the entanglement comes from the interband contribution (green curve with diamond markers), especially in the non-perturbative high-harmonic regime, which shows a similar behavior to that shown by the joint contribution of both dynamics (blue curve with squared markers). In contrast, the intraband contribution decays for increasing harmonic orders. On the other hand, from panel (b) we observe that decreasing dephasing times lead to reduced entanglement contributions.

\section{Generation of squeezed and entangled light states using strongly laser driven ground state depleted atoms} \label{SqueezedDepletedAtoms}

Recently, P. Stammer {\it et al.} have theoretically demonstrated \cite{Stammer_PRL_2024} that the process of HHG from atoms driven by strong laser fields can produce intense squeezed light. It was shown that when the driving field is intense enough to significantly deplete the atomic ground state, resulting in dipole moment correlations, the quantum states of both the driving field and the generated high harmonics become entangled and exhibit squeezing. We have further shown that the angle of the quadrature squeezing of the fundamental driving mode can be tuned by varying the carrier-envelope phase (CEP) of the driving laser field, especially for short pulses (2 optical cycles).

Note that the approximation to obtain Eq.~\eqref{QED_ATOMSeq:Displ_OP_no_dipole_correlations} is equivalent to neglecting dipole moment correlations of the electron \cite{sundaram_high-order_1990, SRL22, Sta22,Stammer_PRL_2024} such that the final field state is given by a product of coherent states. To account for the effects of ground state depletion, we solve the dynamics for higher orders of the exact interaction Hamiltonian $H_{I}(t)=-d(t) E_{Q}(t)$ instead of the approximate Hamiltonian $H'_{I}(t)=-\expval{d(t)} E_{Q}(t)$. Going beyond the linear term of the field operator $E_{Q}(t)$ will lead to squeezing in the field modes. Additionally, all field modes will become entangled due to the mixing of the field operators $a_{q}^{(\dagger)}$ from different modes. Therefore, we expect to observe signatures of squeezing and entanglement between the modes when dipole moment correlations are no longer neglected.
\par Solving the TDSE introducing the interaction Hamiltonian $H_{I}(t)$, the evolution of the initial light state reads (more information can be found in ~\cite{Stammer_PRL_2024})
 
\begin{equation}\label{GEN_SQ_ENTeq:Evol_init_state}
    \ket{\Phi(t)} = \textbf{D}[\chi]e^{-\frac{1}{2}\expval{Q^{2}(t)}}\ket{\{0_{q}\}}
\end{equation}

\noindent where $Q(t) = \int_{0}^{t} \dd{t'} E_{Q}(t') [d(t') - \expval{d(t')}]$. The additional contribution is quadratic in $Q(t)$ and thus in the electric field operator and therefore leads to squeezing and mixing between the filed modes. Focusing on the fundamental mode of the driving field, we find \cite{Stammer_PRL_2024} that $\expval{Q^{2}} = 2 \textit{A} P_{\psi}^{2}$, with $P_{\psi} = \frac{i}{\sqrt{2}}(a^{\dagger}e^{i\psi} - a e^{ - i\psi})$ denoting the momentum quadrature. The exact expression for the constant $\textit{A}$ can be found in \cite{Stammer_PRL_2024}. Finally, the field state of the fundamental mode after the process of HHG is given by (in the laboratory frame)

\begin{equation}\label{GEN_SQ_ENTeq:Fund_after_HHG}
    \ket{\Phi_{1}} = D[\alpha] D[\chi_{1}] S(\psi) \ket{0}
\end{equation}

\noindent which is a high photon number squeezed coherent state, with the squeezing operator given by

\begin{equation}\label{GEN_SQ_ENTeq:Sq_operator}
    S(\psi) = \exp[-\textit{A} P^{2}_{\psi}]
\end{equation}

\noindent Variation of the squeezing phase $\psi$, allows us to consider squeezing along different directions in phase space.
\par
\begin{figure}
    \centering
    \includegraphics[width=0.8 \columnwidth]{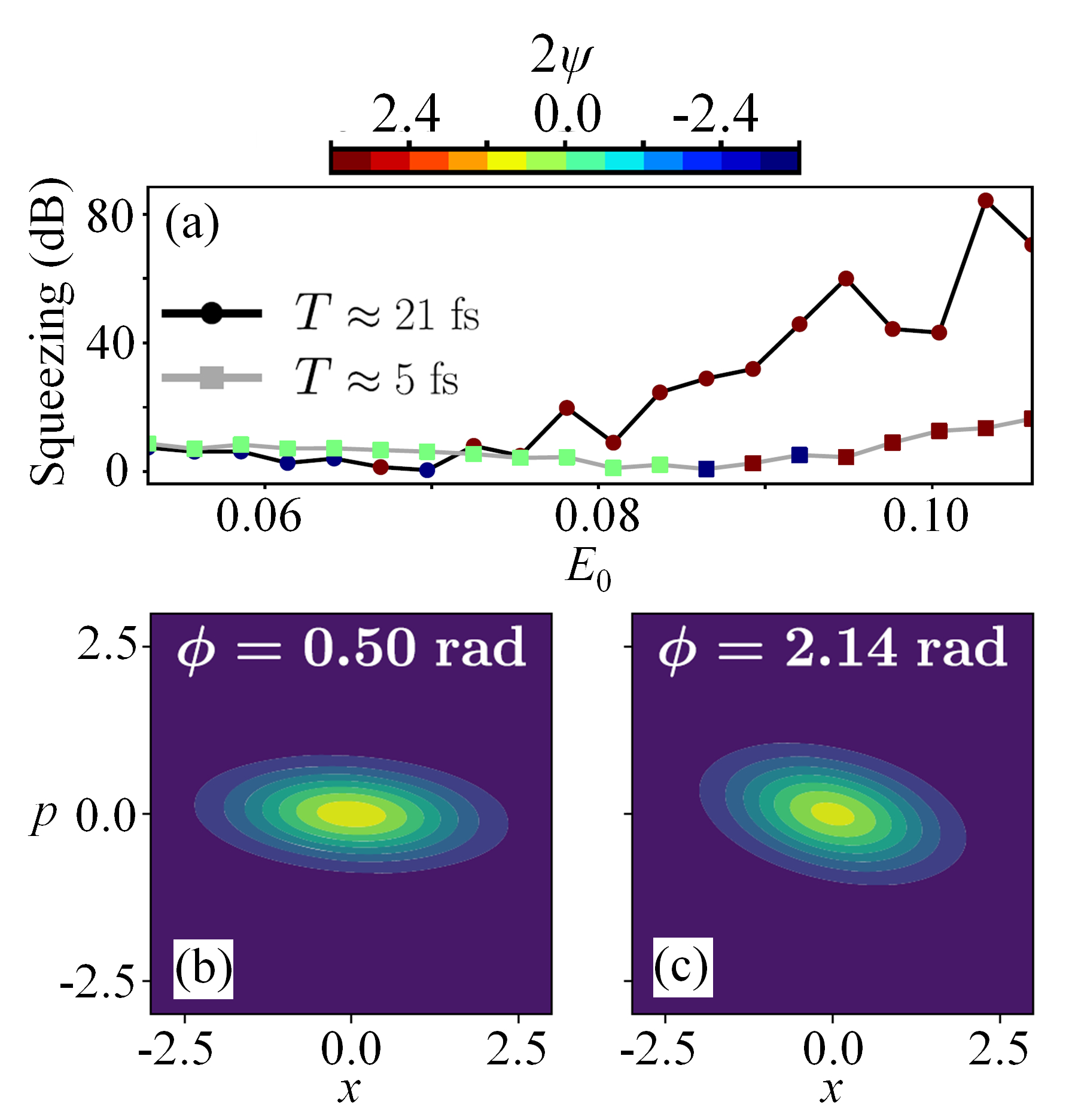}
    \caption{Generation of squeezing IR (800 nm) light states in intense laser--atom interactions. (a) Calculated squeezing parameter $r$ (with dB=$10\log_{10}\exp{2|r|^{2}}$) as a function of the electric field strength $E_{0}$ (in atomic units) of the driving IR field for IR pulse duration 21 fs and 5 fs. (b) and (c) Calculated $W(x,p)$ for two different CEP values $\phi$. The calculations have been conducted for H atoms with a number of atoms in the interaction region $N_{at}=5\times10^{13}$. $\psi$ is the squeezing phase. The Figure has been reproduced from ref. ~\cite{Stammer_PRL_2024}}
    \label{fig20:Squeezed_gas}
\end{figure}

 
Taking into account that in the interaction region we have $N_{at}$ atoms independently contributing to the HHG process we can write the total squeezing power via the squeezing parameter $r \equiv - \text{A} N_{at}$ in units of $[\text{dB}] = 10 \log_{10}(e^{2\abs{r}^{2}})$. In Fig.~\ref{fig20:Squeezed_gas}(a) we show the squeezing of the fundamental mode for increasing field strength for two different pulse durations of the driving laser \cite{Stammer_PRL_2024}. Also, in Fig.~\ref{fig20:Squeezed_gas} (b),(c) we show the rotation of the Wigner function induced by variation of the CEP ($\phi$) highlighting the relation between the CEP and the squeezing phase $\psi$.

\section{Generation of non-classical and entangled light states using excited atoms} \label{Methods8}

Recent theoretical works conducted by S. Yi {\it et al.,} \cite{ Misha_ArXiv_2024} and J. Rivera--Dean {\it et al.,} \cite{Javier_ExcitedAtoms_2024} have shown that the high-harmonic generation (HHG) process, induced by the interaction of intense  IR laser fields with excited media, can be used for generating non-classical and entangled light states. 

\begin{figure}
    \centering
    \includegraphics[width=0.8 \columnwidth]{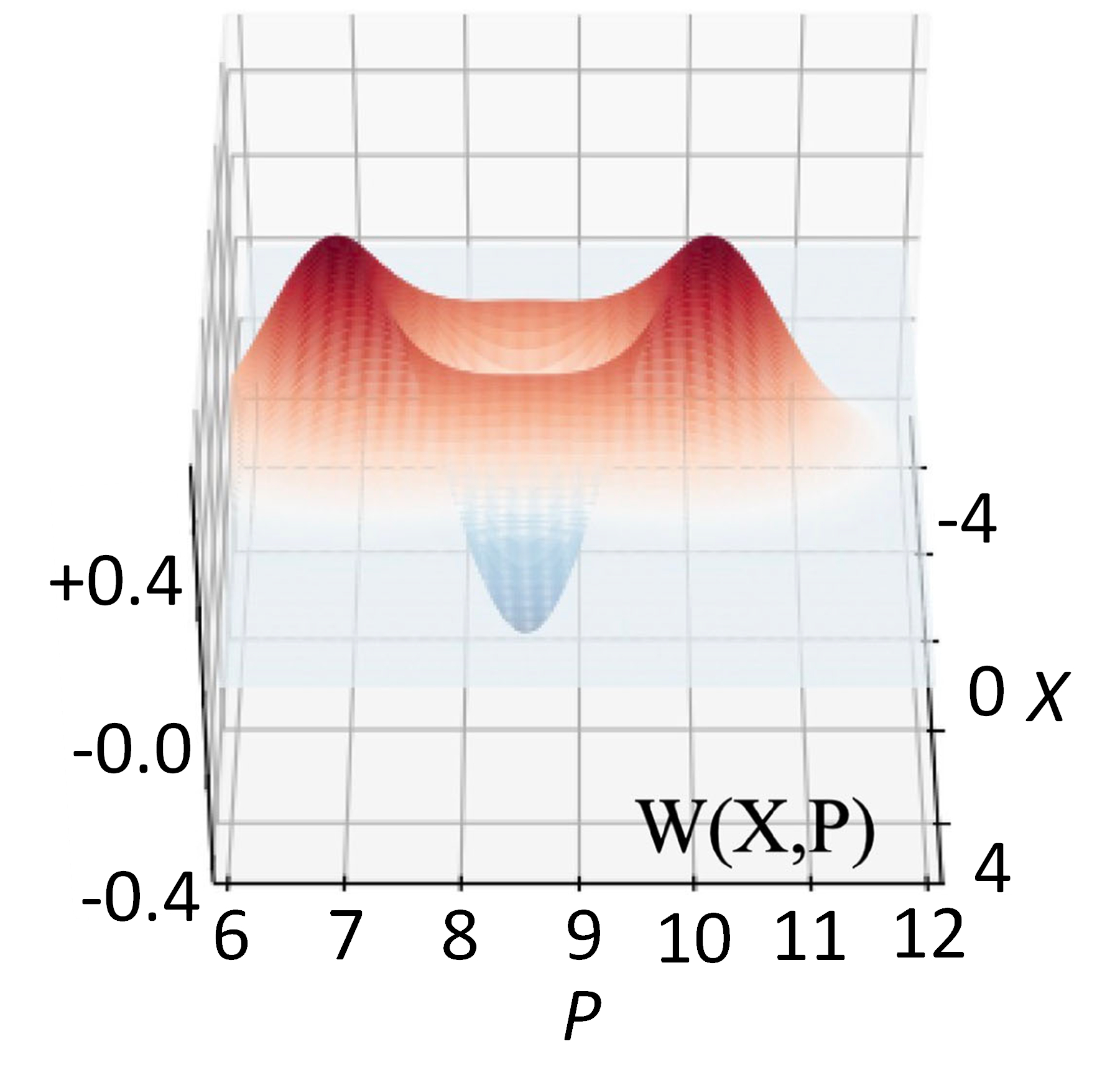}
    \caption{Calculated Wigner functions of the 3rd harmonic generated by resonant media. The Figure has been reproduced from ref. ~\cite{Misha_ArXiv_2024}.}
    \label{fig20:NonClassical_Harmonics_ResonantMedia}
\end{figure}

In the work of S. Yi {\it et al.,} \cite{Misha_ArXiv_2024}, this can be achieved when building quantum correlations between the generated harmonics and the material used to generate them. To develop these correlations the system should undergo a non-adiabatic excitation between laser dressed states and the different laser dressed states should exhibit different non-linear responses. Under these excitation conditions, the emitted harmonics are entangled and correlated to the excited state of the material system. Nonlinear media that are nearly resonant with at least one of the harmonics seem to be promising for the controlled generation of highly entangled quantum light states. \PT{Fig.~\ref{fig20:NonClassical_Harmonics_ResonantMedia} shows an example of the theoretically calculated Wigner function of the 3rd harmonic generated under such conditions.} 

\begin{figure}
    \centering
    \includegraphics[width=0.8 \columnwidth]{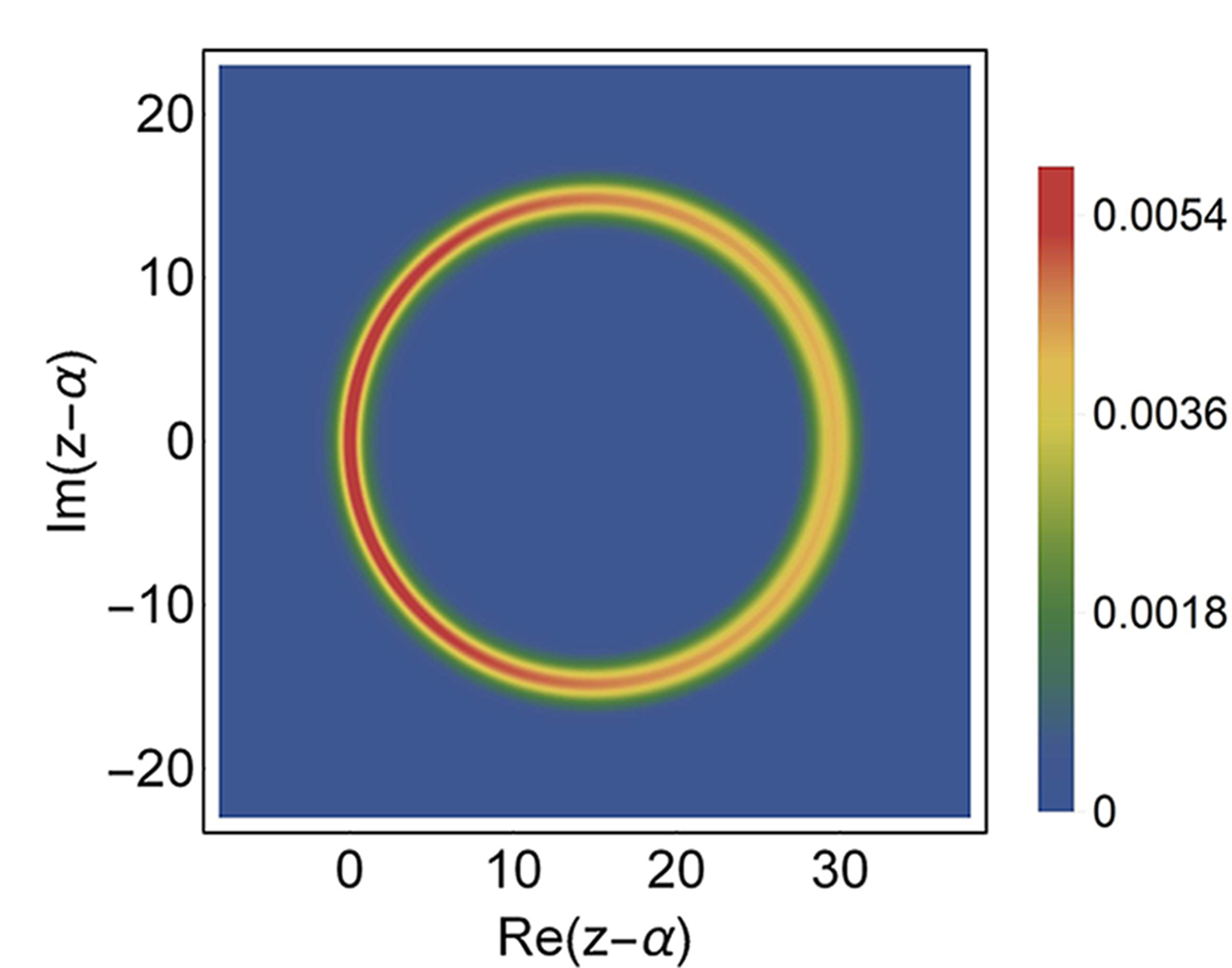}
    \caption{Calculated Wigner function of the IR laser field after the interaction with high-density free electrons generated through the ionization of a target gas. The Figure has been reproduced from ref. ~\cite{Andrianov_2024}. However, this Wigner function does not exhibit quantum signatures.}
    \label{fig21:NonClassical_HighDensityGas}
\end{figure}

The work of J. Rivera--Dean {\it et al.,} \cite{Javier_ExcitedAtoms_2024}, theoretically investigates the generation of non--classical light states using the HHG induced by the interaction of intense IR laser field with atoms excited in first excited state. It has been found that such HHG conditions lead to the generation squeezed light states in both the driving IR field as well as low-order harmonic modes. Finally, in a recent theoretical study ~\cite{Andrianov_2024}, it has been shown how the interaction between a strong laser field and the high-density free electrons generated through the ionization of a target gas by the field, influences the quantum state of the field. (Fig.~\ref{fig21:NonClassical_HighDensityGas}). However, we emphasize that the deviation from the Gaussian Wigner function does not imply the presence of quantum signatures in the field modes due to the lack of negativities.

\section{Generation of non-classical and entangled light states using many body systems} \label{Methods9}

Recent theoretical works have shown that the HHG process induced by strongly laser-driven quantum correlated many body systems, can also serve as a resource for generating non-classical and entangled light states, or light states that significantly deviate from the Gaussian properties of a pure coherent state. An example is the theoretical work of Pizzi \emph{et al.}~\cite{PGR23}, where it has been shown that the HHG process induced by the interaction of intense laser fields with correlated atoms leads to the emission of radiation with non-Gaussian features that deviate from coherent states. Additional works relevant to this issue can be found in the review article of Ref. \cite{Tzallas_ROPP_2024}. An additional, example in this category is the very recent theoretical work of C. S. Lange {\it et al.} \cite{Lange_PRAe-eCorr_2024}. In this study the authors, using the Fermi-Hubbard model, have theoretically demonstrated the generation of non--classical light when the matter system was in the Mott-insulating phase, where the coupling between different sites in the chain becomes non--vanishing. They also highlight the importance of accounting for electron-electron interactions for predicting quantum properties of HHG radiation. 

\begin{figure}
    \centering
    \includegraphics[width=0.9 \columnwidth]{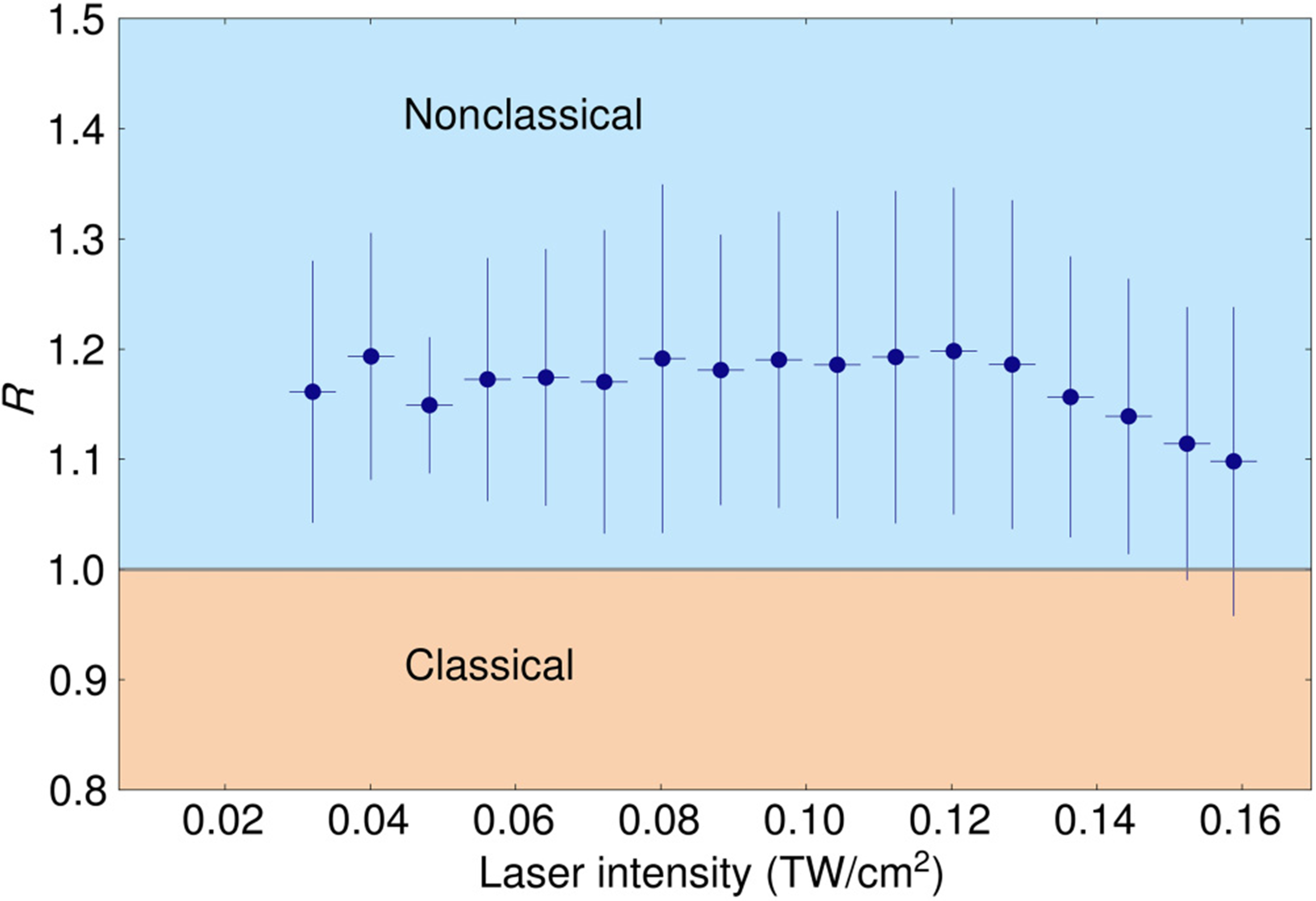}
    \caption{Dependence $R$ on the intensity of the IR pulse used to generate the harmonics in an GaAs crystal. A value of $R > 1$ corresponds to the violation of the classical limit (shown in gray line) given by the Cauchy-Schwarz inequality. The Figure has been reproduced from Ref. ~\cite{Hamed_PRXQ_2024}.}
    \label{fig22:Hamed}
\end{figure}

Furthermore, recent experiments in solids \cite{Hamed_PRXQ_2024, Merdji_multimodeSqueezed_Arxiv2024} have indicated the presence of two-mode squeezing features between low-order harmonics \PT{and reported a violation of Cauchy-Schwarz inequality in the measured intensity correlation functions $g^{(2)}$ (Fig.~\ref{fig22:Hamed}) showing the presence of entanglement between the two modes. The authors have obtained these results by measuring the intensity correlation of the 3rd and 5th harmonic generated by the interaction of an IR pulsed laser with GaAs crystal. The violation of Cauchy-Schwarz inequality was quantified by the parameter $R \equiv[g^{(2)}_{ij}]^2/(g^{(2)}_{ii} g^{(2)}_{jj})$ with $i,j=3,5$}. The results seem to be a consequence of Bloch oscillations of electrons within the conduction band of the solid as has been demonstrated in a recent theoretical work of I. A. Gonoskov {\it et al.}~\cite{Gonoskov_PRB_2024}, conducted using fully quantized approaches. \\

\section{Perspectives}\label{Applications}

\begin{figure}
    \centering
    \includegraphics[width=0.9 \columnwidth]{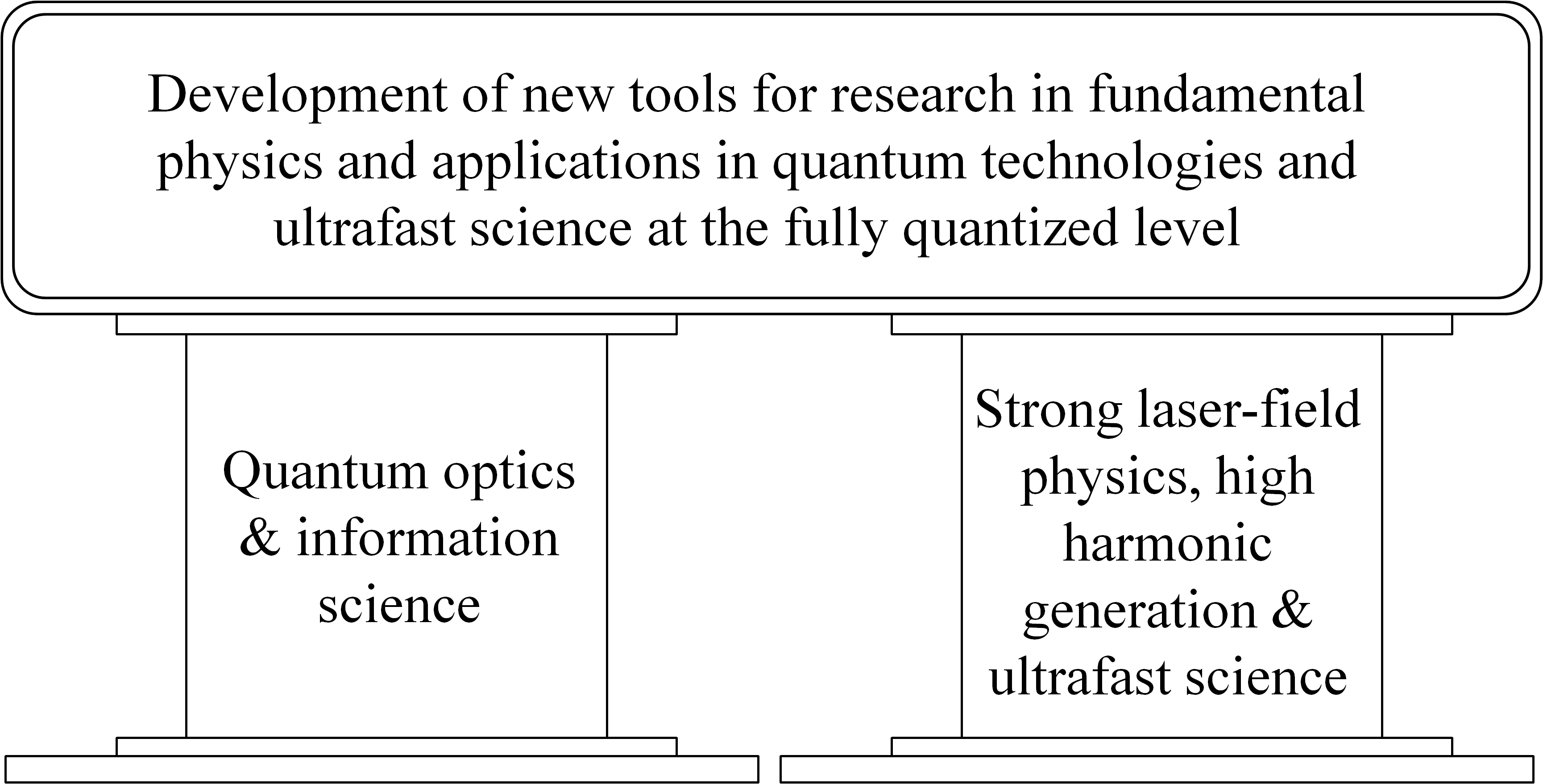}
    \caption{The main pillars and a perspective of the QED in intense laser--matter interaction.}
    \label{fig23:Perspective}
\end{figure}

The synergy between quantum optics, quantum information science and strong laser-field physics aims to merge tools and methods from these fields to create new techniques and approaches for research in fundamental science, as well as novel applications in quantum information and ultrafast science at the fully quantized level (Fig. \ref{fig23:Perspective}) (see examples in refs.~\cite{Tzallas_ROPP_2024, SRM2023, CDF24}).

An immediate outcome of this research field are the recently developing methods where high photon number non--classical light sources have been used in non-linear optics. The importance of this research direction has been highlighted in the past in a mini review article by Th. Lamprou {\it et al.}, in Ref.~\cite{Lamprou2020}. For example, in the recent past, notable achievements have been done on the generation of intense squeezed light sources~\cite{Maria_OL_2016, Spasibko2017, Manceau2019}. Interaction of matter with squeezed light sources leads to a significant enhancement of the rates of the non-linear excitation processes. For example in the case of multiphoton excitation, where for interactions with coherent light states $R_{c}\propto F^q$, for squeezed light states is $R_{s}\propto g^{(q)}F^q$, where $q$ is the non-linearity of the process, $F$ is the photon flux of the driving field and $g^{(q)}=(2q-1)!!$ includes the double-factorial. 

\begin{figure}
    \centering
    \includegraphics[width=1.0 \columnwidth]{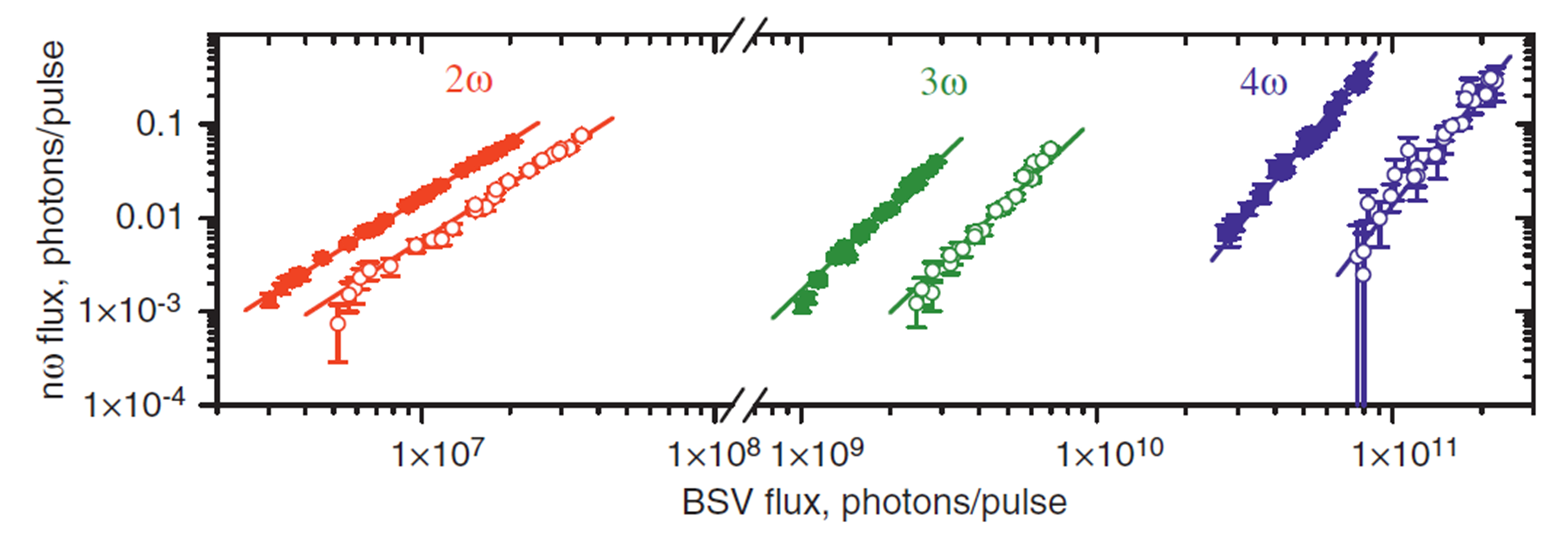}
    \caption{Measured dependence of the 2nd (red), 3rd (green), and 4th (blue) harmonic photon yield on the photon flux with the driving bright squeezed vacuum light field (solid squares) and pseudocoherent light (empty circles). The solid lines show the theoretical fits with the 2nd-, 3rd-, and 4th-order functions corresponding to the power dependence for each harmonic order. The Figure has been reproduced from ref. ~\cite{Spasibko2017}.}
    \label{fig24:BSV_PRL}
\end{figure}

\begin{figure}
    \centering
    \includegraphics[width=1.0 \columnwidth]{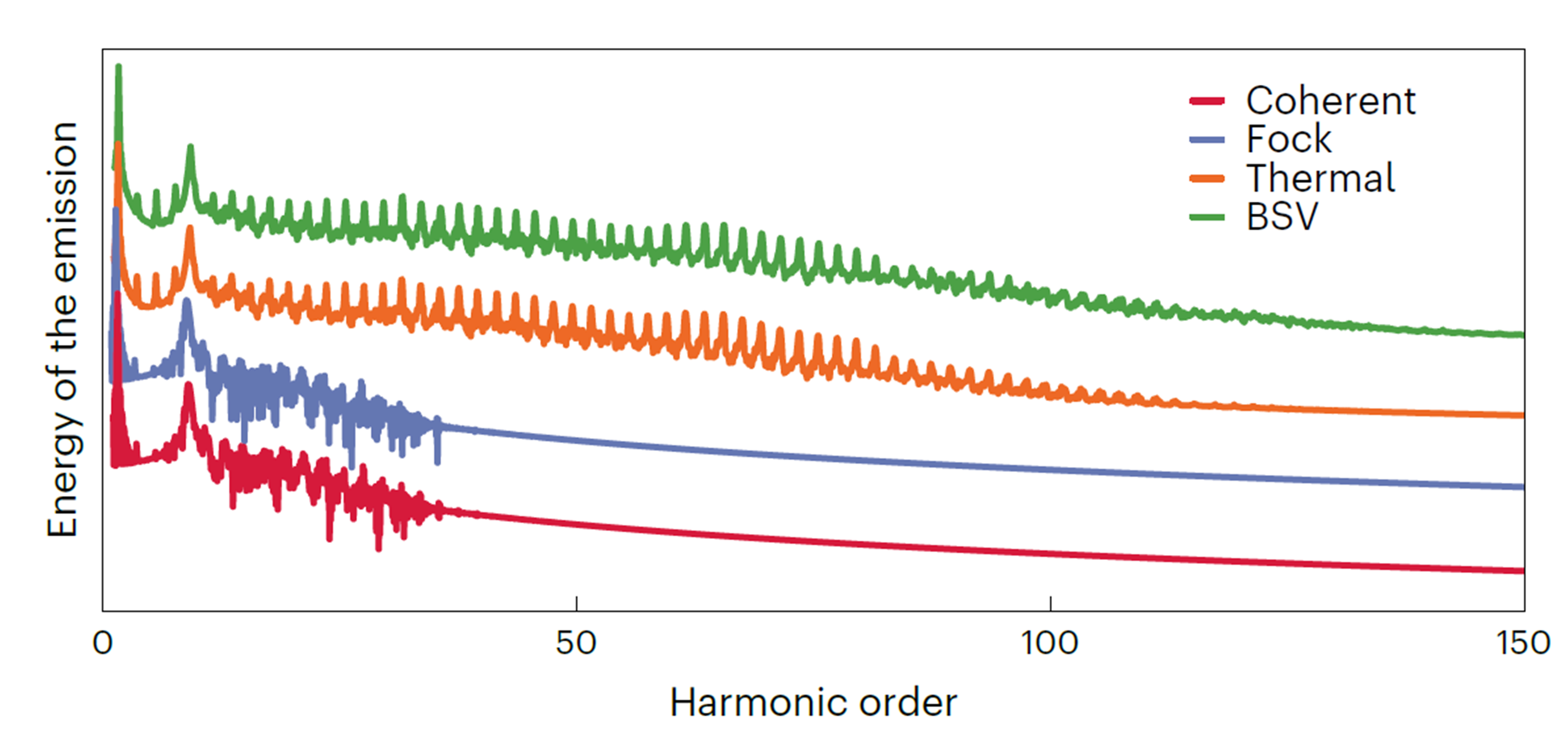}
    \caption{Calculated HHG spectrum in case of using as a driving IR field state a coherent (red line), Fock (blue line), thermal (orange line) and bright squeezed vacuum (green line). The Figure has been reproduced from ref. ~\cite{Ido_NatPhoton_2023}.}
    \label{fig25:HHG_BSV}
\end{figure}

These light sources have been used for the generation of low-order harmonics in an optical crystal \cite{Spasibko2017} (Fig.~\ref{fig24:BSV_PRL}), theoretical and experimental investigations on multiphoton electron emission \cite{Heimerl_NatPhys2024}, and have triggered theoretical investigations concerning the influence of the statistical properties of light in atomic spectroscopy \cite{Mouloudakis2020}.  Also, they have triggered novel theoretical investigations concerning the influence of driving fields different than coherent laser light in the HHG process \cite{Tzur_NatPhot2023, Ido_NatPhoton_2023, Wang_PRR2024, Stammer_NatPhys_2024, stammer2024absence,rivera-dean_non-classicality_2025} and atomic ionization \cite{Liu_PRL_2025}. Fig.~\ref{fig25:HHG_BSV} shows the theoretically calculated dependence of the harmonic spectrum on the photon statistics of the driving IR field. Recent experiments \cite{Vampa_ArXiv_2024, Maria_NatPhys_2024} using bright squeezed vacuum sources in semiconductor solids have shown the generation of harmonics with photon energy above the energy band gap of the material.

\begin{figure}
    \centering
    \includegraphics[width=1.0 \columnwidth]{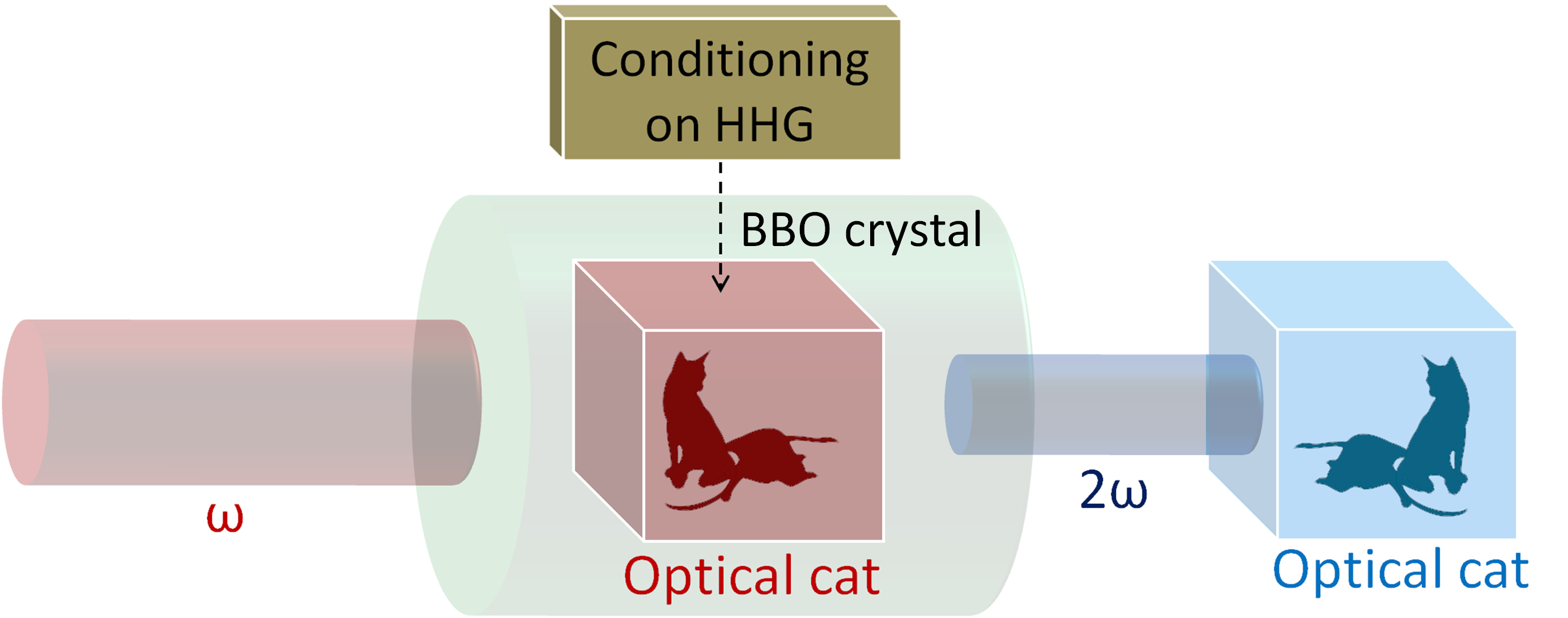}
    \caption{A schematic showing the operation principle used in ref. ~\cite{Lamprou_PRL_2025} for implementing intense IR optical ``cat'' states in non-linear optics. An intense IR optical ``cat'' state of frequency $\omega$ (created via conditioning on HHG) was used to drive the second harmonic ($2\omega$) generation process in an optical BBO crystal through a frequency up-conversion process. The $2\omega$ also depicts the quantum features of an optical ``cat'' state.}
    \label{fig26:NonLinear_CAT}
\end{figure}

Finally, a recent experimental and theoretical achievement \cite{Lamprou_PRL_2025} in high photon-number optical ``cat'' state engineering has brought optical ``cat'' states into the realm of nonlinear optics. In this work, an intense IR optical ``cat'' state (created via conditioning on HHG--see section \ref{Methods6}) was used to drive the second harmonic generation process in an optical crystal through a frequency up-conversion process. This process has been used for quantum state characterization of the optical IR ``cat'' state and the generation of a second harmonic which depicts non-classical features (\PT{Fig.~\ref{fig26:NonLinear_CAT}}). Such findings open new ways in quantum light engineering in different spectral regions.

\section{Challenges}\label{Challenges}

\PT{The aforementioned works constitute an important step towards this novel and thriving research direction. Yet, as in any new research direction, there are questions and issues that need to be addressed. Below, we provide some of those which, in our view, are worth to be mentioned.

\begin{enumerate}[I)]
\item The quantum nature of the HHG outputs, and its potential implementations in quantum information science, depends on finding ways to properly measure the quantum state of the harmonics. This can be achieved by utilizing the correct diagnostic and state characterization methods. Although the traditional methods, such as quantum tomography~\cite{LR09,Leonhardt_book_1997,Bachor_book_2019} and photon correlation measurements~\cite{glauber1963coherent, glauber1963quantum, Glauber2006, HBT56}, are well-known and mature enough for the quantum state characterization of a light field in the IR spectral range, this is not the case for light fields in the VUV-XUV spectral range. The state characterization in this spectral range is challenging as it requires the development of the proper optical elements (such as 50:50 $\%$ beam splitters) and photon detectors with quantum efficiency $>0.7$. Additionally the minimization of decoherence effects is crucial. The development of new methods along the lines of Ref. \cite{Lamprou_PRL_2025} may serve towards this direction.

\item The entanglement certification of a multi-mode light state remains an outstanding challenge in quantum information theory~\cite{gurvits_classical_2004}. Even for bipartite systems of dimension $3 \times 3$, the problem is open: widely used criteria such as the logarithmic negativity presented earlier provide necessary but insufficient conditions for separability~\cite{horodecki2009quantum}. This underscores a need to develop tailored entanglement measures or witnesses that are both theoretically rigorous and experimentally practical~\cite{Huber_EntanglementCertif_NatRevPhys_2019}. Efforts to address this challenge must emphasize as well advances in analytical and numerical methods, especially in the context of strong-field physics where quantum optical states naturally have high-photon numbers, and where potentially non-Gaussian features can emerge~\cite{walschaers_non-gaussian_2021}.
Finally, we note that in general it is not clear how to describe multi-photon (many) quantum states numerically in a tractable manner. However, most of the papers and observations predict generation of the multimode squeezed states, i.e. multimode Gaussian states (i.e., states with the Gaussian Wigner function). For such states there are operational criteria to determine entanglement, and even bound entanglement (i.e., entanglement of states with positive partial transpose) \cite{Giedke1,Giedke2,Giedke3}. These criteria rely on the measurement of the correlation (covariance) matrix, which provides all necessary information. They can also be applied to non-Gaussian states, although there they provide only necessary condition for separability (i.e., absence of entanglement).

\item From a theoretical point of view in the characterization of strong-field physics from a quantum optical perspective, the most instant challenge is to unify various numerical approaches, approximations, and, perhaps most importantly, the relation to experimental results. At this stage many groups in this field use their own techniques, some quantum, some semi-quantum, as well as other levels of approximations. Below, we briefly describe some of this attempts:

\begin{itemize}
\item In initial descriptions (cf.~\cite{LCP21}) discrete modes of EM field were used and a time-dependent coupling constant was introduced, to describe the effect of pulsed interaction. With respect to the electron dynamics, the main approximation was inspired by the SFA~\cite{ lewenstein-theory-1994}: in the final state only contributions from the ground state and continuum states for atoms were considered---with inclusion in some instances of excited states as in molecules~\cite{rivera-dean_quantum_2023}---and ``ground'' valence band states and conductance continuum for simple solids \cite{rivera-dean_bloch_2023}. Under these conditions and within the low depletion limit, the final state of the EM field corresponds to a trivial product of coherent states. The correlations (expressing energy conservation) have to be put by hand (via the wave-packet modes $\dyad{\tilde n}$ introduced above). On the other hand, moving beyond the low depletion limit, this method allowed describing the generation of multimode squeezed states \cite{Stammer_PRL_2024}.

\item Generation of multimode entanglement and squeezing was observed in the experiments of Refs.~\cite{Hamed_PRXQ_2024, Merdji_multimodeSqueezed_Arxiv2024} performed in solid-state systems. In these papers violation of Cauchy-Schwarz inequality was the main tool to certify the presence of quantumness. Such experiments may serve as examples for further investigating multimode entanglement and its dependence on the driving laser field, using certification protocols along the lines of Ref.~\cite{Huber_EntanglementCertif_NatRevPhys_2019}.

\item The works of Refs. \cite{Tzur_PRR2024,  Maria_NatPhys_2024} are focused on HHG driven by bright squeezed light. Here, the theoretical approach employs quasi-probability distributions, such as Glauber’s $P$-distribution, or generalized positive $P$-distribution. While the developed methods have been quite powerful to understand how the HHG spectrum characteristics get affected by quantum light both in atomic~\cite{Kaminer_squeezing,rivera-dean_non-classicality_2025} and solid-state systems~\cite{Maria_NatPhys_2024,Gothelf_ArXiv_2025}, or how it modifies the electronic trajectories~\cite{Tzur_NatPhot2023,rivera-dean_non-classicality_2025}, they typically rely on approximations to simplify the analytical calculations. Furthermore, as noted in Ref.~\cite{Tzallas_NatPhot_2023}, investigations taking into account propagation effects (including multiphoton processes, ionization effects and low order harmonic generation) of the radiation in the HHG medium remain unknown. Such investigations can provide important information about the conversion efficiency of the HHG process and the survival of the quantum features of the harmonics exiting the HHG medium.

\item The work of Ref.~\cite{Moiseyev_ArXivConditioning_2024} describes the IR photon fluctuation after the HHG process using classical Floquet theory. This work constitutes an interesting approach to mimic the QED results using classical Floquet theory. Yet, it does not describe the full QED nature of the process, but makes reliable predictions on the ability to infer on the harmonic spectrum from counting the driving photons. Thus, answering any potential question of ``how much of these IR photon fluctuations can be captured from quantum or classical statistics ?'' remains a challenging task.

\item In the recent work of Ref.~\cite{Rubio_QEDHHG_2025}, the authors analyze a model with a single mode of the electromagnetic field. While this approach requires generalization to many modes, it makes a nice connection of the theory to one of the most studied areas for Quantum Optics, as is cavity QED \cite{Haroche_book_2006, Booklarson}, and also provides novel and efficient numerical tools to describe laser-matter interaction driven by intense quantum light.

\item Studies of HHG in strongly correlated systems have been so far realized only with classical driving fields (cf.~\cite{Alcala_2022, Tyulnev_2024, Zhang_2024}). Here only small systems may be studied with exact diagonalization methods, and other numerical approximate, but feasible methods must be used. We used for instance generalized pairing theory for d-wave superconductors \cite{Alcala_2022}, and/or a simplified phenomenological mean-field model for this material, based on which the HHG spectra is calculated through the time-dependent Schr\"{o}dinger equation \cite{Zhang_2024}. How to generalize these solution to the QED case is an open problem and a challenging task. Recent efforts have been done in this direction~\cite{Lange_PRAe-eCorr_2024} in the context of 1D Fermi-Hubbard models, demonstrating the presence of non-classical light on specific phases of the material. Recently, the authors of this work have studied the same system but using a series of hierarchical approximations~\cite{lange_hierarchy_2025}, providing quantitative insights about the validity of the approximations mentioned thus far when studying strongly correlated systems.

\item In the theoretical study of Ref.~\cite{Liu_PRL_2025}, the authors have investigated the atomic ionization (single and double) in the presence of intense squeezed light field. Atomic double ionization, a fundamental process for investigating electron correlation effects, could be strongly influenced by the quantum state of light. Various quantum states, such as phase-squeezed coherent states and bright squeezed vacuum states, can lead to significant modifications in ionization probabilities and correlated electron dynamics. Exploring these effects would not only enhance our understanding of light-matter interactions but also shed light on the role of quantum fluctuations in atomic and strong-field physics, potentially opening new ways for quantum control of multiple ionization processes.
      
\end{itemize}

\item Theoretical simulations \cite{Stammer_ArXivConditioning_2024} performed under conditions similar to the experiment, have demonstrated how post-selection schemes can generate coherent state superpositions such as those presented in Ref.~\cite{LCP21, RLP22}. These simulations have shown that post-selection can be used to generate high-fidelity ($>99\%$) optical ``cat'' states~\cite{Stammer_EnergyConservation2024}. However, due to experimental limitations in implementing these approaches, the experimental observation of such high-fidelity optical ``cat'' states remains pending. Hence, the conditioning methods from Ref.~\cite{TKG17} require further improvement. Additionally, the simulations of Ref.~\cite{Stammer_ArXivConditioning_2024} highlight the potential of the conditioning method to generate a variety of non-classical light states with distinct quantum features. The experimental observation of the rich manifold of these states remains open.

\end{enumerate}
}

\section{Conclusions}

In this topical review, after introducing the fundamentals of quantum optics, we discussed recent progress in the fully quantized description of intense laser–matter interactions and the approaches used to generate high-photon-number, non-classical, and entangled light states from the far-IR to XUV. We explored how such light states can be generated through the HHG process in strongly laser-driven atoms, molecules, semiconductor solids, and excited media, and how the quantum properties of the light after interaction are related to many-body quantum correlations. Finally, we provided a perspective on this emerging research direction, highlighting applications that arise from the synergy between strong laser-field physics, quantum optics, and quantum information science. As an example, we discussed recent findings concerning the applicability of these light states in non--linear optics. The findings reported here are of general importance, as they open the way to a novel quantum nonlinear spectroscopy, based of the interplay between the quantum properties of light with that of quantum matter. However, as in any newborn research field, there are still open questions and challenges. Here, we discussed some of those which in our view are worth to be mentioned.

\section*{Data availability statement}
No new data were created or analyzed in this study.

\section*{Acknowledgements}
 
P.T. acknowledges the Hellenic Foundation for Research and Innovation (HFRI) and the General Secretariat for Research and Technology (GSRT) under grant agreement CO2toO2 Nr.:015922, the European Union’s HORIZON-MSCA-2023-DN-01 project QU-ATTO under the Marie Skłodowska-Curie grant agreement No 101168628, the LASERLABEUROPE V (H2020-EU.1.4.1.2 grant no.871124) and ELI--ALPS. ELI--ALPS is supported by the EU and co-financed by the European Regional Development Fund (GINOP No. 2.3.6-15-2015-00001). H2020-EU research and innovation program under the Marie Skłodowska-Curie (No. 847517). M.L. acknowledges the Government of Spain (Severo Ochoa CEX2019-000910-S and TRANQI), Fundació Cellex, Fundació Mir-Puig, Generalitat de Catalunya (CERCA program) and the ERC AdG CERQUTE. ERC AdG NOQIA; MCIN/AEI (PGC2018-0910.13039/501100011033, CEX2019-000910-S/10.13039/501100011033, Plan National FIDEUA PID2019-106901GB-I00, STAMEENA PID2022-139099NB, I00, project funded by MCIN/AEI/10.13039/501100011033 and by the EU Next Generation EU/PRTR (PRTRC17.I1), FPI). QUANTERA MAQS PCI2019-111828-2; QUANTERA DYNAMITE PCI2022-132919, QuantERA II Programme co-funded by H2020-EU program (No 101017733); Ministry for Digital Transformation and of Civil Service of the Spanish Government through the QUANTUM ENIA project call-Quantum Spain project, and by the EU through the Recovery, Transformation and Resilience Plan—Next Generation EU within the framework of the Digital Spain 2026 Agenda; Fundació Cellex; Fundació Mir-Puig; Generalitat de Catalunya (European Social Fund FEDER and CERCA program, AGAUR Grant No. 2021 SGR 01452, QuantumCAT \textbackslash{} U16-011424, co-funded by ERDF Operational Program of Catalonia 2014-2020); Barcelona Supercomputing Center MareNostrum (FI-2023-1-0013) funded by the EU. Views and opinions expressed are however those of the author(s) only and do not necessarily reflect those of the EU, European Commission, European Climate, Infrastructure and Environment Executive Agency (CINEA), or any other granting authority. Neither the EU nor any granting authority can be held responsible for them (EU Quantum Flagship PASQuanS2.1, 101113690, EU H2020 FET-OPEN OPTOlogic (No 899794), EU Horizon Europe Program (No 101080086-NeQST); ICFO Internal “QuantumGaudi” project; EU H2020 program under the Marie Sklodowska-Curie (No. 847648); “La Caixa” Junior Leaders fellowships, La Caixa” Foundation (ID 100010434): CF/BQ/PR23/11980043. P.S. acknowledges funding from the European Union’s Horizon 2020 research and innovation programme under the Marie Skłodowska-Curie grant agreement No 847517. M. F. C. acknowledges the National Key Research and Development Program of China (Grant No.~2023YFA1407100), the Guangdong Province Science and Technology Major Project (Future functional materials under extreme conditions - 2021B0301030005) and the Guangdong Natural Science Foundation (General Program project No. 2023A1515010871).

\bibliography{JPB.bib}

\end{document}